\documentclass[preprint,showpacs,titlepage,aps,prd,tightenlines,amsmath,byrevtex,nofootinbib]{revtex4-1}

\usepackage{graphicx}
\usepackage{epsfig}
\usepackage{multirow}

\def\lsim{\raise0.3ex\hbox{$\;<$\kern-0.75em\raise-1.1ex
\hbox{$\sim\;$}}}
\def\gsim{\raise0.3ex\hbox{$\;>$\kern-0.75em\raise-1.1ex
\hbox{$\sim\;$}}}

\newcommand{\epsetau}{$\epsilon_{e\tau}\!$ }
\newcommand{\epsmutau}{$\epsilon_{\mu\tau}\!$ }
\newcommand{\epstautau}{$\epsilon_{\tau\tau}\!$ }
\newcommand{\epsemu}{$\epsilon_{e\mu}\!$ }
\newcommand{\epsmumu}{$\epsilon_{\mu\mu}\!$ }
\newcommand{\epsee}{$\epsilon_{ee}\!$ }

\newcommand{\phimutau}{$\phi_{\mu\tau}\!$ }
\newcommand{\phiemu}{$\phi_{e\mu}\!$ }
\newcommand{\phietau}{$\phi_{e\tau}\!$ }

\newcommand{\be}{\begin{equation}}
\newcommand{\ee}{\end{equation}}
\newcommand{\bea}{\begin{eqnarray}}
\newcommand{\eea}{\end{eqnarray}}
\newcommand{\e}{\epsilon}

\newcommand{\ba}{\begin{array}}
\newcommand{\ea}{\end{array}}

\newcommand{\si}{s_{12}}

\newcommand{\st}{s_{13}}
\newcommand{\ci}{c_{12}}

\begin{document}
\thispagestyle{empty}
\begin{flushright}
{IFT-UAM/CSIC-11-30} \\
{IFIC-UV/CSIC-11-25} \\
{EURONU-WP6-11-33}\\
{IPPP/11/24}\\
{DCPT/11/48}
\end{flushright}
\vspace*{1cm}

\begin{center}
{\Large{\bf  Non-Standard Interactions at a Neutrino Factory: \\ Correlations and CP violation} }\\
\vspace{.5cm}

P.~Coloma$^{\rm a,b}$, A.~Donini$^{\rm b,c}$, J.~L\'opez-Pav\'on$^{\rm d}$ and H.~Minakata$^{\rm e}$ \\
\vspace*{1cm}
$^{\rm a}$ Dep. F\'{\i}sica Te\'{o}rica, Universidad Aut\'onoma de
Madrid, 28049 Madrid, Spain \\
$^{\rm b}$ I.F.T., Universidad Aut\'onoma de Madrid/CSIC, 28049 Madrid,
Spain \\
$^{\rm c}$I.F.I.C., Universitat de Valencia/CSIC, 46071 Valencia, Spain
\\
$^{\rm d}$ I.P.P.P., Department of Physics, Durham University, South Road, \\ Durham, DH1 3LE, U.K. \\
$^{\rm e}$ Department of Physics, Tokyo Metropolitan University, Hachioji, Tokyo 192-0397, Japan \\
\end{center}

\vspace{.3cm}
\begin{abstract}
\noindent

We explore the potential of several Neutrino Factory (NF) setups to
constrain, discover and measure new physics effects due to
Non-Standard Interactions (NSI) in propagation through Earth matter. We
first study the impact of NSI in the measurement of
$\theta_{13}$: we find that these could be large due to strong correlations of
$\theta_{13}$ with NSI parameters in the golden channel, and the
inclusion of a detector at the magic baseline is crucial in order to
reduce them as much as possible. We present, then, the sensitivity of
the considered NF setups to the NSI parameters, paying special
attention
to correlations arising between them and the standard oscillation
parameters, when all NSI parameters are introduced at once. 
Off-diagonal NSI parameters could be tested down to the level of $10^{-3}$, whereas
the diagonal combinations $(\epsilon_{ee} - \epsilon_{\tau\tau})$ and $(\epsilon_{\mu\mu}-\epsilon_{\tau\tau})$
can be tested down to $10^{-1}$ and $10^{-2}$, respectively. 
The possibilities of observing CP violation in this context are also
explored, by presenting a first scan of the CP
discovery potential of the NF setups to the phases $\phi_{e\mu}, \phi_{e
\tau}$ and $\delta$.
We study separately the case where CP violation comes only from
non-standard sources, and the case where it is entangled
with the standard source, $\delta$. In case $\delta$ turns out to be CP
conserving, the interesting possibility of
observing CP violation for reasonably small values of the NSI
parameters emerges.

\end{abstract}

\pacs{14.60.Pq,14.60.Lm}

\maketitle

\newpage

\section{Introduction}
\label{sec:intro}

From the results of neutrino oscillation experiments 
\cite{atm-review,solar-review,reactor-review} we now know that neutrinos 
have masses and there is flavor mixing in the lepton sector \cite{MNS}. 
However, the leptonic mixing is still not completely understood. In the three-family $\nu$-mass enriched Standard Model ($\nu$SM), 
there are still three unknown oscillation parameters:  $\theta_{13}$, the CP-violating phase $\delta$
and the mass hierarchy $\textrm{sgn}(\Delta m_{31}^2)$. 
Several reactor  \cite{DCHOOZ,Daya-Bay,RENO} and accelerator \cite{T2K,NOVA}  neutrino experiments 
are currently running, or will start running in the near future, to search for positive signals of non-zero 
$\theta_{13}$. Strategy for exploration of the remaining two unknown parameters in the leptonic mixing, 
$\delta$ and the mass hierarchy, heavily depends upon whether they succeed or fail to detect non-zero $\theta_{13}$. 

If $\sin^2 2\theta_{13} \lsim 10^{-2}$, the ongoing and the near future experiments would face difficulties
 to observe clear evidence for $\theta_{13}$. 
In this case, a new generation of experiments will be needed, beyond doubt, 
to determine the remaining unknowns in the leptonic mixing. 
The candidates for such facilities are the High-Energy Neutrino Factory (HENF) \cite{Geer,De-Rujula}, 
the Beta-Beam \cite{Zucchelli,Bouchez}, and upgraded conventional beams (\textit{a.k.a.} SuperBeams; see, for example, Refs.~\cite{T2K,MEMPHYS,LBNE-DUSEL}).
The main motivation for a facility as ambitious and technologically demanding as the HENF is commonly considered to be the search for a very small $\theta_{13}$ 
($\sin^2 2\theta_{13} \lsim 10^{-4}$) \cite{Cervera:2000kp} and, consequently, the possibility to measure $\delta$ and the mass hierarchy in this regime, something beyond the reach of both Beta-Beams and Super-Beams \cite{Bandyopadhyay:2007kx}. 

On the contrary, if $\theta_{13}$ is relatively large so that the forthcoming experiments are able to see its effects, alternative strategies and scenarios for further exploration of the leptonic mixing could be possible. Such a possibility of $\theta_{13}$ being of the order of the Chooz limit \cite{CHOOZ,Palo-Verde,K2K-bound,MINOS-bound} has been recently suggested by some global analyses \cite{Fogli:2008jx,Schwetz:2008er,Maltoni:2008ka,GonzalezGarcia:2010er,Schwetz:2011qt,KamLAND-13,Abe:2010hy} but it is still controversial.
If confirmed, possible strategy and scenarios for further exploration of lepton mixing would have to be changed. A possibility is that facilities such as upgraded SuperBeams or Beta-Beams could be exploited to measure the standard oscillation parameters. Another possibility is 
that different designs for the Neutrino Factory (NF)  can be exploited in order to achieve the same goal. Such a scheme is, for example, represented by the so-called Low-Energy Neutrino Factory \cite{nufact-lowE1,nufact-lowE2}. 
However, the physics case for a HENF scheme can be well motivated even in the case of a relatively large $\theta_{13}$, given its enormous 
accuracy. In particular, the HENF could be re-designed in order to look for both the $\nu$SM parameters and possible New Physics (NP) beyond $\nu$SM. 
These include, for example, Non-Standard neutrino Interactions (NSI) \cite{Wolfenstein,Valle,Guzzo,Grossman,Berezhiani}, 
effects of non-unitarity of the leptonic mixing matrix \cite{Antusch:2006vwa,FernandezMartinez:2007ms}, 
and new phenomena due to light sterile neutrinos \cite{Pontecorvo:1967fh}. 
Notice that, for a relatively small $\theta_{13}$ ($\sin^2 2\theta_{13} \lsim 10^{-2}$), the NP parameters discussed above can spoil the sensitivity of the HENF
to standard oscillation parameters, a possibility that must be studied.
Recent works dedicated to the study of the potential of HENF in this context include, for example, 
Refs.~\cite{Kopp:2007mi,Ribeiro:2007ud,Kopp:2008ds,Winter:2008eg,Gago:2009ij,Goswami:2008mi,Donini:2008wz}.
For an extensive list of remaining references see, {\em e.g.}, Ref.~\cite{Minakata:2009gh}. 

In the present work, we will attempt
a complete study of the NSI effects in neutrino propagation.
Since this kind of NSI can be regarded as a new ``effective matter potential", a HENF 
with very long baselines seems to be the optimal facility to study their effects. 
Due to an advanced minimum-searching algorithm in multi-parameter space, 
the MonteCUBES software \cite{Blennow:2009pk}, we are able to include, for the first time, 
all NSI parameters in propagation (including their associated CP-violating phases) at the same time into our analysis. 
Therefore, we will pay special attention to correlations appearing when the whole set of NSI parameters is introduced at once in the simulations, a problem that has never 
been addressed in the literature. First, we will study the potential effect of these NSI 
in the measurement of the standard parameters $\theta_{13}$ and $\delta$. Then, we will examine the sensitivity to the NSI parameters, studying their correlation with the whole set of parameters, standard and NSI ones. Finally, we will explore the new avenues of CP violation coming from NSI and the relation among the different CP-phases involved.

Intensive efforts have been devoted to find out which of the proposed future neutrino oscillation facilities is the best option. The {\em International Scoping Study} (ISS) ended 
with the publication of three reports on the status of HENF and their competitors in 2007~\cite{Bandyopadhyay:2007kx,ISS-detector,ISS-accelerator}. 
Currently, the study is being continued by the {\em International Design Study for a Neutrino Factory} (IDS)~\cite{IDS}, which has proposed a ``baseline" HENF setup which optimizes the performance of the NF for measuring the $\nu$SM parameters: it consists of a neutrino beam obtained from the decay of 25 GeV muons aimed at two magnetized iron neutrino detectors (MIND) located at 4000 km and 7500 km. This setup makes use of two channels:  the ``golden channel'' \cite{Cervera:2000kp}, $\nu_{e} \rightarrow \nu_{\mu}$, and the muon disappearance channel $\nu_\mu \to \nu_\mu$, together with their CP conjugate ones. We will study the sensitivities than can be achieved at this particular setup in the context of NSI. 

As stressed before, however, new HENF designs can be explored to optimize searches for NP beyond the $\nu$SM parameters. In particular, higher neutrino energies turn out to be better in order to study effects of NSI in propagation \cite{Ribeiro:2007ud}, since these can be regarded as a kind of generalized matter effect. We will also study, therefore, a simple modification of the setup described above by increasing the parent muon energy to 50 GeV. 
In the context of NSI sensitivity studies, a quite similar setup was previously examined in Refs.~\cite{Ribeiro:2007ud,Kopp:2008ds,Winter:2008eg,Gago:2009ij}. 

In this paper, we also study a different setup with a composite detector consisting of a MIND and a Magnetized Emulsion Cloud Chamber (MECC) located at 4000 km from the source, 
with doubled statistics with respect to IDS-inspired proposals in which each of the two baselines receives half of the available neutrino flux. The MECC is an enlarged and more sophisticated version of the emulsion detector used in the ongoing OPERA experiment \cite{OPERA}, which aims at detecting $\tau$ particles with much higher efficiency. 
Using this setup, it is possible to measure the NSI parameters using two additional channels: $\nu_{e} \rightarrow \nu_{\tau}$, the ``silver channel'' \cite{Donini:2002rm}, and ${\nu}_{\mu} \rightarrow {\nu}_{\tau}$, the ``discovery channel''~\cite{Donini:2008wz}. In order to overcome the strong suppression due to the small $\nu_\tau N$ cross-section, a parent muon energy of 50 GeV is adopted also in this case. 
Notice that for $\nu$SM parameter searches, it was shown that this setup is outperformed by the baseline IDS setup defined above (see Refs.~\cite{Huber:2006wb,Bandyopadhyay:2007kx}). In Ref.~\cite{Kopp:2008ds} this analysis was extended to the case of NSI in propagation when one NSI parameter is turned on at a time, with similar results (see, also,
Ref.~\cite{Bernabeu:2010rz} for a summary on the performance of this setup in these cases). We want to check in this paper if these results hold in the case in which all NSI 
parameters are turned on simultaneously, \textit{i.e.} in a case in which correlations between $\nu$SM and NSI parameters are taken into account.

The paper is organized as follows. In section \ref{sec:NSI}, the formalism for NSI is presented, and the main dependences of the probabilities on the parameters are introduced; in section \ref{sec:statandsetups} we introduce the statistical approach we have used, and the details for the three NF-based setups we have studied; section \ref{sec:golden} is dedicated to study the sensitivities to $\theta_{13}$ (in presence of NSI) and to \epsemu and \epsetau, which are achieved mainly through the $P_{e\mu} $ and $P_{e\tau}$ channels; section \ref{sec:disappearance} is devoted to the study of \epsmutau and $\epsilon_{\alpha\alpha}$, whose sensitivities are achieved mainly through the $P_{\mu\mu}$ channel; in section~\ref{sec:CP} we study the CP discovery potential of the three setups in presence of NSI; and finally, we conclude in section~\ref{sec:conclusions}.
In the Appendix we show approximate expressions for the oscillation probabilities $P_{e\mu},P_{e\tau},P_{\mu\mu}$ and $P_{\mu\tau}$ in matter with constant density.

\section{Neutrino Oscillations with Non-Standard Interactions}
\label{sec:NSI}

NSI can be studied from a top-down or a bottom-up approach. In the top-down approach, a given NP model is studied, and the corresponding 
set of low-energy effective operators are derived systematically. Being these operators derived from  a fundamental theory, their coefficients are related and (usually) 
stringent bounds exist between them. 
In the bottom-up approach, on the other hand, all the effective four-fermion operators which can 
affect neutrino oscillations are included in the analysis, and the experimental 
bounds are used to constrain their coefficients independently of the model. This latter approach is clearly model 
independent, in the sense that no assumption is made regarding the model of NP behind and, consequently, 
the possible relations between operators. Bounds obtained in this way are looser than in the 
top-down approach, but apply to a wide variety of high-energy extensions of the Standard Model.

In the bottom-up approach, NSI that modify neutrino production, propagation and detection processes must be included~\cite{Grossman,GonzalezGarcia:2001mp,Ota:2001pw,Huber:2002bi}. Such a large number of new parameters in the analysis, however, makes it extremely difficult to extract any useful information from the results. 
It is a standard strategy, thus, to separate the study of NSI in neutrino propagation in matter from the NSI effects in production and detection. 
The latter can be studied using near detectors~\cite{Tang:2009nm,Alonso:2010wu}\footnote{
Notice, however, that near detectors will necessarily put bounds on the combination of NSI in production and detection processes. Several near detectors with different sources and/or target materials could be used to disentangle them (see Ref.~\cite{Antusch:2010fe}).
}.
This is the strategy that we follow in this paper. We introduce all the operators that modify neutrino propagation in matter at once, adding nine new parameters 
(six moduli and three phases)
to the existing $\nu$SM parameter space. A complete analysis with such a huge number of parameters 
is  extremely demanding, from both the numerical and analytical point of view. For this reason the
effects of NSI in matter propagation have been widely explored in the literature 
turning on only one new parameter in the analysis, or two at most (one modulus and one phase)~\cite{Kopp:2008ds,Winter:2008eg,Gago:2009ij}. 
In this work, we try to achieve two main goals: 
(1) We attempt a first complete phenomenological analysis of the potential of HENF to constrain all the NSI parameters which can contribute to propagation in matter; 
(2) We illuminate complicated correlations between the effects of NSI and $\nu$SM CP violating phases. 
We believe that the complete treatment of effects of NSI in matter propagation
is an important step toward a better understanding of possible physics beyond the $\nu$SM. 

In the model independent approach, constraints on NSI parameters in propagation are very mild, generically at $\mathcal{O}(10^{-1})$ or even order unity~\cite{Davidson:2003ha,Biggio:2009nt}. However, from the theoretical point of view, such large values of the NSI parameters are not really expected. 
This is easily understood if one tries to find a model of NP responsible for NSI effects without enlarging the low-energy SM particle content. 
Effects of NP at high energies manifest at low energies through an infinite tower of non-renormalizable effective operators of dimension $d>4$ which are invariant under the SM gauge group. 
These are weighted by inverse powers of the NP scale $\Lambda$:
\bea
\label{L}
{\cal L}^{eff}={\cal L}_{SM}\,+\,\frac{1}{\Lambda}\,{\delta \cal L}^{d=5}\,+\,\frac{1}{\Lambda^2}\,{\delta \cal L}^{d=6}+\, \dots \, ,
\eea
where ${\cal L}_{SM}$ is the SM Lagrangian which contains all $SU(3)_c\times SU(2)_L \times U_Y(1)$ invariant operators of dimension $d\le4$. The factors $1/\Lambda^{d-4}$ 
appear to suppress the effective operators which generate neutrino masses and produce NSI effects at low energy. 
Moreover, the necessary requirement of gauge invariance of the new operators under the SM gauge group~\cite{Berezhiani,Davidson:2003ha,Gavela:2008ra,Antusch:2008tz,Biggio:2009kv} 
leads to another remarkable point: the effective operators which generate neutrino NSI may be tightly related to their analogues  in the charged lepton sector, 
which are much more constrained experimentally. 
Therefore, it is very hard to construct a feasible model giving large NSI effects in neutrino oscillations.

Following the model independent approach, NSI in neutrino propagation (from here on, we will refer to them simply as NSI)
are described through the inclusion of the following four fermion effective operators:
\begin{eqnarray}\label{eq:L}
\delta {\cal L}_{\rm NSI} = -2\sqrt{2}\,G_F
\sum_{f,P}\varepsilon^{fP}_{\alpha\beta} \left(
\overline{\nu_\alpha}\gamma^\mu P_L \nu_\beta \right) \left(
\overline{f}\gamma_\mu P f \right) \ ,
\end{eqnarray}
where $G_F$ is the Fermi constant, $f$ stands for the index running over fermion species in the
Earth matter, $f = e, u, d$, $P$ stands for the projection operators $P_L \equiv\frac{1}{2}(1-\gamma_5)$ or $P_R \equiv \frac{1}{2}(1 + \gamma_5 )$, and $\alpha,\beta = e, \mu, \tau$. 
Notice that from neutrino oscillations we have no information on the separate contribution of 
a given operator with coefficient $\varepsilon_{\alpha\beta}^{fP}$, but only on their sum over flavours and chirality.
The effects of these operators appear in the neutrino evolution equation, in the flavour basis\footnote{
If production or detection NSI were present, though, the effective production and detection flavour 
eigenstates would not coincide with the standard flavour ones~\cite{Langacker:1988up}.
}, 
as:
\begin{eqnarray} 
i \frac{d}{dt} \left( \begin{array}{c} 
                   \nu_e \\ \nu_\mu \\ \nu_\tau 
                   \end{array}  \right)
 = \left[ U \left( \begin{array}{ccc}
                   0   & 0          & 0   \\
                   0   & \Delta_{21}  & 0  \\
                   0   & 0           &  \Delta_{31}  
                   \end{array} \right) U^{\dagger} +  
                  A \left( \begin{array}{ccc}
            1 + \epsilon_{ee}     & \epsilon_{e\mu} & \epsilon_{e\tau} \\
            \epsilon_{e \mu }^*  & \epsilon_{\mu\mu}  & \epsilon_{\mu\tau} \\
            \epsilon_{e \tau}^* & \epsilon_{\mu \tau }^* & \epsilon_{\tau\tau} 
                   \end{array} 
                   \right) \right] ~
\left( \begin{array}{c} 
                   \nu_e \\ \nu_\mu \\ \nu_\tau 
                   \end{array}  \right)\, ,
\label{evolution_equation}
\end{eqnarray}
where $\Delta_{ij}=\Delta m^2_{ij}/2E$, $U$ is the lepton flavor mixing matrix,
$A\equiv 2 \sqrt 2 G_F n_e$ and $\epsilon_{\alpha\beta} \equiv (1/n_e) \sum_{f,P} n_f 
 \varepsilon_{\alpha\beta}^{fP}$, with $n_f$ the $f$-type fermion number density. 
The three diagonal entries of the modified matter potential are real parameters. Only two of them affect neutrino oscillations: we will consider
the combinations $\epsilon_{ee}- \epsilon_{\tau\tau}$ and $\epsilon_{\mu\mu}- \epsilon_{\tau\tau}$, subtracting $\epsilon_{\tau\tau} \times \bf{I}$ from the Hamiltonian. 
The three complex NSI parameters $\epsilon_{e\mu},\epsilon_{e\tau}$ and $\epsilon_{\mu\tau}$ will be parametrized as
\footnote{This is the prescription used in the MonteCUBES software. In the section devoted to CP violation, though, the prescription is precisely the opposite, $\epsilon_{\alpha\beta} \equiv |\epsilon_{\alpha\beta}| e^{i \phi_{\alpha\beta}}$. 
} 
$\epsilon_{\alpha\beta} = |\epsilon_{\alpha\beta}| e^{-i \phi_{\alpha\beta}}$.

In order to understand the impact of different NSI parameters in various oscillation channels it is useful to obtain approximate analytical expressions for the oscillation probabilities. 
In Ref.~\cite{Kikuchi:2008vq} approximate formul\ae ~were derived for all the oscillation probabilities up to order $\varepsilon^2$ ($\varepsilon^3$ for the golden channel) by making a perturbative expansion in $\Delta m^2_{21}/\Delta m^2_{31} \equiv \varepsilon$ and $\epsilon_{\alpha\beta} \sim \theta_{13} \sim \varepsilon$. 
In the Appendix we present the approximate expressions for $P_{e\mu},P_{e\tau},P_{\mu\mu}$ and $P_{\mu\tau}$ up to second order in $\varepsilon$
expanding in $\Delta m^2_{21}/\Delta m^2_{31}, \epsilon_{\alpha\beta}, \theta_{13}$ and $\delta\theta_{23}\equiv \theta_{23}-\pi/4$, too.

Let us review very briefly the main conclusions which can be extracted from this analytical study: 
\begin{itemize}
\item 
Up to second order in $\varepsilon$, $P_{e\mu}$ and $P_{e\tau}$ depend only on \epsemu and \epsetau but not on the rest of the NSI parameters 
(see Eqs.~\ref{Pemu} and~\ref{Petau} in the Appendix). The precise determination of these two NSI elements is only possible in the golden and the silver channels. 
However, it is well-known that already in the $\nu$SM  case a combination of data, either from different oscillation channels or different baselines, is needed in order to avoid the well known degeneracy problem~\cite{BurguetCastell:2001ez,Minakata:2001qm,Fogli:1996pv}. 
This problem is even more difficult to solve in presence of NSI, because four extra parameters (2 moduli and 2 phases) appear 
simultaneously in the golden and silver channels and severe correlations are expected to exist, not only between NSI parameters but also between them and the $\nu$SM ones.
\item 
The $P_{\mu\mu}$ and $P_{\mu\tau}$ oscillation probabilities show a leading ${\cal O}(\varepsilon)$ dependence on the real part of $\epsilon_{\mu\tau}$ 
(which provides a very high sensitivity to this parameter), in addition to the usual quadratic dependence on $\epsilon_{e\mu}$ and $\epsilon_{e\tau}$ as in the golden and silver channels. On the other hand, the sensitivity to the imaginary part of $\epsilon_{\mu\tau}$ is expected to be much worse, since it comes only through ${\cal O}(\varepsilon^2)$ terms in the probability.
The dependence on the diagonal combination $(\epsilon_{\mu\mu}-\epsilon_{\tau\tau})$ appears in $P_{\mu\mu}$ and $P_{\mu\tau}$ at ${\cal O}(\varepsilon^2)$, too. 
Terms proportional to $\delta \theta_{23} (\epsilon_{\mu\mu}-\epsilon_{\tau\tau})$ lead to important correlations between these two parameters. 
\item 
The dependece on $(\epsilon_{\mu\mu}-\epsilon_{\tau\tau})$ and 
$\epsilon_{\mu\tau}$ in $P_{\mu\mu}$ and $P_{\mu\tau}$ is the same. Therefore, the $\nu_\mu \to \nu_\tau$ channel may be useful only because it adds further statistics at the detector. However, the sensitivities to  $\epsilon_{\alpha\alpha}$ and $\epsilon_{\mu\tau}$ are not limited by statistics, since the disappearance channel alone already provides enough events at the detector. As a consequence, the sensitivities to these parameters are mainly achieved through the $P_{\mu\mu}$ channel.
\item 
The dependence on the diagonal combination $(\epsilon_{ee}-\epsilon_{\tau\tau})$ appears at 
third order in $\varepsilon$ in the oscillation probabilities. Therefore, it is hard to expect a good
 sensitivity to this parameter. Moreover, as we can see in Eq.~(\ref{evolution_equation}), 
when all NSI parameters vanish except for the combination $(\epsilon_{ee}-\epsilon_{\tau\tau})$, 
$A\,(\epsilon_{ee}-\epsilon_{\tau\tau})$ can be interpreted as a small perturbation on the 
standard $\nu$SM matter effect. Therefore, our sensitivity to $(\epsilon_{ee}-\epsilon_{\tau\tau})$ 
will be ultimately limited by uncertainties of the earth matter density. 

\end{itemize}

In view of the features listed above, in the following we are going to distinguish two different groups of oscillation parameters: 
(i) $\theta_{13}$, $\epsilon_{e\mu}$ and $\epsilon_{e\tau}$, that will be studied in Sec.~\ref{sec:golden}, and
(ii) $(\epsilon_{ee}-\epsilon_{\tau\tau})$, $(\epsilon_{\mu\mu}-\epsilon_{\tau\tau})$ and $\epsilon_{\mu\tau}$, that will be studied in Sec.~\ref{sec:disappearance}. 
This classification is the natural consequence of the fact that, in practice,  $P_{e\mu}$ and $P_{e\tau}$ are sensitive to (i), while the sensitivity to (ii) comes 
mainly from $P_{\mu\mu}$.  
The only possible exception to this classification is $(\epsilon_{ee}-\epsilon_{\tau\tau})$, for which the golden channel also plays an important role. Because of this structure, strong correlations between (i) and (ii) are not expected, as it has indeed been found in our
numerical simulations. For this reason, we will study in Sec.~\ref{sec:CP}  the CP discovery potential of the HENF for all the parameters belonging to (i) simultaneously (as strong correlations are expected between them), whilst neglecting parameters belonging to (ii).

\section{The statistical approach and the setup}
\label{sec:statandsetups}

In this section, we first introduce the statistical approaches used to perform the numerical analyses of Secs.~\ref{sec:golden} and \ref{sec:disappearance} (see Sec.~\ref{subsec:montecubes}),
and of Sec.~\ref{sec:CP} (see Sec.~\ref{subsec:CPfrequentist}).
After that, we recall the input values for the $\nu$SM parameters and the current bounds on the NSI parameters (that are included as priors in the algorithm) and describe
the marginalization procedure (Sec.~\ref{subsec:Mprocedure}). Eventually, we define the three HENF experimental setups which we are going to study (Sec.~\ref{subsec:NF-setups}).

\subsection{The statistical procedure used in Secs.~\ref{sec:golden} and \ref{sec:disappearance}}
\label{subsec:montecubes}

It is well-known that, in order to sample a $N$-dimensional parameter space through $\chi^2$ grids with $n$ samplings per parameter, a total of $\mathcal{O}(n^N)$ evaluations of the expected number of events are required. When only  three-family oscillations are considered, the computation can become heavy (if all $\nu$SM parameters are taken into account) 
but is still affordable within the standard frequentist approach. When the NSI parameters are also taken into account, however, the number of parameters to be fitted simultaneously increases considerably and the computation time required to perform the standard minimization procedure becomes too large. A different approach must therefore be used
if we want to sample a huge number of parameters with limited computational resources. The way out is suggested by noticing that most of the points belonging to the $\chi^2$ grids that are computed in the standard approach are useless, as they are very far from the $\chi^2$ minimum. For this reason the standard technique used to 
sample large multi-dimensional manifolds is to rely on efficient (either deterministic or stochastic) algorithms that search for the global minimum and then start to sample the 
region near the minimum to determine its size and shape. Most of the algorithms used fall into the category of Markov Chain Monte Carlo (MCMC): using these class of 
algorithms, the number of evaluations required for the algorithm to converge and sample properly the desired distribution grows polynomially with $N$, ${\cal O}(N^k)$, with $k$ some integer. We have followed this approach to scan the NSI parameter space in Secs.~\ref{sec:golden} and \ref{sec:disappearance}, using the MonteCUBES  (``Monte Carlo Utility Based Experiment Simulator") software \cite{Blennow:2009pk} that contains a C library plug-in to implement 
MCMC sampling into the GLoBES~\cite{globes} package. It, thus, benefits from the flexibility of GLoBES in defining different experiments while implementing an efficient scanning of large parameter spaces. 

Parameter determination through MCMC methods are based on Bayesian inference. The aim is to determine the probability distribution function of the different model parameters $\theta$ given some data set $d$, \textit{i.e.}, the \emph{posterior} probability $P(\theta\mid d)$. From Baye's theorem we have:
\begin{equation} 
\label{eq:bayes}
{\cal P} = P(\theta\mid d) = \frac{P(d\mid \theta)P(\theta)}{P(d)}\equiv\frac{L_d(\theta)\pi(\theta)}{M} \, .
\end{equation}
The \emph{likelihood} $L_d(\theta)=P(d\mid\theta)$ is the probability of observing the data set $d$ given certain values of the parameters $\theta$. The prior $\pi(\theta)=P(\theta)$ is the probability that the parameters assume the value $\theta$ regardless of the data $d$, that is, our previously assumed knowledge of the parameters. Finally, the \emph{marginal} probability $M$ is the probability $P(d)$ of measuring the values $d$. It does not depend on the parameters $\theta$, and therefore it can be regarded as a normalization constant\footnote{In the limit of infinite statistics, it can be shown that the Bayesian probability distribution is maximized by the same set of parameters $\theta$ that minimize the $\chi^2$ function in the 
frequentist approach.}. Notice that the $\chi^2$ functions defined in GLoBES provide the logarithm of the likelihood of the data $d$ following a Poisson distribution normalized to the distribution with mean $d$. Therefore, the actual probability density sampled by MonteCUBES is the posterior probability $P(\theta \mid d)$:
\[ \mathcal{P}= \mathrm{exp} \left [-\frac{\chi^2(\theta)}{2} \right ] \mathrm{exp} \left [ -\frac{\chi_P^2(\theta)}{2} \right ] ,   \]
where $\chi_P^2(\theta)=-2 \mathrm{ln}\pi(\theta)$. This probability distribution is equivalent to a Boltzmann weight with temperature $T = 1$ and energy $E = \chi^2 (\theta) + \chi^2_P (\theta)$.

We have used ten MCMC chains in all our simulations. 
The convergence of the whole sample improves as $R \rightarrow 1$, with $R$ being the ratio between the variance in the complete sample and the variance for each chain. We have checked that the chains have reached proper convergence in all cases better than $R-1=2.5 \times 10^{-2}$. 

A typical problem when a minimization algorithm different from the complete computation of the multi-dimensional grid is applied is the possible presence of local minima
or of multiple global minima  (``degeneracies"). In both cases, if the minima are deep enough the algorithm will get stuck there and sample a region that does not correspond 
to the global minimum or will not be able to identify the presence of degenerate minima.
The MonteCUBES package includes a method to identify local minima by increasing the temperature $T$ of the chain so that the likelihood is modified to $\mathcal{P}\propto\mathcal{P}^{1/T}$. This procedure flattens the likelihood distribution, making it possible for the chains to jump from a local minimum to another. 
The temperature and step sizes are then decreased in successive steps and thus the different chains get stuck around different minima, unable to move through the disfavored regions when $T$ is too low. After this, the points where the different chains have stopped are compared to decide how many different minima the chains have fallen into. Finally, new steps are added with the correct length in the direction between the degeneracies. As a result, the algorithm is able to jump between minima and sample all of them properly.

For the implementation of the NSI probabilities in matter, we use the non-Standard Interaction 
Event Generator Engine (nSIEGE) distributed along with the MonteCUBES package. 

The definition of confidence level (CL) in a multi-dimensional MCMC algorithm need to be clarified, as it approaches the standard definition only in the limit of infinite statistics. 
What is done in practice with a MCMC is to generate a given number of points distributed stochastically in the multi-dimensional parameter space around the global minimum (minima).
 After that, projections of that multi-dimensional ``cloud'' of points are performed onto any desired plane chosen accordingly to the variables under study in the analysis. The projection 
 over a given plane corresponds to marginalization over the parameters that are no longer retained. After projection, the two-dimensional plane is divided into cells and the number
 of points falling into each cell is computed. For small enough cells, the resulting two-dimensional histogram can be approximated by a smooth surface, for which 
  slices can be drawn for the desired CL  ($68\%$, $90\%$, or $95\%$, in this paper). 

\subsection{The statistical procedure used in Sec.~\ref{sec:CP}}
\label{subsec:CPfrequentist}

To explore the CP discovery potential we need a different statistical procedure from the one defined in Sec.~\ref{subsec:montecubes}.
The reason is the following: a MCMC, as described above, explores the region which is close to the global minimum (or to degenerate minima) sampling
with good accuracy the $\chi^2$ distribution around that point(s). This is the right procedure to follow if we are exploring the sensitivity that 
a facility has to some particular observable.
Consider the particular case of the sensitivity to $\theta_{13}$ in the $(\theta_{13},\delta)$ plane with marginalization over the rest of $\nu$SM and NSI parameters (see Sec.~\ref{sec:golden}).
 In this case, the MCMC algorithm scans the multi-dimensional surface corresponding to a given choice of the input parameters (with the particular choice $\bar \theta_{13} = 0$
 for $\theta_{13}$) and a contour at a given CL of the region compatible with vanishing $\theta_{13}$ is drawn, after projecting over the $(\theta_{13},\delta)$ plane. 
When we compute a discovery potential, on the other hand, we first fix the parameters to be tested and draw the corresponding CL contours. Then, we check if the condition we want to fulfill is satisfied or not at a given CL (in the case of the CP discovery potential, the condition is that the contours drawn for a given set of CP violating input parameters do not touch any CP conserving point of the parameter space). Eventually, we repeat the procedure again and again varying the input parameters.
If the grid density is large enough, the distribution of the input parameters that satisfy the required condition is smooth and a``CL contour'' can be drawn.

If we were to use the MCMC approach to compute a discovery potential, then, we should run the algorithm as many times as the points in the grid that we want to test. In this case, the total time required to compute the discovery potential goes as $n^{N_{g}} \times N^k$, with $n$ the number of points to be tested for one parameter
and $N_g$ the dimensionality of the grid. If $n$ is large the MCMC cannot be used and the standard frequentist approach must be adopted instead. 
The drawback of the frequentist approach is that, in order to keep the computational time from being rapidly divergent, we cannot marginalize over the whole $\nu$SM  and NSI
parameter space. For this reason, in Sec.~\ref{sec:CP} we will not marginalize over atmospheric and solar parameters and will consider fixed inputs for the NSI parameters.

The procedure that has been used in this work to determine the CP-discovery potential is outlined below: 
\begin{enumerate}
\item
We first compute the number of expected events at the detector(s): 
$N(\theta_{13},\left\{\phi \right\}) $, where $\left\{\phi \right\}\equiv \left\{\delta,\,\phi_{e\mu},\, \phi_{e\tau} \right\} $. 
\item 
After having computed grids of number of events as a function of $(\theta_{13};\{ \phi \})$, we compute the $\chi^2$ as follows: 
\begin{equation}
\chi^2 (\theta_{13}, \bar \theta_{13};  \{ \phi \},  \{ \bar \phi \} ) = \sum_{\rm polarities, bins} \frac{(N (\theta_{13}; \{\phi \})-N (\bar \theta_{13}; \{ \bar \phi \})^2}{
\left (  N (\bar \theta_{13}; \{ \bar \phi \} )^{1/2} + f N (\bar \theta_{13};  \{ \bar \phi \}) \right )^2}
\label{eq:chi2CPV}
\end{equation}
with $f$ an overall systematic error. In all the plots given in this section, we assume the overall systematic error for the MIND detector as $f_\mu = 0.02$, and the one for the $\tau$-signal as $f_\tau = 0.05$. No background has been considered to compute Eq.~(\ref{eq:chi2CPV}).
\item
We compute, then, for any input ($\bar \theta_{13};  \{ \bar \phi \}$), the $\chi^2$ function defined in Eq.~(\ref{eq:chi2CPV}) at the eight CP-conserving (CPC) points: 
\begin{equation}
 \{ \phi \}_{CPC} = (0,0,0);(0,0,\pi);(0,\pi,0);(\pi,0,0);(0,\pi,\pi);(\pi,0,\pi);(\pi,\pi,0); (\pi,\pi,\pi) \, , \nonumber
\end{equation}
taking the smallest $\chi^2$ value found. Using this procedure, we obtain the five-dimensional surface: 
\begin{equation}
\label{eq:chi2CPCmin}
\chi^2_{CPC} (\theta_{13}, \bar \theta_{13};  \{ \bar \phi \}) = \min_{ \{ \phi \}_{CPC}} \left ( \chi^2 (\theta_{13}, \bar \theta_{13};  \{ \phi \}_{CPC},  \{ \bar \phi \} )  \right )
\end{equation}
This procedure generalizes to the case of three simultaneously active phases the procedure outlined in Ref.~\cite{Winter:2008eg}, where only $\delta$ and $\phi_{e\tau}$ were considered.
\item
We distinguish, then, between two cases depending on the value of $\bar\theta_{13}$: 
\begin{itemize}
\item 
The first possibility is that $\theta_{13}$ is already measured by the time the HENF is built. In this case, we can use the $\chi^2$ function (\ref{eq:chi2CPCmin}) computed at  $\theta_{13} = \bar \theta_{13}$ to see the region of the phase parameter space where CP violation can be distinguished from the CP conservation hypothesis.

\item The second possibility stands for a very small (or even vanishing) $\theta_{13}$. In this case, it is also necessary to marginalize over $\theta_{13}$ since possible CP-conserving solutions can be found for a given CP-violating input  $(\bar \theta_{13}; \{\bar \phi\} )$ at a different $\theta_{13}$ (what in the standard three-family oscillation scenario is called an ``intrinsic degeneracy" \cite{BurguetCastell:2001ez}). In order to take these degeneracies into account, we minimize the $\chi^2$ over $\theta_{13}$:
\begin{equation}
\nonumber
\chi^2_{\theta,CPC} ( \bar \theta_{13};  \{ \bar \phi \}) = \min_{\theta_{13}} \left ( \chi^2_{CPC} (\theta_{13}, \bar \theta_{13};  \{ \bar \phi \})  \right ).
\end{equation}
However, notice that in this case the only information we have on $\bar\theta_{13}$ is an upper bound, $\theta_{13}\leq 3^\circ$, approximately. Therefore, marginalization over $\bar\theta_{13}$ in the allowed range is also required here:
\begin{equation}
\label{eq:chi2CPCth13th13barmin}
\chi^2_{\theta,\bar \theta,CPC} (  \{ \bar \phi \}) = \min_{\bar \theta_{13}} \left ( \chi^2_{\theta, CPC} ( \bar \theta_{13};  \{ \bar \phi \})  \right ).
\end{equation}
\end{itemize}
\item Eventually, we draw the three-dimensional surfaces corresponding to $\chi^2 = 11.34$. 
These contours represent the area of the phases parameter space in which CP violation can be distinguished from CP conservation at the 99\% CL for 3 d.o.f.'s. Results will be shown 
for both cases in which $\theta_{13}$ is known, using Eq.~(\ref{eq:chi2CPCmin}), or unknown, using Eq.~(\ref{eq:chi2CPCth13th13barmin}). 
\end{enumerate}

\subsection{Input parameters and marginalization procedure}
\label{subsec:Mprocedure}

Unless otherwise stated, the input values taken in this work for the atmospheric and the solar parameters are: 
$\Delta \bar m_{21}^2=7.59\times10^{-5}\mathrm{eV}^2 $, 
$\bar\theta_{12}=34^\circ $, 
$\Delta \bar m^2_{31}=2.45\times10^{-3}\mathrm{eV}^2 $,
$\bar\theta_{23}=45.5^\circ $ \cite{Schwetz:2011qt}. 
In all the simulations, the matter density has been taken according to the PREM density profile~\cite{PREM} assuming a $5\%$ error. 
The sign of the atmospheric mass difference, $sgn(\Delta m_{31}^2)$, has been chosen to be positive throughout the paper and marginalization over it will not be considered.

In all the simulations presented in Secs. \ref{sec:golden} and \ref{sec:disappearance} we have marginalized over the whole set of $\nu$SM parameters.
A gaussian prior distribution centered on the input values given above with variance $\sigma = 0.08 (0.03)$ has been assumed for the atmospheric (solar) parameters. 
On the other hand, marginalization over $\theta_{13}$ and $\delta$ has also been performed assuming a flat prior distribution.

In Secs. \ref{sec:golden} and \ref{sec:disappearance} we have also marginalized over the NSI parameters $\epsilon_{\alpha \beta}$, that is, over both their moduli 
$\vert \epsilon_{\alpha \beta} \vert$ and phases $\phi_{\alpha \beta}$. Gaussian priors, in agreement with the bounds computed in Ref.~\cite{Biggio:2009nt}, are taken into account for all the moduli of the NSI parameters around their input values, which have been set to zero throughout the next two sections. 
Notice that for the NSI phases no prior knowledge has been taken into account ({\em i.e.}, $\pi(\theta) = 1$), since we do not have any information 
about these phases yet. 
We will refer the above procedure involving $\nu$SM and NSI parameters as the {\em ``standard marginalization procedure''} hereafter in this paper.

\subsection{Neutrino Factory setups}
\label{subsec:NF-setups}

At a NF, intense $\nu_{e}$ and $\nu_{\mu}$ beams are available as decay products of muons  (with both polarities) circulating in storage ring(s). As a consequence, a total of twelve different oscillation channels could in principle be studied. The {\em International Design Study for a Neutrino Factory} \cite{IDS}, as already mentioned, has undertaken the task of defining the optimal setup to have good sensitivity to $\theta_{13}$, $\delta$, and to the neutrino mass hierarchy, ${\rm sign}(\Delta m^2_{23})$.
This resulted in what we will refer to as the IDS25 setup hereafter: a HENF with a muon beam energy of $25$ GeV, and $10^{21}$ useful muon decays per year aimed at two identical 50 kton MIND detectors located at two different baselines to look for $\nu_{\mu}$ appearance events\footnote{
Notice that, in the latest IDS design, the MIND located at the intermediate baseline is 100 kton.}. 
The detector located at $ L = 4000$ km (which will be referred to as
``intermediate baseline" from now on, to distinguish it from the short baseline where a near detector will be located\footnote{
We will not consider such a near detector in our analysis, though.}
) is optimized to have sensitivity to the CP-violating phase $\delta$. It, however, suffers from a severe degeneracy problem \cite{BurguetCastell:2001ez,Minakata:2001qm,Fogli:1996pv},  that can be solved locating an additional
``far" detector at the so-called ``magic" baseline (7500 km) \cite{BurguetCastell:2001ez,Huber:2003ak}. This second detector increases significantly the potential of the NF to measure the mass hierarchy, taking advantage of matter effects.

If a HENF is to be optimized to detect effects of NSI, several issues must be understood:
which energy and baseline would be the best;  how and to what extent the synergy between two detectors help; and, if there are any ways of optimization in order to achieve good sensitivities to both $\nu$SM and NSI parameters. Some of these problems were addressed in \cite{Ribeiro:2007ud}, concluding that a setting similar to IDS25 but with higher muon energies (such as 50 GeV) would be preferred to look for NSI. This is easily explained by the fact that, 
since NSI in propagation are introduced as an effective matter potential, an increase in the average neutrino energy will improve the relative significance of the NSI with respect to the leading standard oscillations in vacuum. 
This was indeed confirmed in Ref.~\cite{Kopp:2008ds} where, however,  it was found that the improvement with respect to the 25 GeV setup was not very large
(see, also, Ref.~\cite{Gago:2009ij}). All of these works were performed within the ad-hoc assumption of having only one $\epsilon_{\alpha \beta}$ at a time.
One of the goals of this paper is indeed to check if the results obtained in these analyses survive  
when correlations between the various NSI parameters are taken into account.
Armed with MonteCUBES, we examine this problem by comparing the sensitivity to NSI of the IDS25 and of a variant of the IDS setup with the same detectors but with parent 
muon energy $E_{\mu} = 50$ GeV when all NSI parameters are turned on simultaneously. The new setup, for obvious reasons, will be called as IDS50.

We will also check the potential of the two detectors to reduce the strong correlations between $\nu$SM and NSI  parameters and within different NSI parameters, to
test if this detector combination can be optimized or not for NP searches. For this reason,  the performance of the two setups above, in which two identical detectors
looking for $\nu_e\to \nu_\mu$ and $\nu_\mu\to\nu_\mu$ oscillations are located at different baselines, will be compared with the performance of a HENF setup in which two different
detectors, one of which is equipped with $\tau$-identification capability (that could, thus, profit of the $\nu_e\to \nu_\tau$ and $\nu_\mu\to\nu_\tau$ channels), are located at one single baseline at $4000$ km. Initially, $L=2000$ and $L=3000$ km were also considered as alternative baselines. They give similar results, although the $4000$ km performs slightly better, and therefore will not be considered here.
This setup will be called 1B50. Notice that the parent muon energy for the 1B50 setup has to be large enough to overcome the smallness of the $\nu_\tau N$ cross-section due to the $\tau$ production threshold below 4 GeV~\cite{Lipari:1994pz} and get larger statistics at the detector. For this reason, also for this setup we fix $E_\mu = 50$ GeV. 
Notice that the advantage of aiming at one site is two-fold: on one side, we avoid the technical difficulties of aiming one of the beams at a detector located al $L=7500 $ km (with a tilt angle of the storage ring of $\sim 36^\circ$ \cite{ISS-accelerator}); on the other side, as only one storage ring is needed all muon decays are aimed to the same site, therefore doubling the statistics at the detector.

The characteristic features of the three setups  (IDS25, IDS50 and 1B50) are resumed in Tab.~\ref{tab:setups}, where we remind the parent muon energy, the detectors location and technologies and the neutrino flux aiming at each detector per year. For all setups we consider 5 years of data taking for each muon polarity\footnote{
We assume that the experiment is run on the two polarities separately. This means that we are considering a total number of useful muon decays per baseline and polarity of $5\times5\times10^{20}=2.5\times10^{21}$. Notice that this is completely equivalent to consider 10 years of data taking per polarity but with $2.5\times10^{20}$ useful muon decays per baseline, year and polarity. This last option corresponds to the setup where muons of both polarities are circulating at the same time in the decay ring(s).
}.
 
\begin{table}[hbtp]
\renewcommand{\arraystretch}{1.7}
\begin{center}
\begin{tabular}{|c|c|c|c|}
\hline 
  & IDS25 & IDS50 & 1B50 \\
\hline
\; $E_\mu$ \; & 25 GeV & 50 GeV & 50 GeV \\
 \; $D_1$ \; & \;  MIND@4000 km \; & \; MIND@4000 km \; & \; MIND@4000 km \; \\
\;  $D_2$ \; & \;  MIND@7500 km \; & \; MIND@7500 km \; & \; MECC@4000 km \; \\
\; $\Phi_1$ \; & $5 \times 10^{20}$ & $5 \times 10^{20}$ & $1 \times 10^{21}$ \\
\; $\Phi_2$ \; & $5 \times 10^{20}$ & $5 \times 10^{20}$ & $1 \times 10^{21}$ \\
\hline
\end{tabular}
\end{center}
\caption{\label{tab:setups} The characteristics of the three considered setups. From top to bottom: the parent muon energy $E_\mu$; the technology and location of the two detectors, $D_1$ and $D_2$; the number of useful muon decays per year aiming at each of the two detectors, $\Phi_1$ and $\Phi_2$.}
\end{table}

The characteristic features of the two types of detector are summarized in Tab.~\ref{tab:dets}.
For technical details on these parameters, we address the interested reader to Refs.~\cite{ISS-detector} and \cite{Donini:2008wz,ScottoLavina:2008zz}.

\begin{table}[hbtp]
\renewcommand{\arraystretch}{1.7}
\begin{center}
\begin{tabular}{|c|c|c|c|c|}
\hline 
  &$\sigma(E)$ & $f_S$ & $f_B$ & Mass\\
\hline
MIND & \; $0.55 \sqrt{E}$ \; & \; 2.5\%\;  & \; 20\% \;  & \; 50 kton \; \\
MECC & \; $0.2 E$ \; & \; 15\% \; & \, 20\% \; & \; 4 kton \; \\
\hline
\end{tabular}
\end{center}
\caption{\label{tab:dets} Main characteristics of the two detectors technologies.
From left to right: energy resolution, $\sigma(E)$; systematic error over the signal, $f_S$; systematic error over the background, $f_B$; detector mass.
}
\end{table}

In the analyses of Secs.~\ref{sec:golden} and \ref{sec:disappearance}, data have been distributed in bins of the reconstructed neutrino energy with the following size: 
$\Delta E_\nu = 1$ GeV for $E_\nu \in [1,10]$ GeV; $\Delta E_\nu = 2.5$ GeV for $E_\nu \in [10,15]$ GeV; $\Delta E_\nu = 5$ GeV for $E_\nu \geq 15$ GeV.
This binning applies to all setups and detector technologies.
In Sec.~\ref{sec:CP}, on the other hand,  data have been distributed in bins of equal size: $\Delta E_\nu = 5$ GeV for the IDS25; $\Delta E_\nu = 10$ GeV for 
the IDS50, 1B50 (both for the MIND and the MECC technologies).

\begin{figure}[h!]
\begin{tabular}{c}
  \includegraphics[scale=0.7,angle=0]{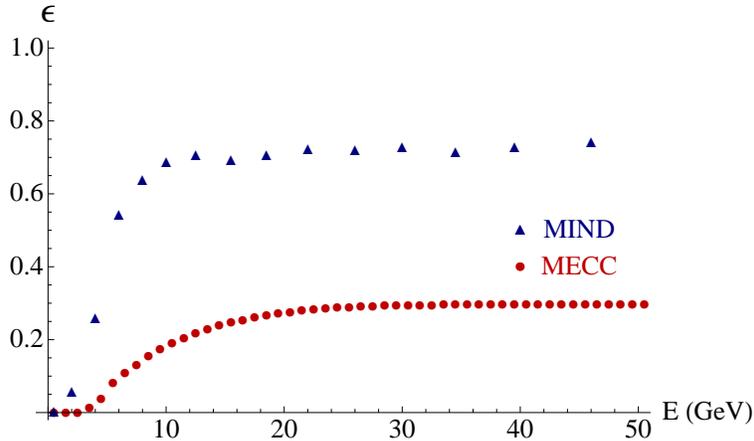} 
 \end{tabular}
\caption{\label{fig:effs} Efficiency of the MIND (blue triangles) and MECC (red circles) detectors as a function of the neutrino energy. The MIND efficiency has been taken from Ref.~\cite{ISS-detector}. The MECC efficiency corresponds to the ECC efficiency~\cite{Autiero:2003fu} multiplied by a factor of five (see text for details). 
}
\end{figure}

The efficiencies for the MIND and the MECC detector technologies as a function of the reconstructed neutrino energy are shown in Fig.~\ref{fig:effs}. 
The $\nu_\mu$ identification efficiency at MIND has been taken from Ref.~\cite{ISS-detector}. The $\nu_\tau$ identification efficiency at MECC corresponds to the efficiency of the ECC for the silver channel $\nu_e\to \nu_\tau$ 
as computed in Ref.~\cite{Autiero:2003fu} multiplied by a factor of five to take into account the capability of the MECC to look for taus not only through their decay into muons (as for the ECC) but also into electrons and hadrons (see a detailed discussion in Ref.~\cite{Donini:2008wz} regarding this point). The cross-sections have been taken from Refs.~\cite{Messier:1999kj,Paschos:2001np}.

Table~\ref{tab:events} shows the number of events in the golden (silver) channel per kton$\times$year at a MIND (MECC) detector with perfect efficiency, located at $4000$ km from the source, for a 50 GeV NF. Normal hierarchy has been assumed, and results are shown for $\delta=\pm 90^\circ$, and for two different values of $\theta_{13}=0,3^\circ$. In order to illustrate the effect when NSI are included in the analysis, we show the total number of events at the detector also in presence of NSI, $\epsilon_{e\mu}=\epsilon_{e\tau}=10^{-2}$. The two NSI CP violating phases have been set to zero. As it can be seen from the table, the number of events in the silver channel is very small for vanishing $\theta_{13}$.

\begin{table}[hbtp]
\renewcommand{\arraystretch}{1.7}
\begin{center}
\begin{tabular}{|c|c|c|c|c|c|}
\hline 
\multirow{2}{*}{$\delta $} & \multirow{2}{*}{Channel}  & $\theta_{13}=0$    &  $\theta_{13}=0$  &  $\theta_{13}=3^\circ$  &  $\theta_{13}=3^\circ$  \cr
  &    & $\epsilon_{\alpha\beta}=0$   & $\epsilon_{e\mu}=\epsilon_{e\tau}=10^{-2}$   &  $\epsilon_{\alpha\beta}=0$  &  $\epsilon_{e\mu}=\epsilon_{e\tau}=10^{-2}$ \\ 
\hline
\hline
%
\multirow{2}{*}{\large{$+90^\circ$} } & $ \nu_e \rightarrow \nu_\mu $     	&   2.75   &    44.93    &   66.70    &   108.54   \cr
\cline{2-6}$  $ & $ \nu_e \rightarrow \nu_\tau $                		&   1.08   &    7.97     &   15.63   &    22.60   \\ 
\hline\hline
 \multirow{2}{*}{\large{$-90^\circ$} } &  $ \nu_e \rightarrow \nu_\mu $ 	&   2.75   &    44.93  &   41.69   &   83.96  \cr
  \cline{2-6} & $ \nu_e \rightarrow \nu_\tau $                 		&   1.08   &    7.97   &   23.77  &   30.56    \\ 
\hline
\hline
\end{tabular}
\end{center}
\caption{\label{tab:events} Total number of events per year for the golden (silver) channel, measured at a 1 kton MIND (MECC) detector with perfect efficiency located at $L=4000$ km from the source, for a 50 GeV NF. Normal hierarchy has been assumed. Results are presented for $\delta=\pm 90^\circ$ and for two different values of $\theta_{13}=0,3^\circ$, with and without including NSI effects in the golden sector. The two NSI CP violating phases have been set to zero: $\phi_{e\mu}=\phi_{e\tau}=0^\circ$. 
} 
\end{table}

\section{Sensitivities achieved mainly through the $\nu_e \rightarrow \nu_\mu$ and $\nu_e \rightarrow \nu_\tau$ channels}
\label{sec:golden}

As we stressed in Sec.~\ref{sec:NSI}, the sensitivity to NSI parameters comes from different oscillation channels depending on the considered parameter. In this section, we study the sensitivities to $\epsilon_{e\mu}$ and $\epsilon_{e\tau}$ which would be achieved mostly through the golden (and, to a lesser extent, also through the silver) channel for the three setups under study. 
Since the sensitivity to the $\nu$SM parameter $\theta_{13}$ (which is the key to the measurement of $\delta$ and of the mass hierarchy, too) is also achieved through the same
oscillation channels, we will examine first the question of how and to what extent the inclusion of NSI affects the sensitivity to $\theta_{13}$ (Sec.~\ref{subsec:theta13}). 
We will, then, study the sensitivities to the moduli of \epsemu and \epsetau as a function of their respective CP violating phases, 
through which some features of the synergy between two detectors/baselines will be illuminated depending upon the settings (Sec.~\ref{subsec:emuetau}). 
To show the effect of the correlations between the NSI parameters and with $\theta_{13}$, the standard marginalization procedure defined in Sec.~\ref{subsec:Mprocedure} will
not be always used. We will specify in each case the procedure adopted. We remind the readers 
that $sgn(\Delta m^2_{31})$ is kept fixed throughout this paper. This point has to be kept in mind when interpreting the results. 

To conclude the preamble, we describe the layout of the figures:
in figures with three columns, the left, middle and right panels correspond to the results for the IDS25, IDS50 and 1B50 settings, respectively. 
Red, green and blue lines correspond to 68\%, 90\% and 95\% 2 d.o.f.'s CL contours, respectively. 
Whenever we depart from the standard format we will give a note in the caption of the corresponding figure to specify the layout.

\subsection{Impact of the NSI on the measurement of $\theta_{13}$}
\label{subsec:theta13}

\begin{figure}[h!]
\hspace{-0.2cm}
\begin{tabular}{ccc}
  \includegraphics[width=0.33\textwidth,angle=0]{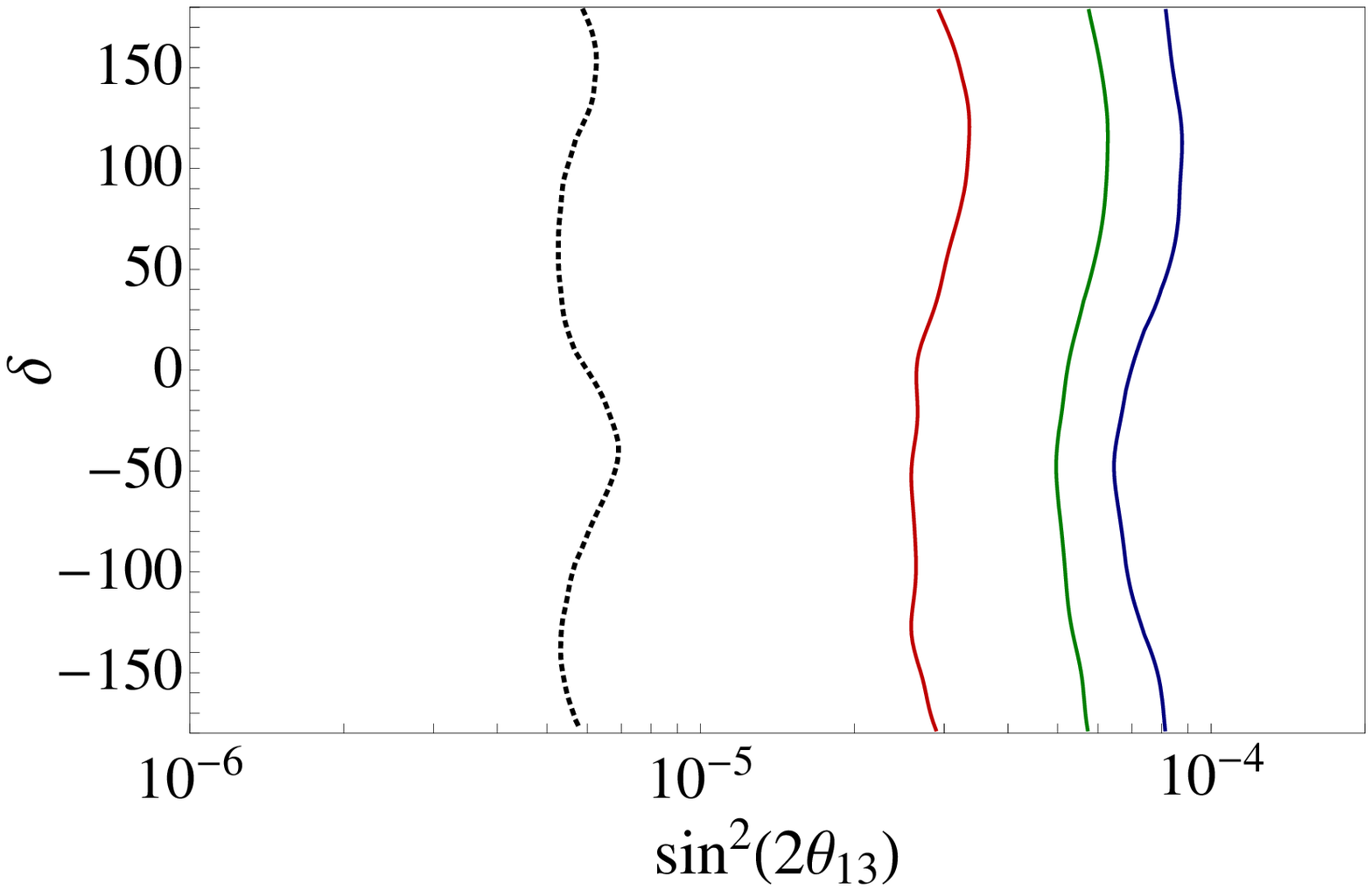} &
    \includegraphics[width=0.33\textwidth,angle=0]{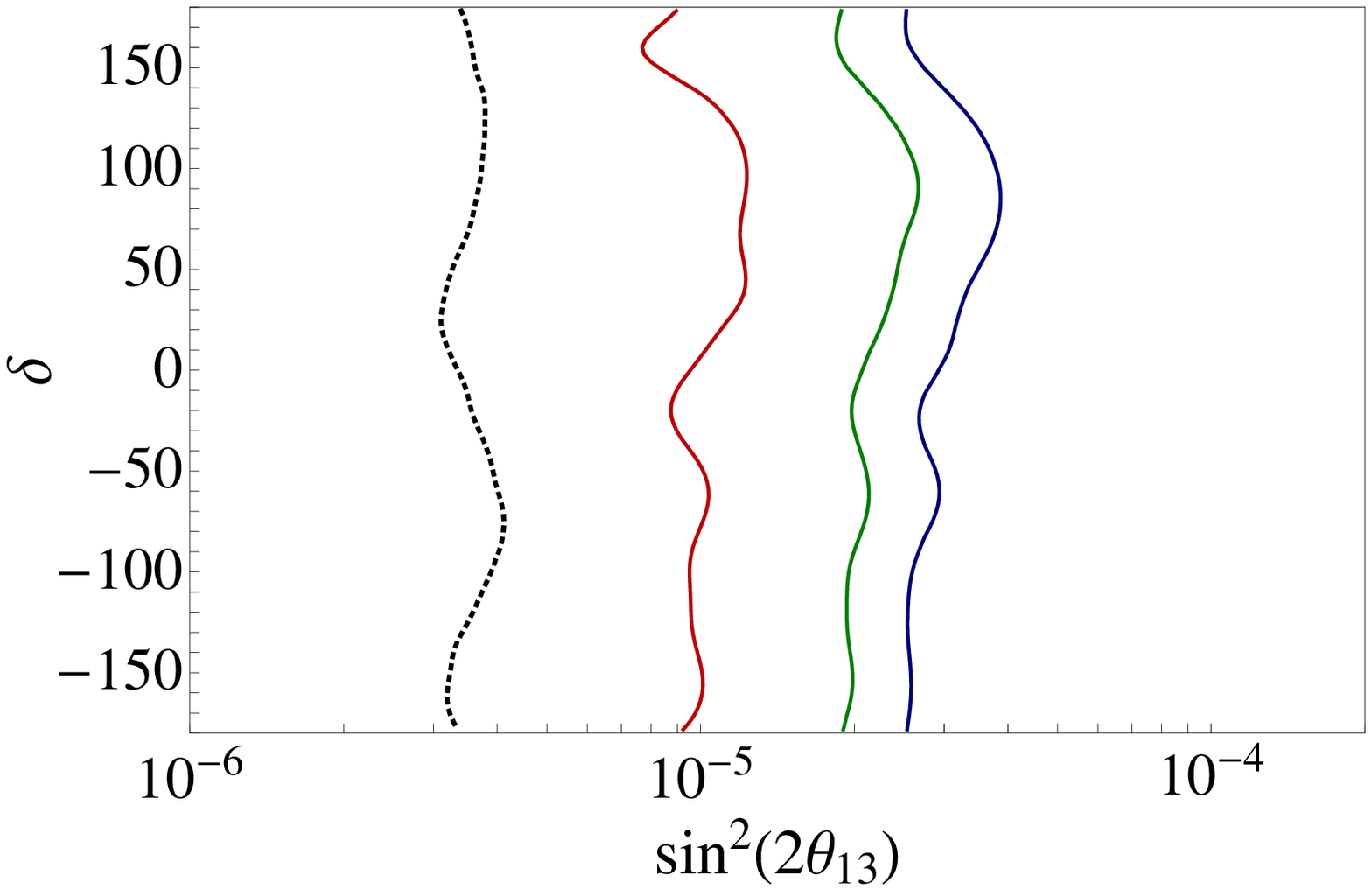} &
      \includegraphics[width=0.33\textwidth,angle=0]{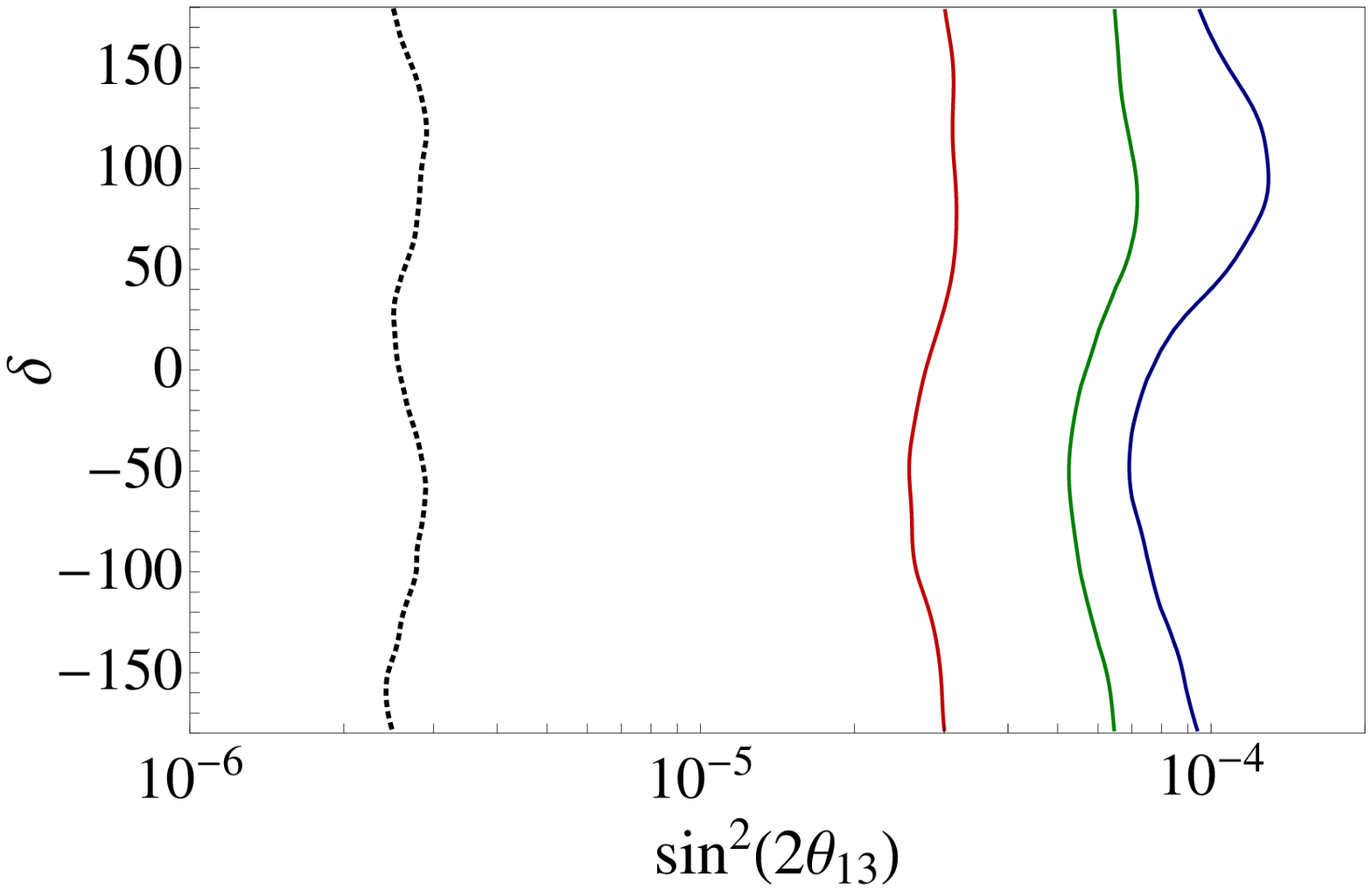} \\
  \includegraphics[width=0.33\textwidth,angle=0]{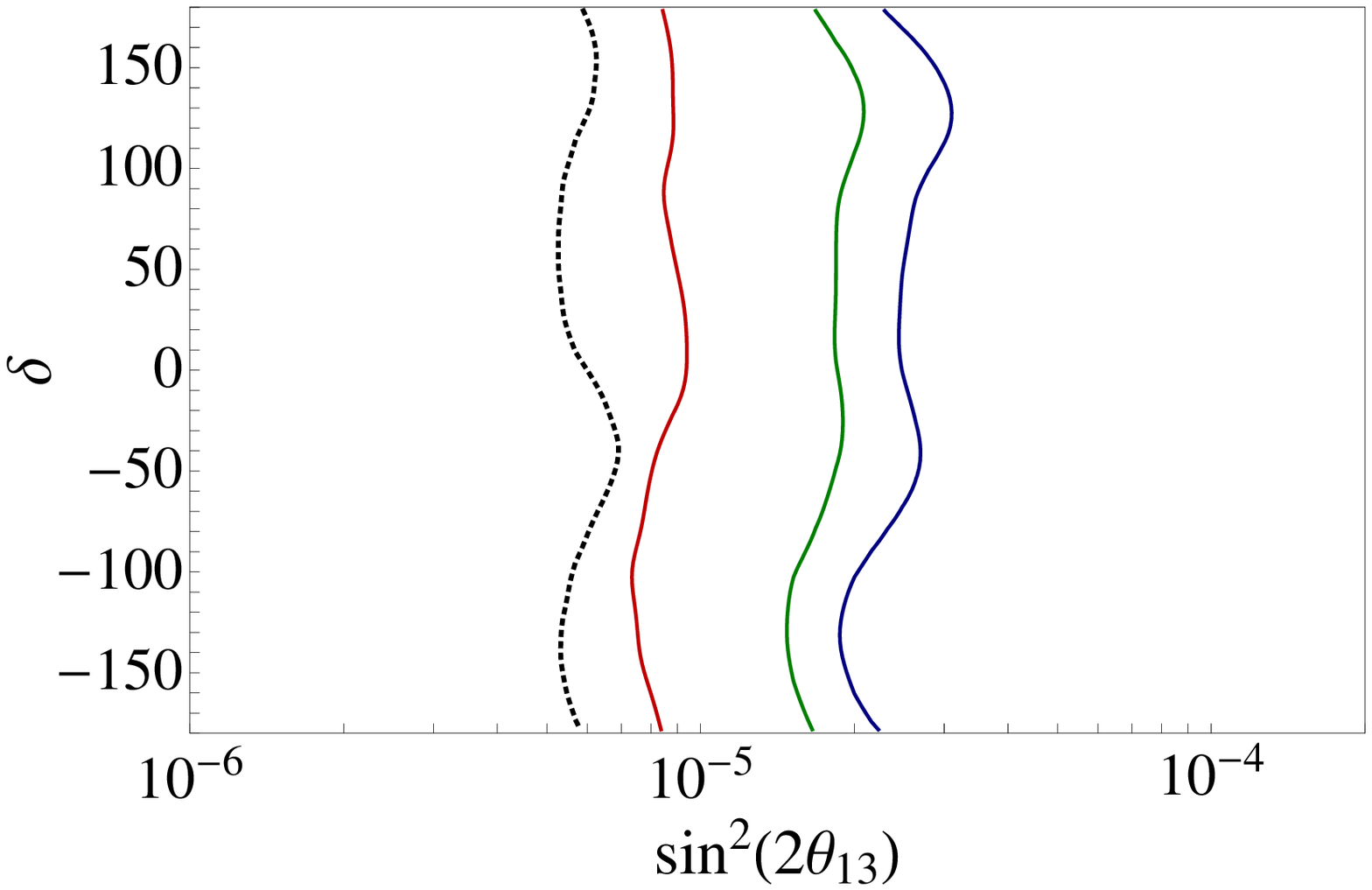} &
   \includegraphics[width=0.33\textwidth,angle=0]{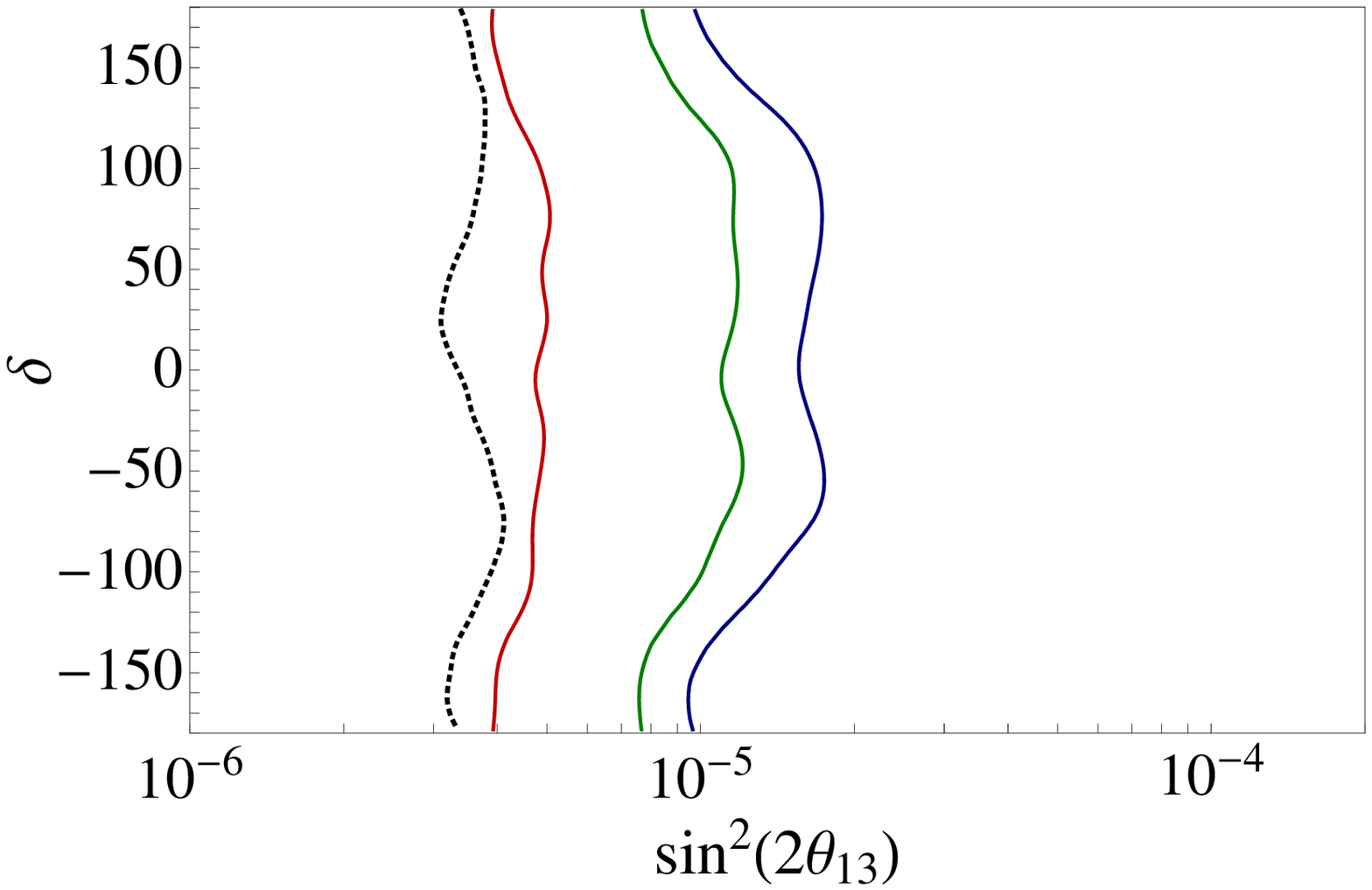} &
  \includegraphics[width=0.33\textwidth,angle=0]{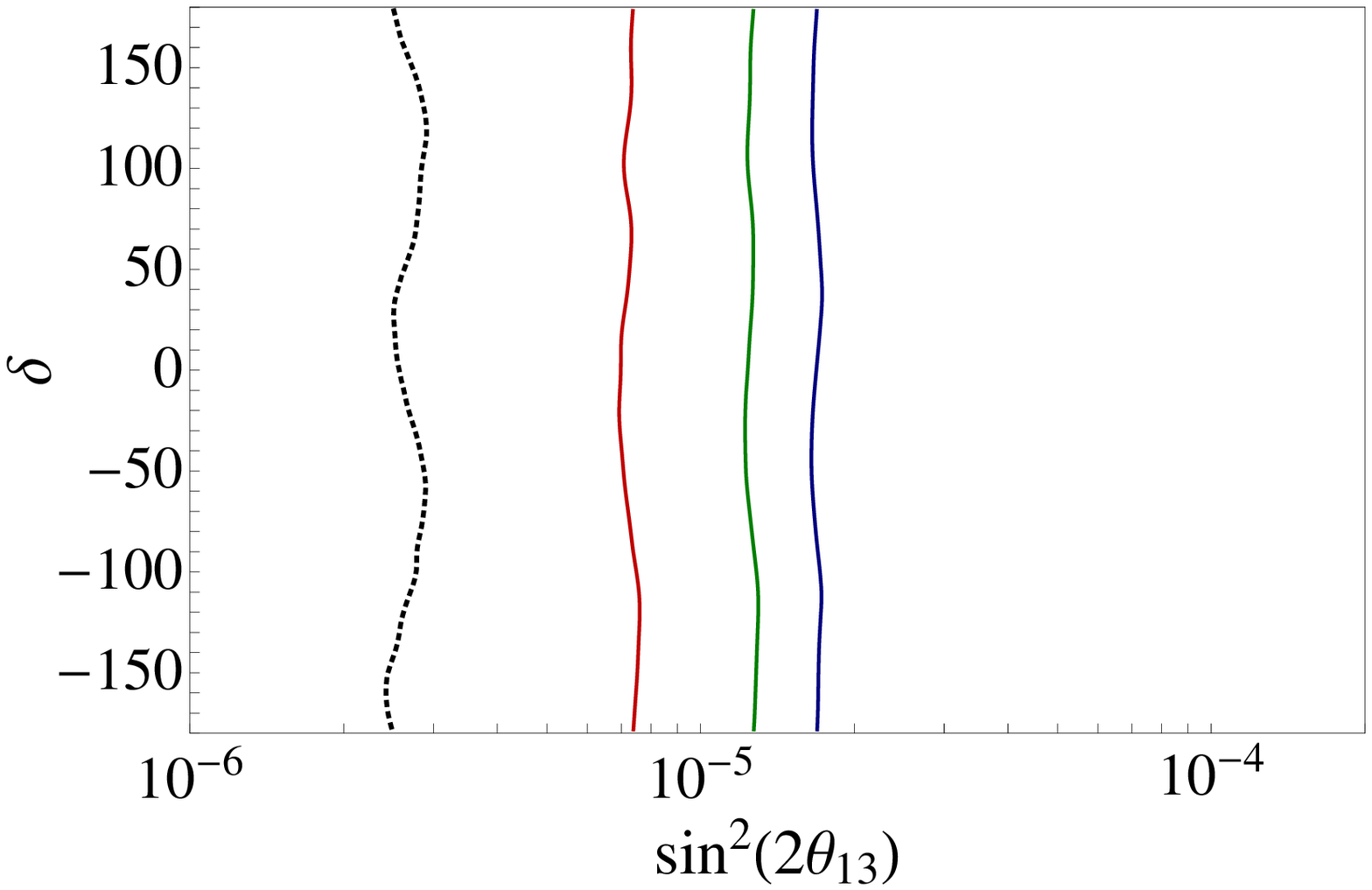} \\
\end{tabular}
\caption{\label{fig:sensdelta_0} 
68\%, 90\% and 95\% CL contours for the sensitivity to $\theta_{13}$ as a function of $\delta$ for the case with NSI, compared to the 68\% CL contour for the sensitivity
to $\theta_{13}$ in the absence of NSI (represented by the black dotted line).
Marginalization was performed over the $\nu$SM parameters, the diagonal NSI parameters $\epsilon_{\alpha\alpha}$ and either
\epsemu and \epsetau (top panels) or \epsmutau (bottom panels).
The left, middle, and right panels show the results obtained for IDS25, IDS50, and 1B50 setups, respectively. 
}
\end{figure}

In Fig.~\ref{fig:sensdelta_0} we show the sensitivity to $\theta_{13}$ as a function of $\delta$ when the NSI parameters, in addition to the $\nu$SM ones, 
are also taken into account during marginalization. In top panels, marginalization over $\epsilon_{\alpha\alpha}$ ($\alpha= e, \mu, \tau$), \epsemu and \epsetau is performed; 
in bottom panels, the marginalization procedure is done over $\epsilon_{\alpha\alpha} $ and $\epsilon_{\mu\tau}$. 
As a reference, we also present the 68\% CL sensitivity to $\theta_{13}$ obtained without considering NSI in the analysis, represented by the dotted black lines\footnote{
We have checked that our results for the $\nu$SM $\theta_{13}$-sensitivity are in reasonable agreement with those reported in the literature 
(see, for example, Ref.~\cite{Bandyopadhyay:2007kx}). 
}.
It can be clearly seen that the impact of the presence of NSI degrees of freedom on sensitivity to $\theta_{13}$ is much more significant in top panels than in bottom panels. When marginalization is performed over \epsemu and $\epsilon_{e\tau}$, the degree of sensitivity loss ranges from a factor 3 (IDS50) to almost an order of magnitude (1B50) with respect to the $\nu$SM result.
On the other hand, the effects of marginalization over $\epsilon_{\alpha\alpha}$ and \epsmutau are quite mild, leading to a sensitivity loss of a factor of 3 (1B50), at most. 
We have checked that the above sensitivity loss comes from the marginalization over $\epsilon_{\alpha\alpha}$, and not over $\epsilon_{\mu\tau}$ 
which is effectively decoupled from $\theta_{13}$.
Approximate decoupling between the two parameter sets ($\theta_{13}$, \epsemu, \epsetau) and ($\epsilon_{\alpha\alpha}$, \epsmutau) is consistent with the expectation from the perturbative analysis \cite{Kikuchi:2008vq} (see Appendix~\ref{sec:expandedP}), as explained in Sec.~\ref{sec:NSI}: in a nutshell, the golden (and silver) channel oscillation probabilities (that dominate the sensitivity to $\theta_{13}$) only depend on \epsemu and \epsetau up to second order in $\varepsilon$. 

A careful comparison between the three upper panels reveals an interesting feature:  the severe impact on the sensitivity to $\theta_{13}$ observed at the 1B50 setup is largely (moderately) overcome in the IDS50 (IDS25) setups. Sensitivity to $\theta_{13}$ at the two baseline settings is robust against inclusion of NSI because they probe 
generalized matter effects at two different distances, a particular type of the synergy between the intermediate and far detectors \cite{Ribeiro:2007ud}. 

In summary, in spite of the fact that the 1B50 setup apparently yields the best sensitivity to $\theta_{13}$ in absence of NSI (which is likely to be due to the doubled flux with respect to the IDS25 and IDS50 setups), its worsening after marginalization over \epsemu and \epsetau, strongly correlated with $\theta_{13}$, is much more severe due to the lack of the magic baseline detector. 

\subsection{Sensitivity to $\epsilon_{e\mu}$ and $\epsilon_{e\tau}$ }
\label{subsec:emuetau}

\begin{figure}[h!]
\begin{tabular}{ccc}
  \includegraphics[width=0.33\textwidth,angle=0]{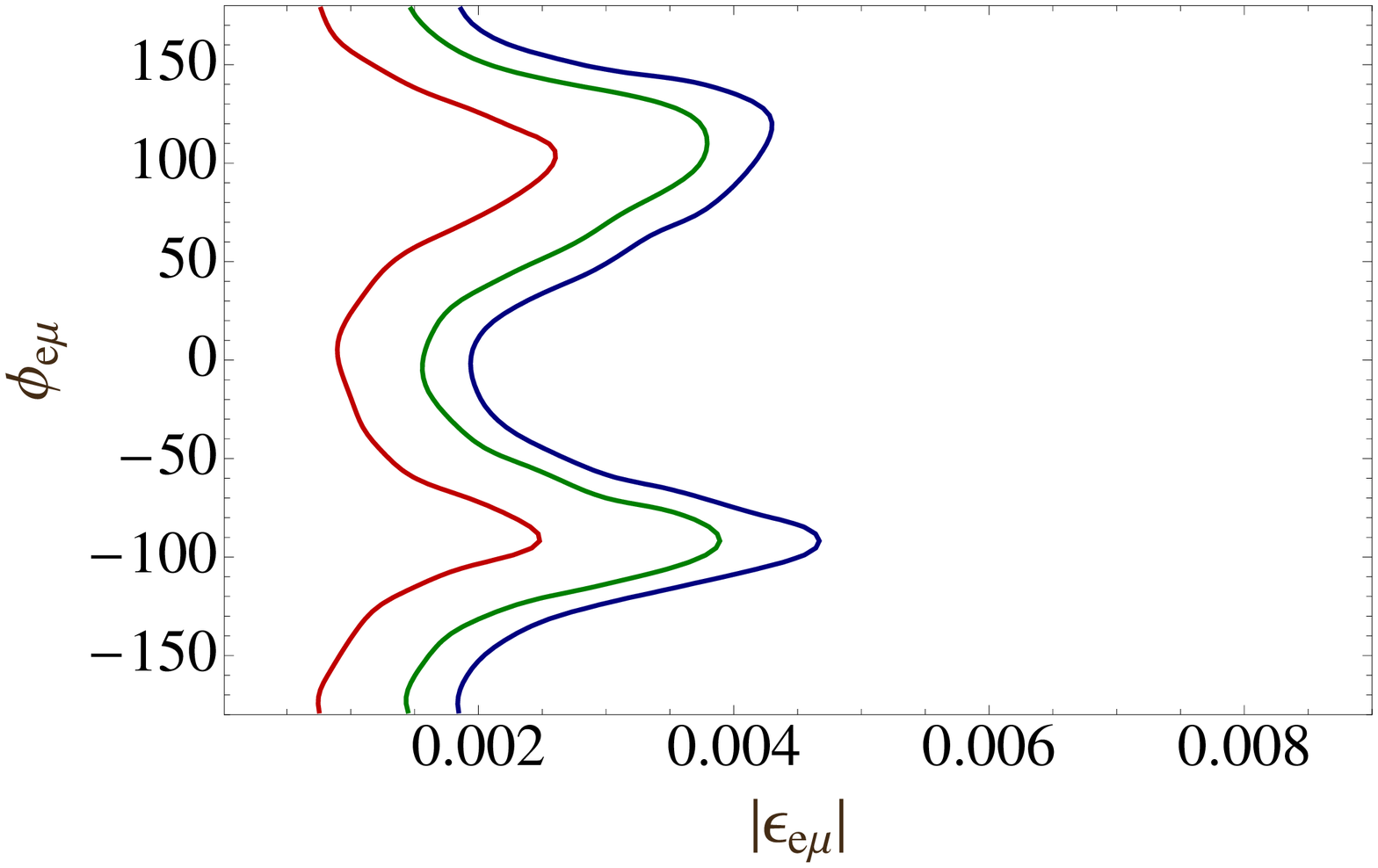} &
  \includegraphics[width=0.33\textwidth,angle=0]{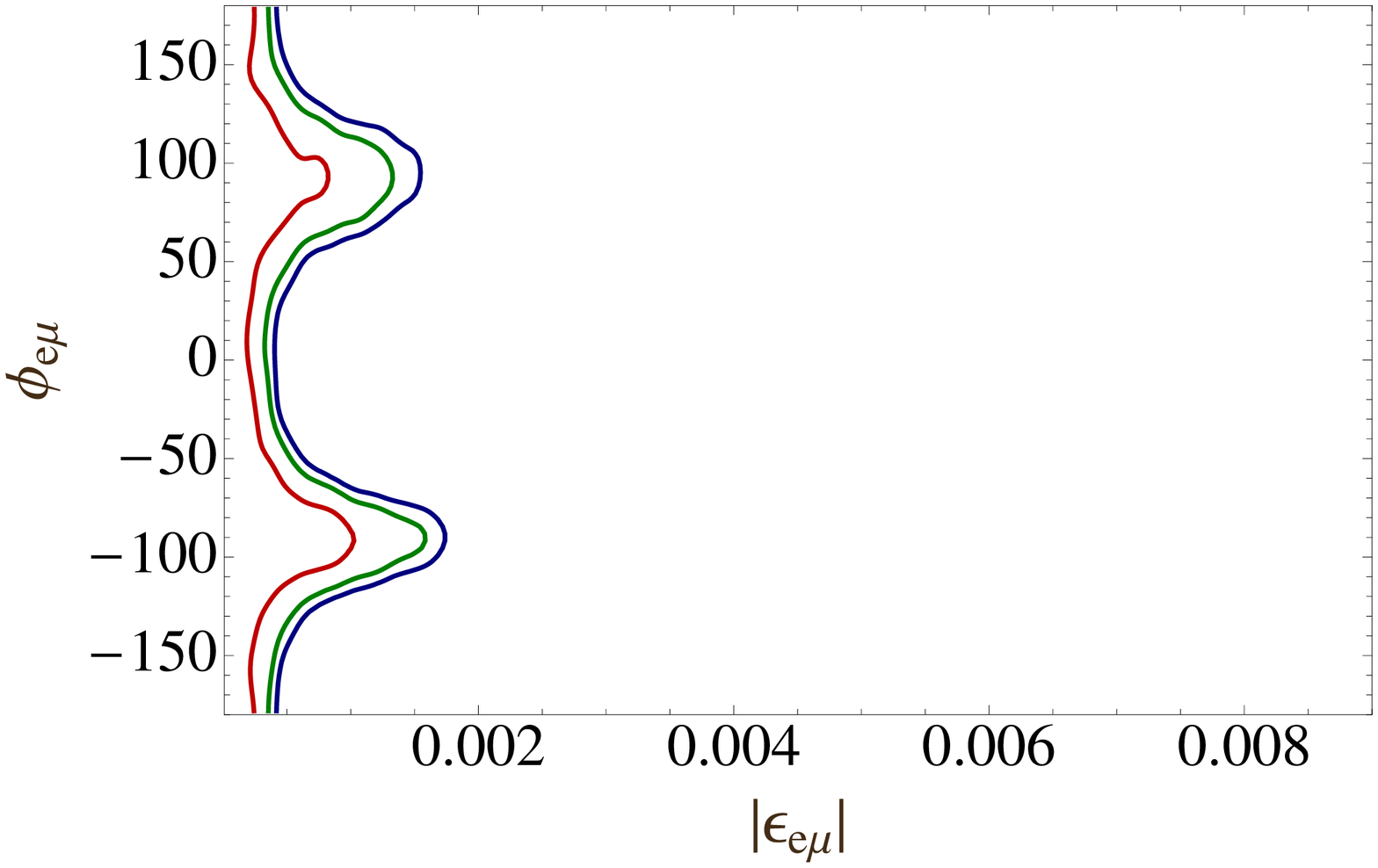} &
  \includegraphics[width=0.33\textwidth,angle=0]{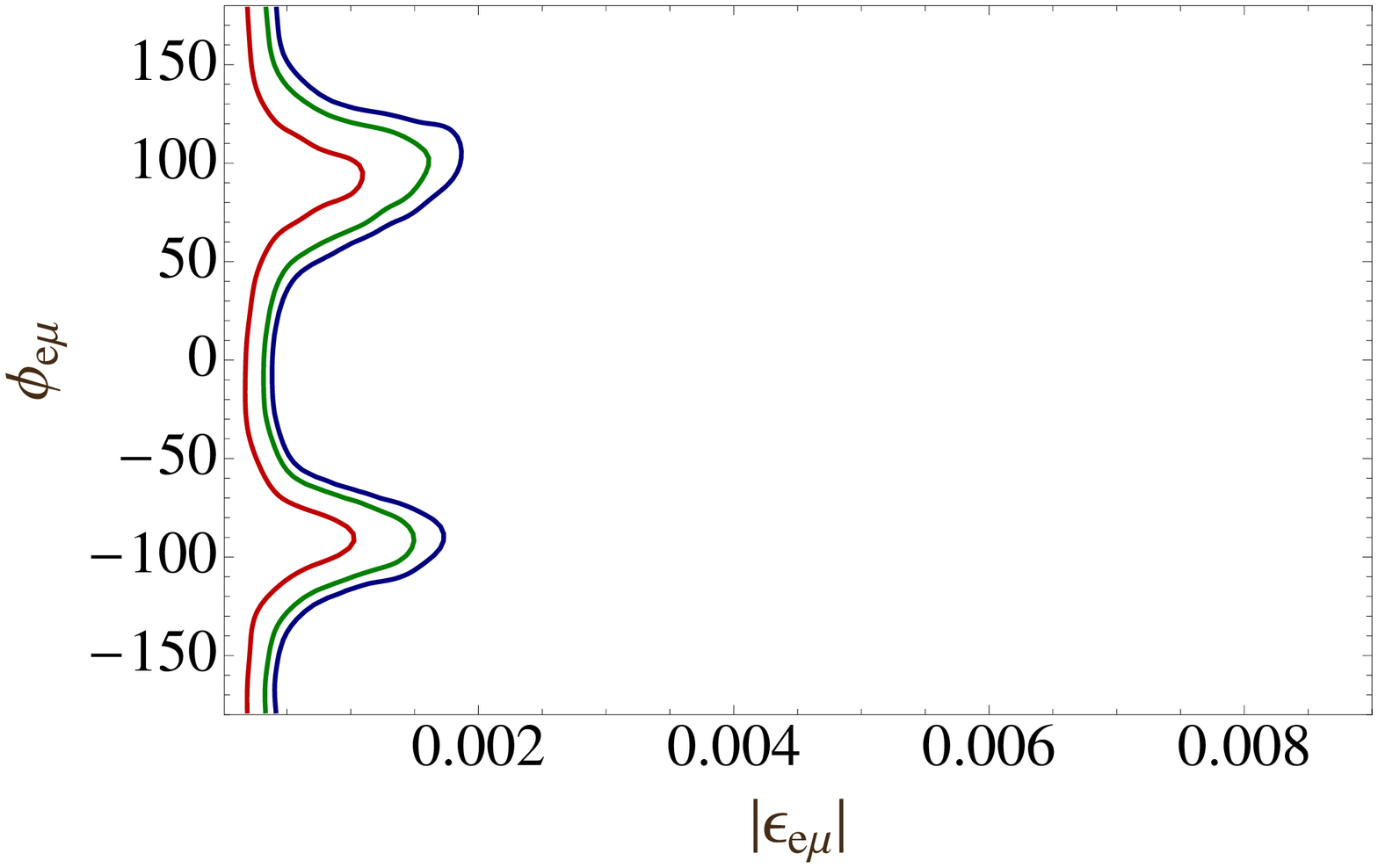} \\
  \includegraphics[width=0.33\textwidth,angle=0]{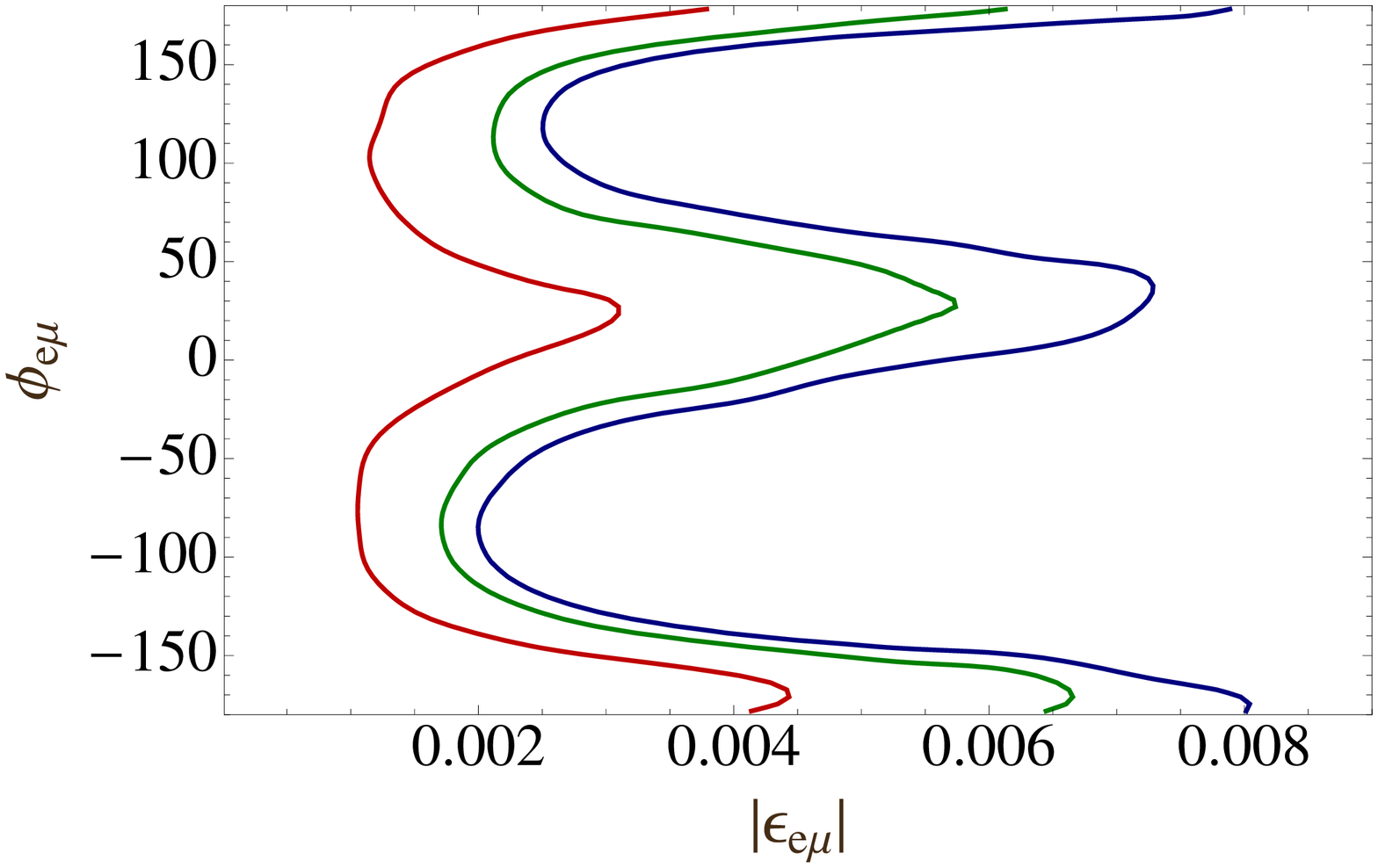} &
  \includegraphics[width=0.33\textwidth,angle=0]{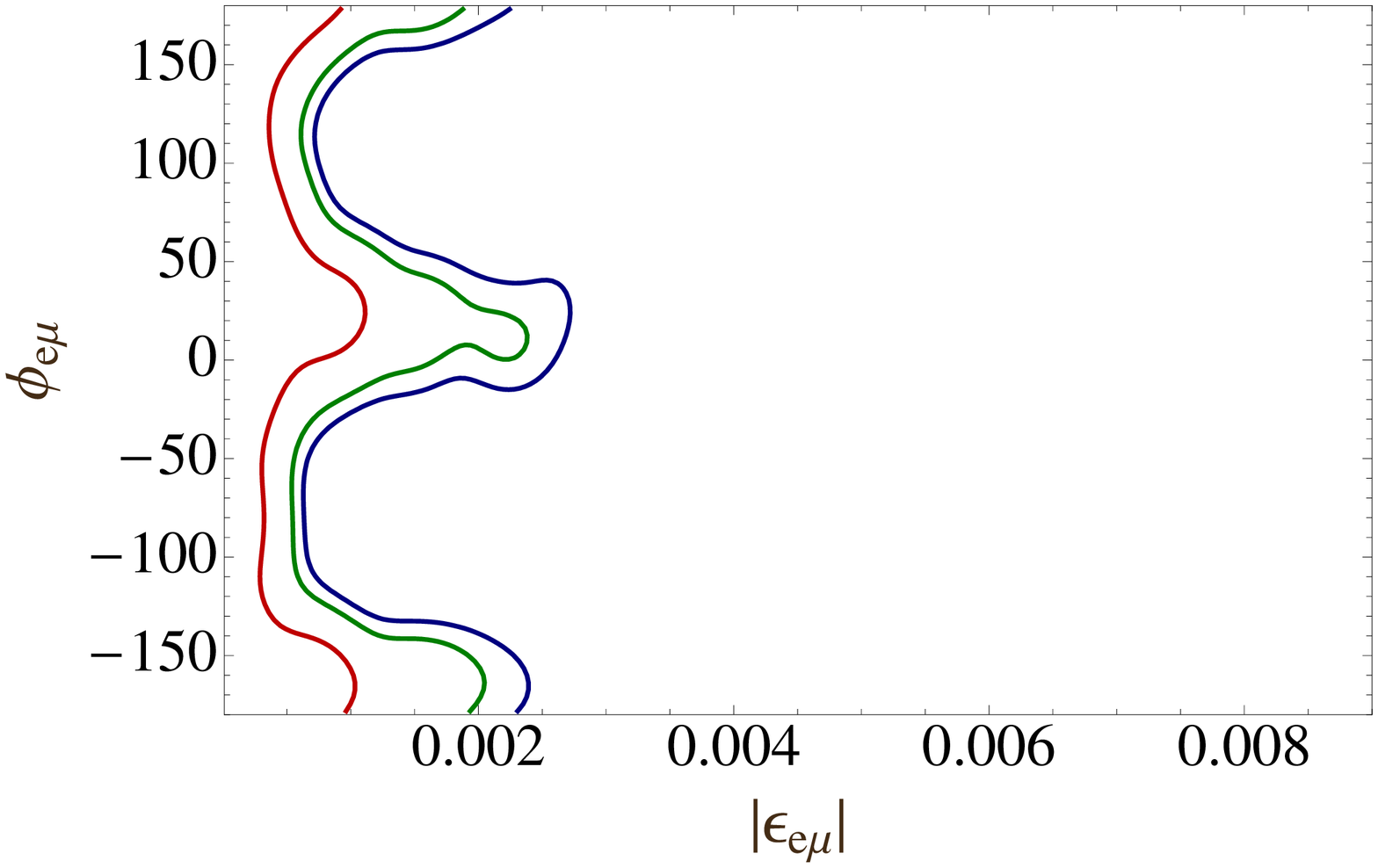} &
  \includegraphics[width=0.33\textwidth,angle=0]{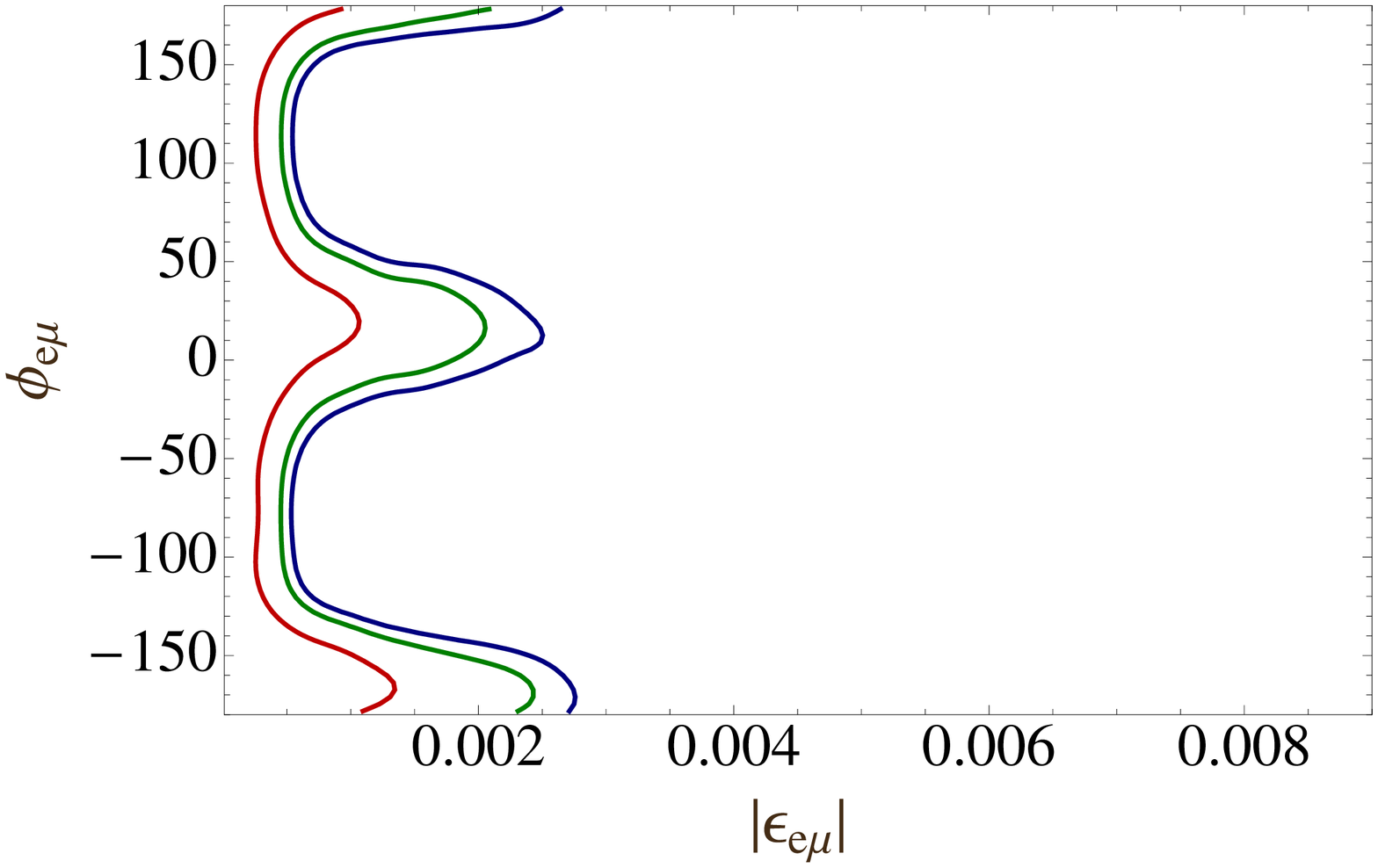} \\
\end{tabular}
\caption{\label{fig:epsemu} 68\%, 90\% and 95\% CL contours for the sensitivity to \epsemu as a function of $\phi_{e\mu}$
for $\bar\theta_{13} = 0$ (upper panels) and $\bar\theta_{13} = 3^\circ; \bar\delta=-\pi/2$ (lower panels). 
Marginalization has been performed over the $\nu$SM parameters, the diagonal NSI parameters $\epsilon_{\alpha\alpha}$ and $\epsilon_{e\tau}$.
The left, middle, and right panels show the results obtained for IDS25, IDS50, and 1B50 setups, respectively. 
}
\end{figure}

\begin{figure}[h!]
\begin{tabular}{ccc}
  \includegraphics[width=0.33\textwidth,angle=0]{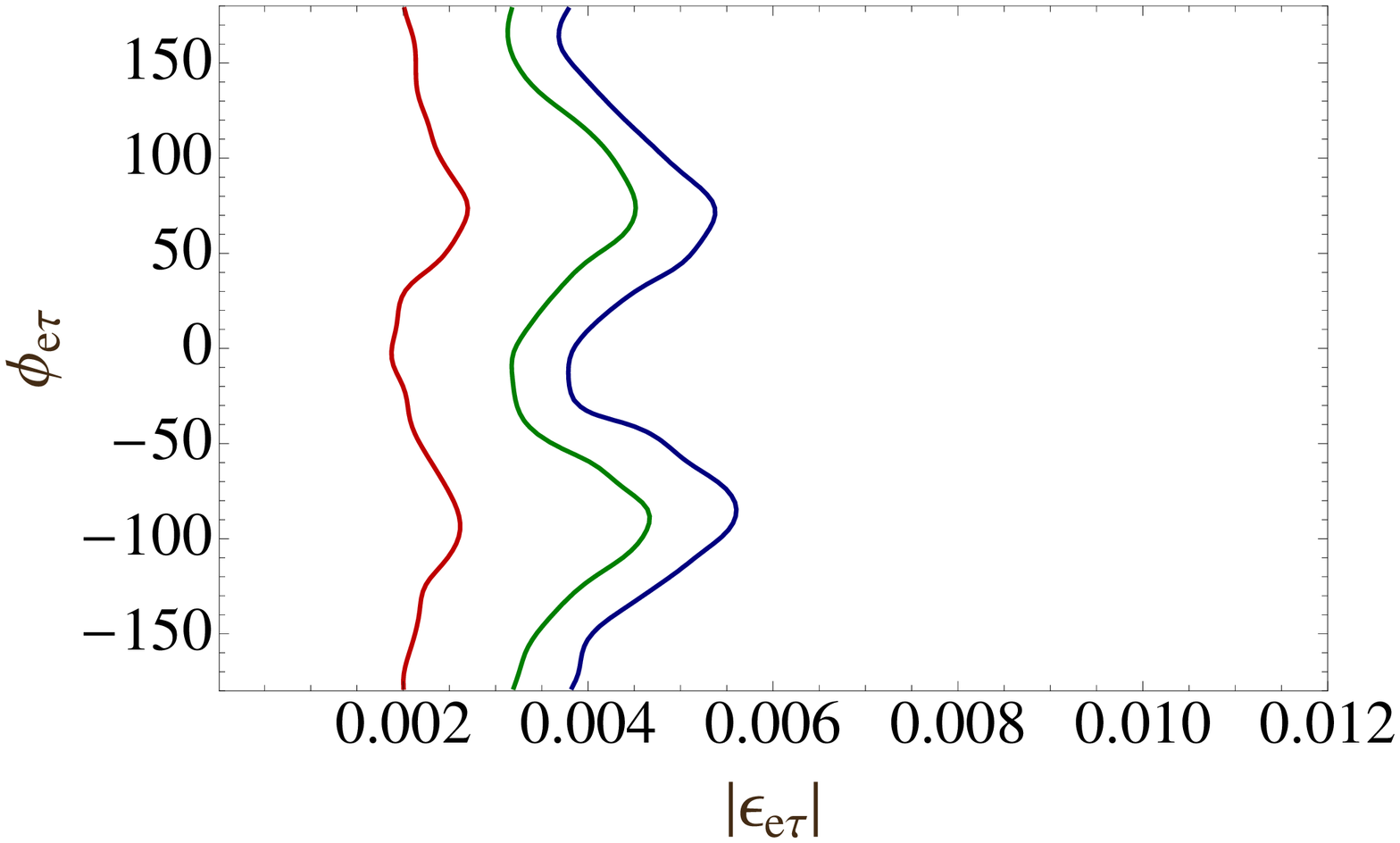} &
  \includegraphics[width=0.33\textwidth,angle=0]{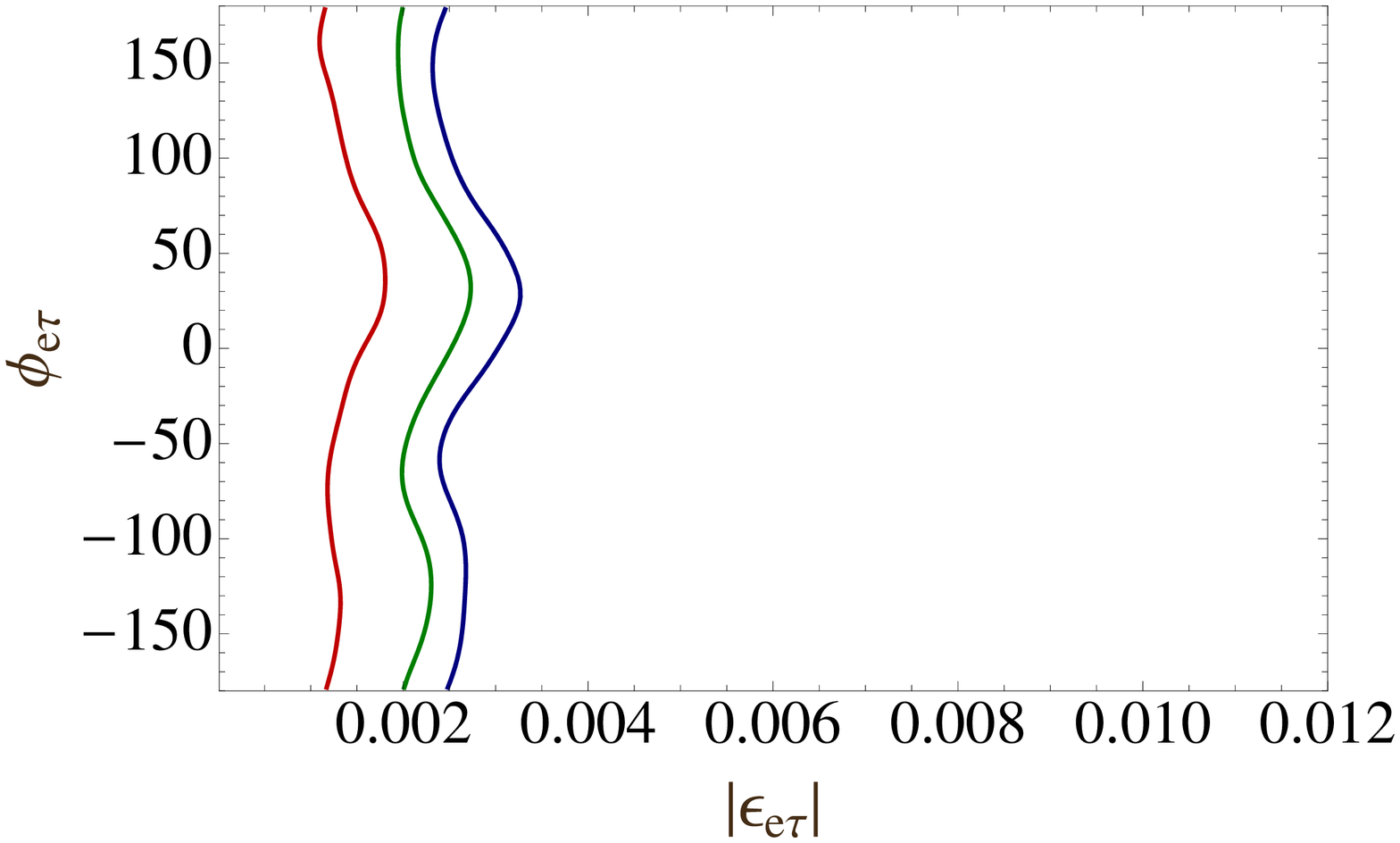} &
  \includegraphics[width=0.33\textwidth,angle=0]{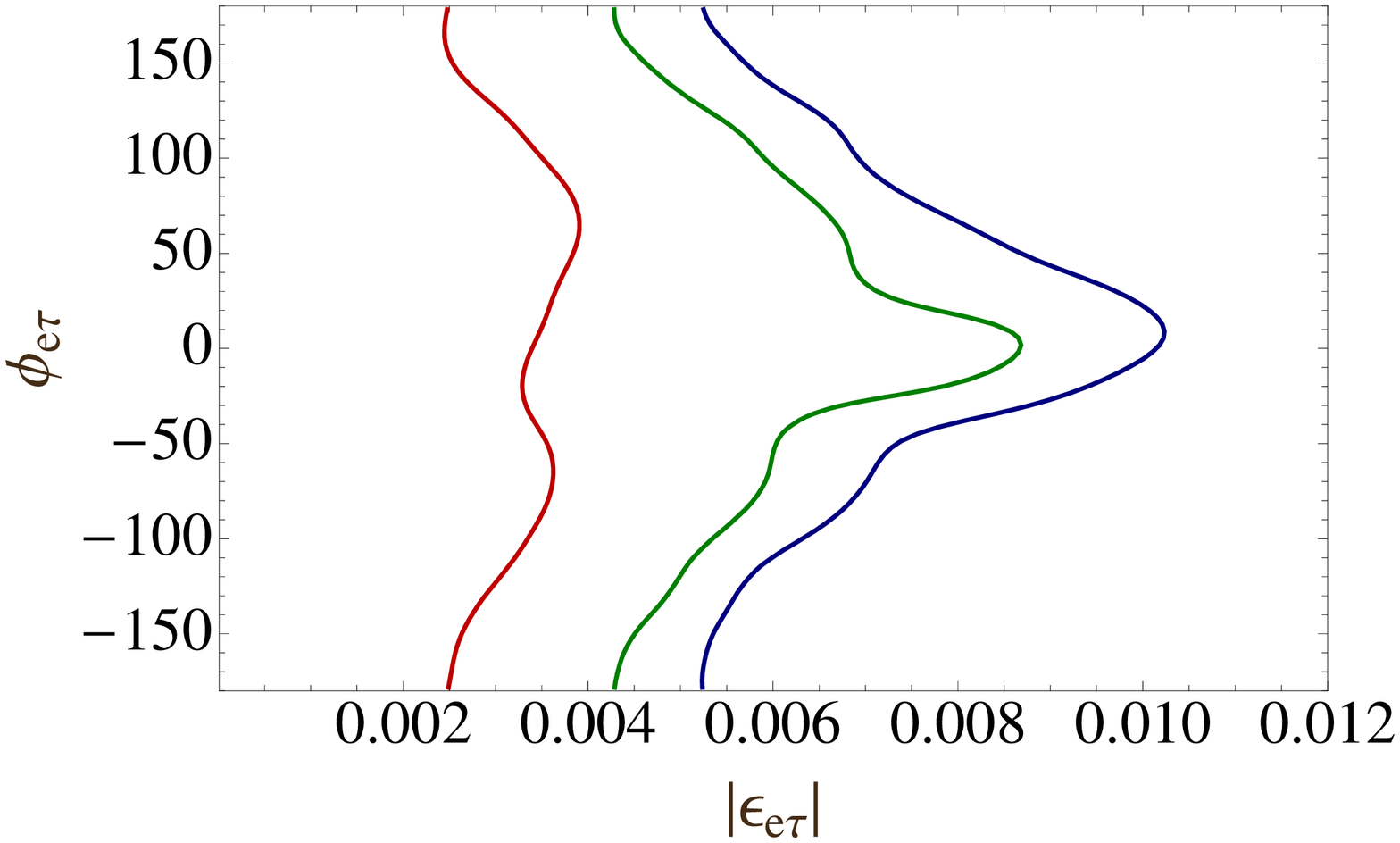} \\
  \includegraphics[width=0.33\textwidth,angle=0]{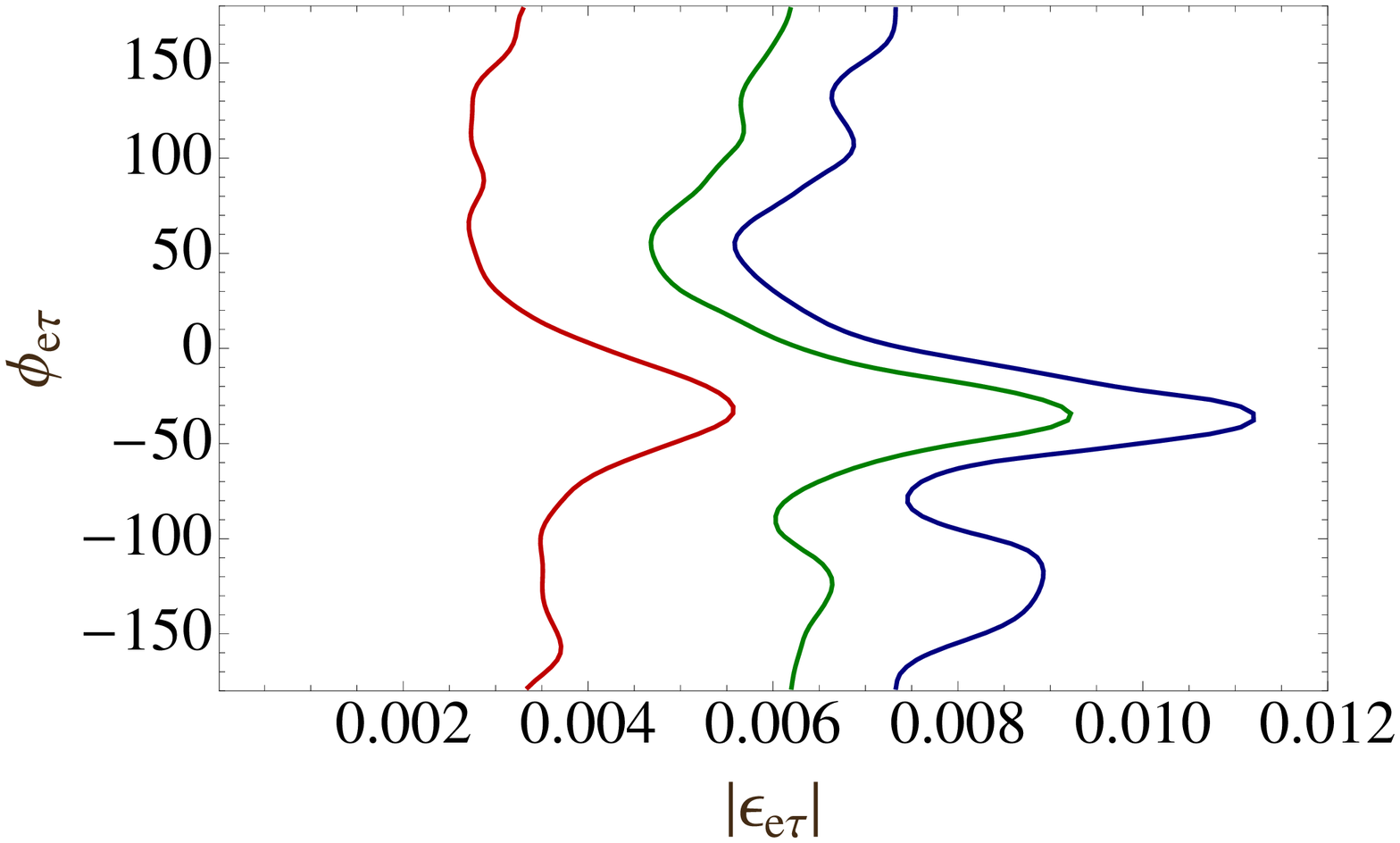} &
  \includegraphics[width=0.33\textwidth,angle=0]{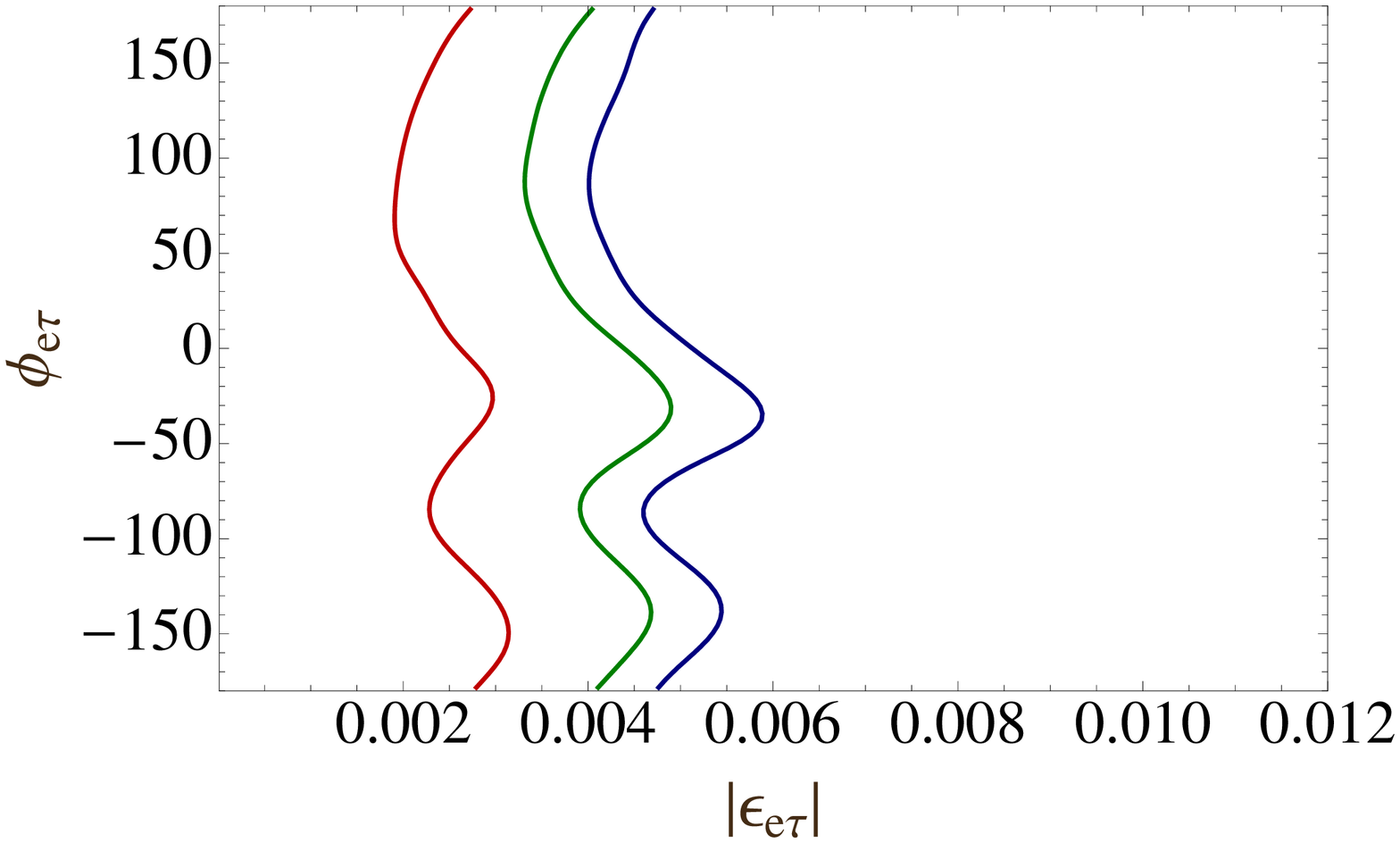} &
  \includegraphics[width=0.33\textwidth,angle=0]{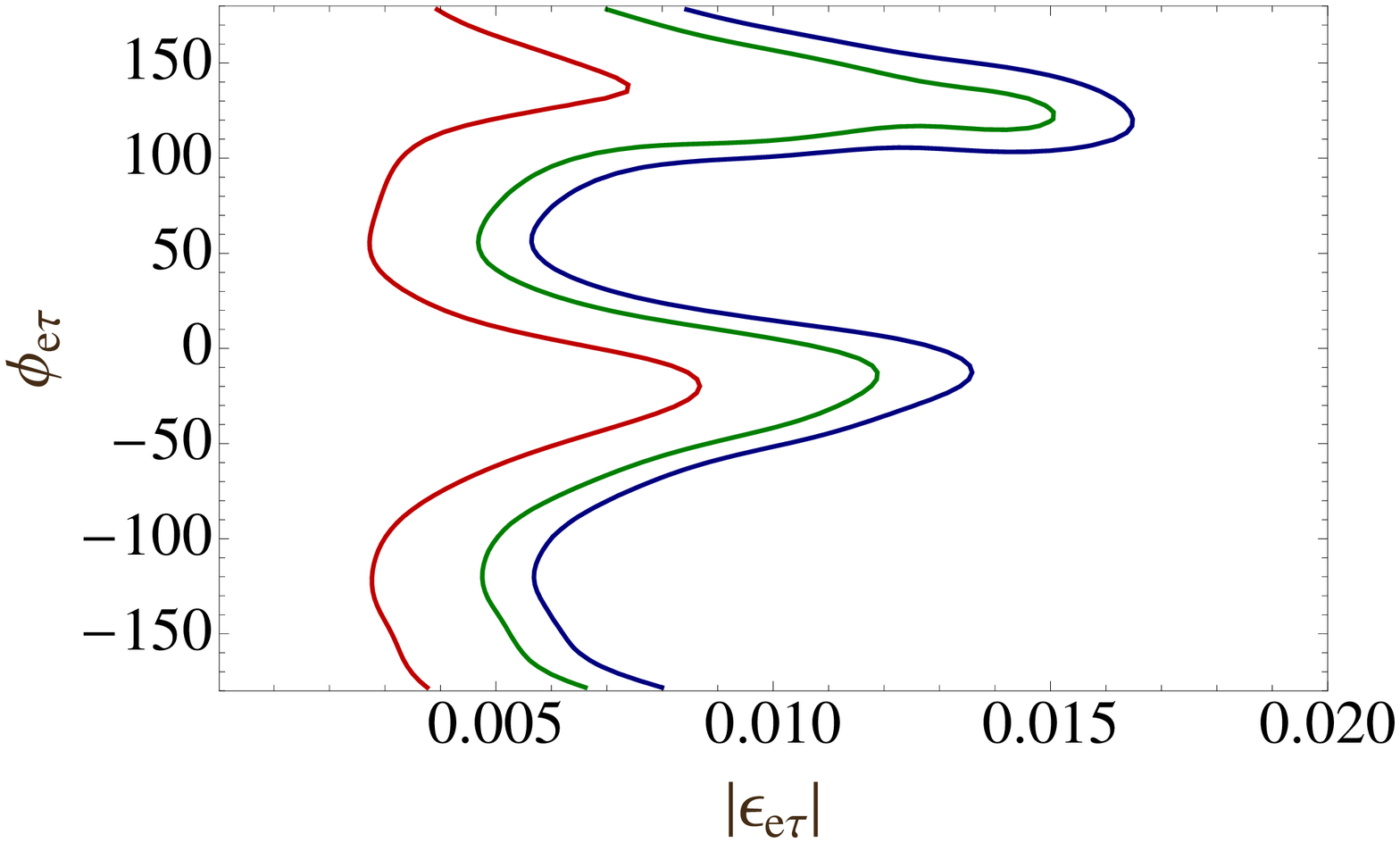} \\
\end{tabular}
\caption{\label{fig:epsetau} 68\%, 90\% and 95\% CL contours for the sensitivity to \epsetau as a function of $\phi_{e\tau}$ 
for $\bar\theta_{13} = 0$ (upper panels) and $\bar\theta_{13} = 3^\circ; \bar\delta=-\pi/2$ (lower panels). 
Marginalization has been performed over the $\nu$SM parameters, the diagonal NSI parameters $\epsilon_{\alpha\alpha}$ and $\epsilon_{e\mu}$.
The left, middle, and right panels show the results obtained for IDS25, IDS50, and 1B50 setups, respectively. 
Notice the different scale for the lower right panel, for which the sensitivity is much worse than for the rest of setups.}
\end{figure}

In Fig.~\ref{fig:epsemu} (\ref{fig:epsetau}) we present the sensitivity to $\vert \epsilon_{e \mu} \vert $ ($\vert \epsilon_{e \tau} \vert $)
for $\bar\theta_{13} = 0$ (upper panels) and $\bar\theta_{13} = 3^\circ;\,\bar\delta=-\pi/2$ (lower panels) as a function of $\phi_{e\mu}$ ($\phi_{e\tau}$), respectively.
For all panels, the standard marginalization procedure defined in Sec.~\ref{subsec:Mprocedure} is carried out, albeit neglecting marginalization over 
$\epsilon_{\mu\tau}$ (which is totally uncorrelated from $\vert \epsilon_{e \mu} \vert $ and $\vert \epsilon_{e \tau} \vert $, as we have checked). 
 
We first discuss the results in Fig.~\ref{fig:epsemu}.  The most important remark is that both high-energy setups (IDS50 and 1B50) present similar performances, 
with sensitivities that are much better than that of IDS25 both for $\bar\theta_{13}=0$ and $\bar\theta_{13}=3^\circ$. 
The fact that both IDS50 and 1B50 give sensitivities to $\vert \epsilon_{e \mu} \vert $ that are extremely similar implies that the improvement with respect to the IDS25 
is due to the increase in energy, in agreement with our expectation for preferring higher energies for NSI searches.    
It can also be seen that the dependence of the sensitivity to $\vert \epsilon_{e \mu} \vert $ on $\phi_{e\mu}$ 
is stronger for $\bar\theta_{13}\neq 0^\circ$ (bottom panels) than in the case of $\bar\theta_{13}=0$. 

In Fig.~\ref{fig:epsetau} we can see that the results obtained for the sensitivity to $\vert \epsilon_{e \tau} \vert $ are quite different from those found for $\vert \epsilon_{e \mu} \vert $. 
For vanishing $\bar\theta_{13}$, it can be seen that the IDS50 setup is better than the IDS25 and the 1B50 by a factor of $\simeq$1.7 and a few, respectively. When $\bar\theta_{13}$ is increased to $\bar\theta_{13} = 3^\circ$, the sensitivity to \epsetau becomes worse by a factor of $\simeq$2 or so, independently on the setups: the relative performance of the three setups remains almost the same as in the case of $\bar \theta_{13} = 0$. Notice that for the sensitivity to $\vert \epsilon_{e \tau} \vert $, unlike for $\vert \epsilon_{e \mu} \vert $, the synergy between the two detectors plays a key role \cite{Ribeiro:2007ud,Kopp:2008ds,Gago:2009ij}, improving the sensitivity up to an order of magnitude when a second detector at the magic baseline is considered. An additional improvement is achieved due to the increase in energy, as expected.

The oscillation probabilities presented in Appendix~\ref{sec:expandedP} can help us to understand further the results presented in Figs.~\ref{fig:epsemu} and \ref{fig:epsetau}. 
In the first line of $P_{e \mu}$ in Eq.~(\ref{Pemu}) we can observe that the two terms proportional to \epsemu appear with the same sign, while the ones proportional 
to \epsetau have opposite sign and tend to cancel. We have checked that, in the energy range relevant for the three setups under consideration, 
the coefficient of the \epsemu term can be one order of magnitude larger than the corresponding coefficient of the \epsetau term. For this reason, 
the golden channel is more sensitive to \epsemu than $\epsilon_{e\tau}$, as it can be seen by comparing Figs.~\ref{fig:epsemu} and \ref{fig:epsetau}.
For the same reason, the \epsemu sensitivity is only mildly affected by marginalization over $\epsilon_{e\tau}$, whereas
 the sensitivity to \epsetau is strongly affected by the marginalization over $\epsilon_{e\mu}$. 
This explains why the detector at the magic baseline plays an important role for the sensitivity to $\vert \epsilon_{e \tau} \vert $, whereas for \epsemu 
the energy is the key parameter independently of the number of baselines. 
Notice that the features of $P_{e \tau}$ are quite the opposite: terms proportional to \epsetau add up, while those proportional to \epsemu tend to cancel 
(see the first line of Eq.~(\ref{Petau}) in the Appendix). One could naively think, then, that a better sensitivity is expected for \epsetau instead of \epsemu in the 1B50 setup 
due to its ability to detect $\tau$'s. This is not the case, however, because the statistics in MECC is poor compared to that in MIND. The good \epsemu sensitivity observed
at the 1B50 depends on the doubled flux at the MIND section of the detector\footnote{
We will see in Sec.~\ref{sec:CP} that the silver channel can, however, play an important role in the discovery of CP violation due to NSI.
}.

A last interesting remark can be drawn in the phase dependence of the results of Figs.~\ref{fig:epsemu} and~\ref{fig:epsetau}.
If we compare the sensitivity contours for $\bar\theta_{13} = 0$ (upper panels) and $\bar\theta_{13} = 3^\circ$ (lower panels), 
we can see a shift of locations of the sensitivity minima. This feature is a result of the complicated correlations between $\delta, \phi_{e\mu}$ and $\phi_{e\tau}$ in the golden channel probability. The key factor for this effect to take place is the CP-violating value we have chosen for $\bar\delta$, which maximizes the effect.

\section{Sensitivities achieved mainly through the $\nu_\mu \rightarrow \nu_\mu$ channel} 
\label{sec:disappearance}

As we have mentioned in Sec.~\ref{sec:NSI}, sensitivity to the rest of the NSI parameters, \emph{i.e.} the diagonal elements $\epsilon_{\alpha\alpha}$ and \epsmutau, 
mainly comes from the $\nu_\mu\to \nu_\mu$ channel\footnote{
The $\nu_\mu \to \nu_\tau$ channel only increases the statistics at the detector. However, the sensitivity to $\epsilon_{\alpha\alpha}$ and $\epsilon_{\mu\tau}$ is not limited by statistics thanks to the disappearance channel. As a consequence, the $\nu_\mu\to \nu_\tau$ channel is not very useful in this context since it does not add any additional information.
}.
In this section, we will study first the expected sensitivity to $\vert \epsilon_{\mu \tau} \vert$ as a function of its CP-violating phase \phimutau (Sec.~\ref{subsec:mutau}). 
The sensitivity to the diagonal NSI parameters will be studied next (Sec.~\ref{subsec:diagonal}), taking particular care to unveil the correlations between 
$\epsilon_{\alpha\alpha}$, $\theta_{13}$ and $\theta_{23}$. 

\subsection{Sensitivity to \epsmutau}
\label{subsec:mutau}

\begin{figure}[h!]
  \includegraphics[width=0.5\textwidth,angle=0]{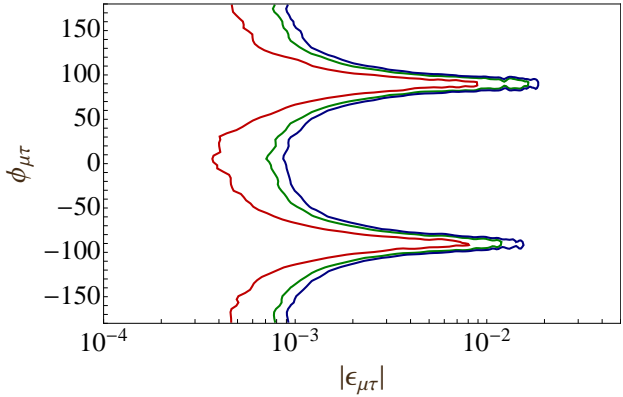}
\caption{\label{fig:epsmutau} 68\%, 90\% and 95\% CL contours for the sensitivity to \epsmutau as a function of $\phi_{\mu\tau}$ for $\bar\theta_{13}=0$. 
Marginalization has been performed over the $\nu$SM parameters, the matter density and the rest of NSI parameters. 
The 50 GeV IDS setup has been assumed. 
}
\end{figure}

In Fig.~\ref{fig:epsmutau} the sensitivity to \epsmutau for $\bar\theta_{13} =0$ is shown only for 
the IDS50 setup, since we have found remarkably similar results for the rest of setups under study. The standard marginalization procedure is employed as usual. 
The most significant feature in this figure is the extremely high sensitivity to the real part of \epsmutau (better than $10^{-3}$). The high sensitivity is driven by the leading NSI correction to the disappearence oscillation probability in Eq.~(\ref{Pmumu}):
\begin{equation}
P_{\mu\mu} = P_{\mu\mu}^\textrm{SI}  - \vert \epsilon_{\mu\tau} \vert \cos \phi_{\mu\tau}  (AL) \sin \left (\Delta_{31}L \right ) + \mathcal{O}(\varepsilon^2) + \dots 
\end{equation}
When $\epsilon_{\mu\tau}$ is mostly imaginary, $\phi_{\mu\tau} \sim \pm 90^\circ$, we observe a significant sensitivity loss of more than an order of magnitude, 
as expected from the fact that the leading dependence on $\text{Im}(\epsilon_{\mu\tau})$ appears at 
${\cal O} (\varepsilon^2)$ in the probability, as seen in Eq.~(\ref{Pmumu}). 
We have explicitly verified that fixing \epsemu and \epsetau during marginalization does not produce any appreciable change in the sensitivity to $\epsilon_{\mu\tau}$. 
This confirms the numerical results presented in Sec.~\ref{subsec:emuetau} where no correlation between \epsemu (or \epsetau) and \epsmutau was found.  
Eventually, we have checked that the \epsmutau sensitivity does not vary significantly for nonzero $\bar\theta_{13}$, in agreement with the discussion in Sec.~\ref{subsec:theta13}.

\subsection{Sensitivity to the diagonal NSI parameters}
\label{subsec:diagonal}

\begin{figure}[h!]
\begin{tabular}{cc}
  \includegraphics[width=0.36\textwidth,angle=0]{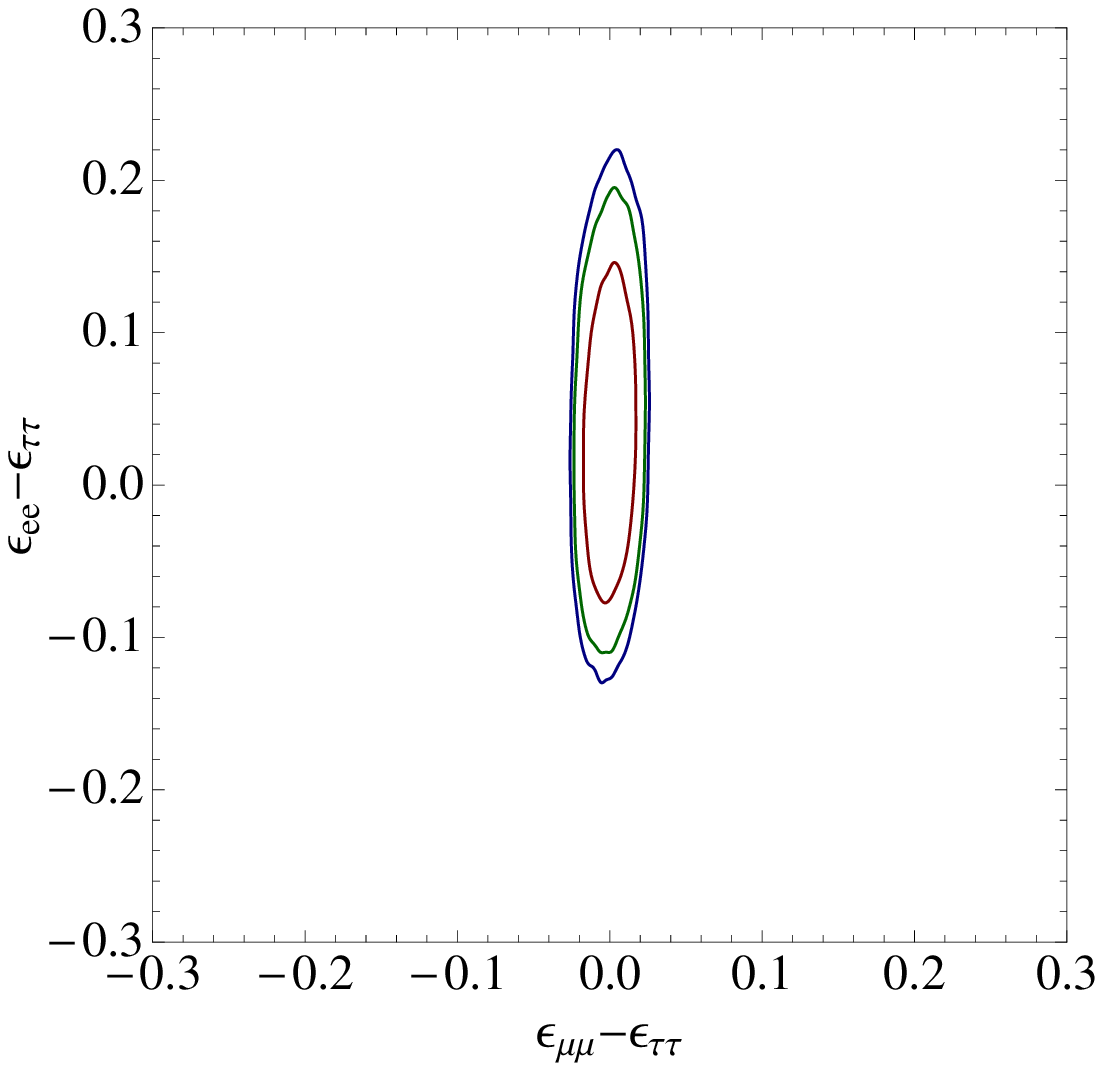} &
  \includegraphics[width=0.36\textwidth,angle=0]{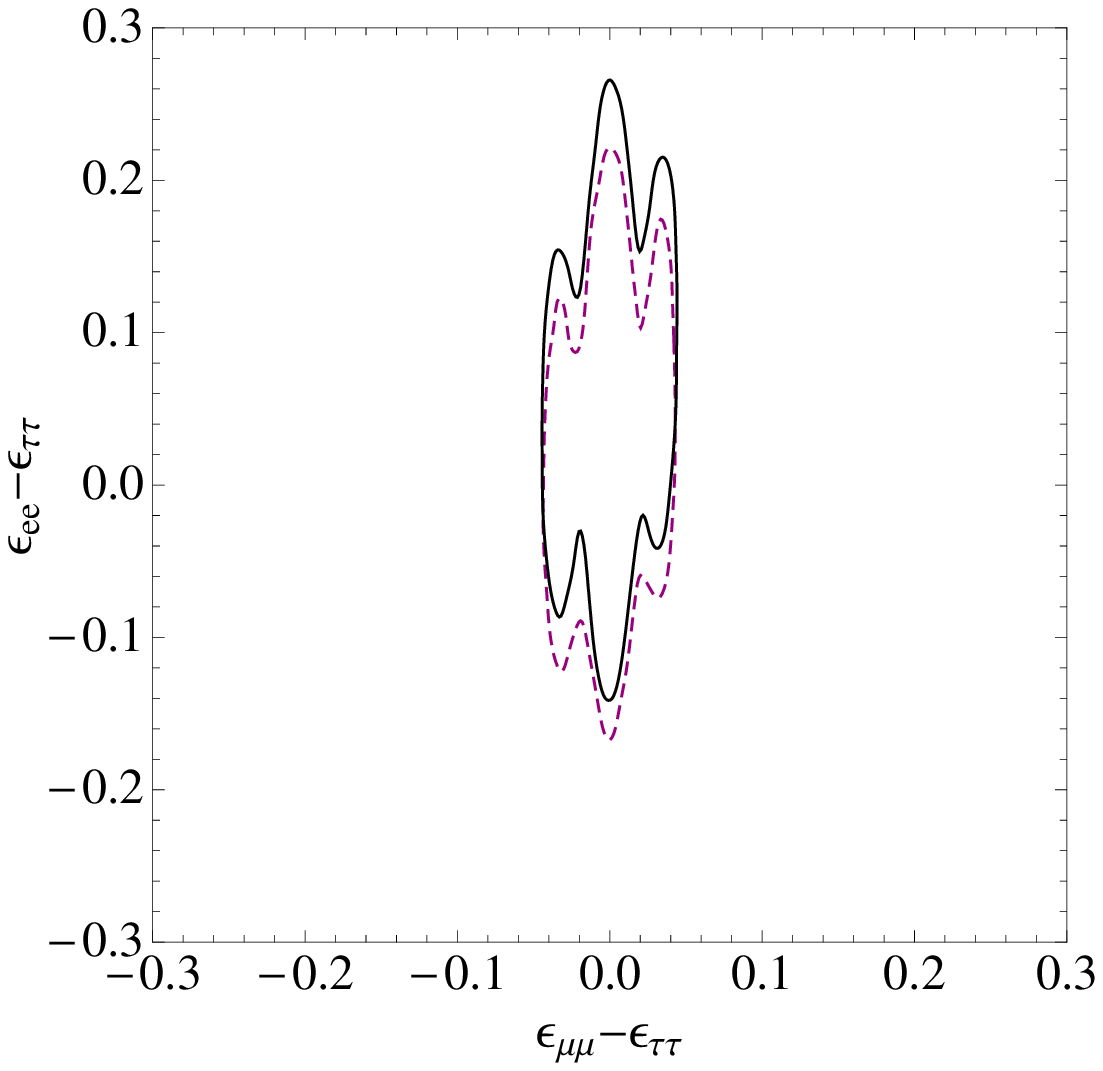} \\
 \includegraphics[width=0.36\textwidth,angle=0]{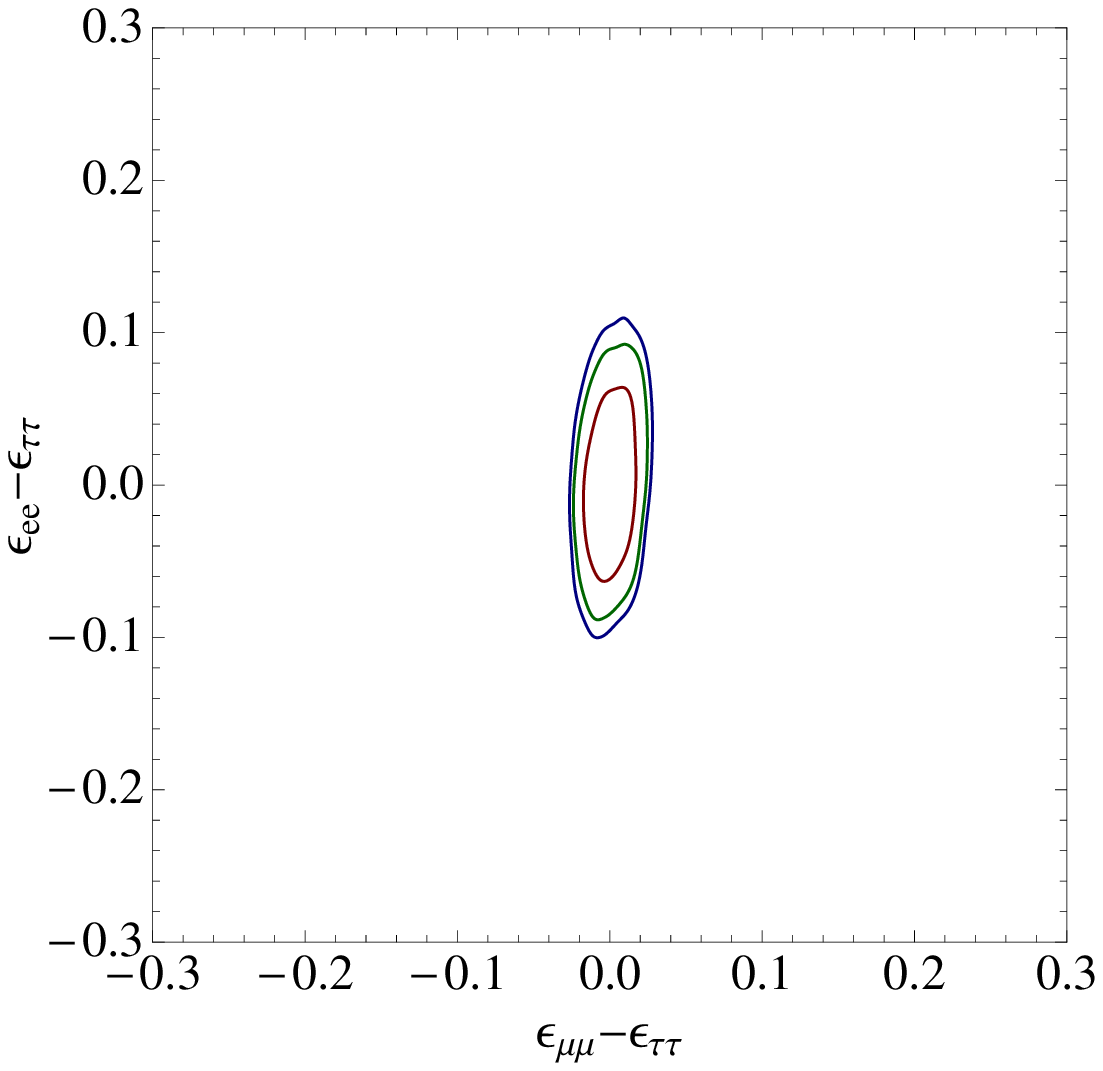} &
  \includegraphics[width=0.36\textwidth,angle=0]{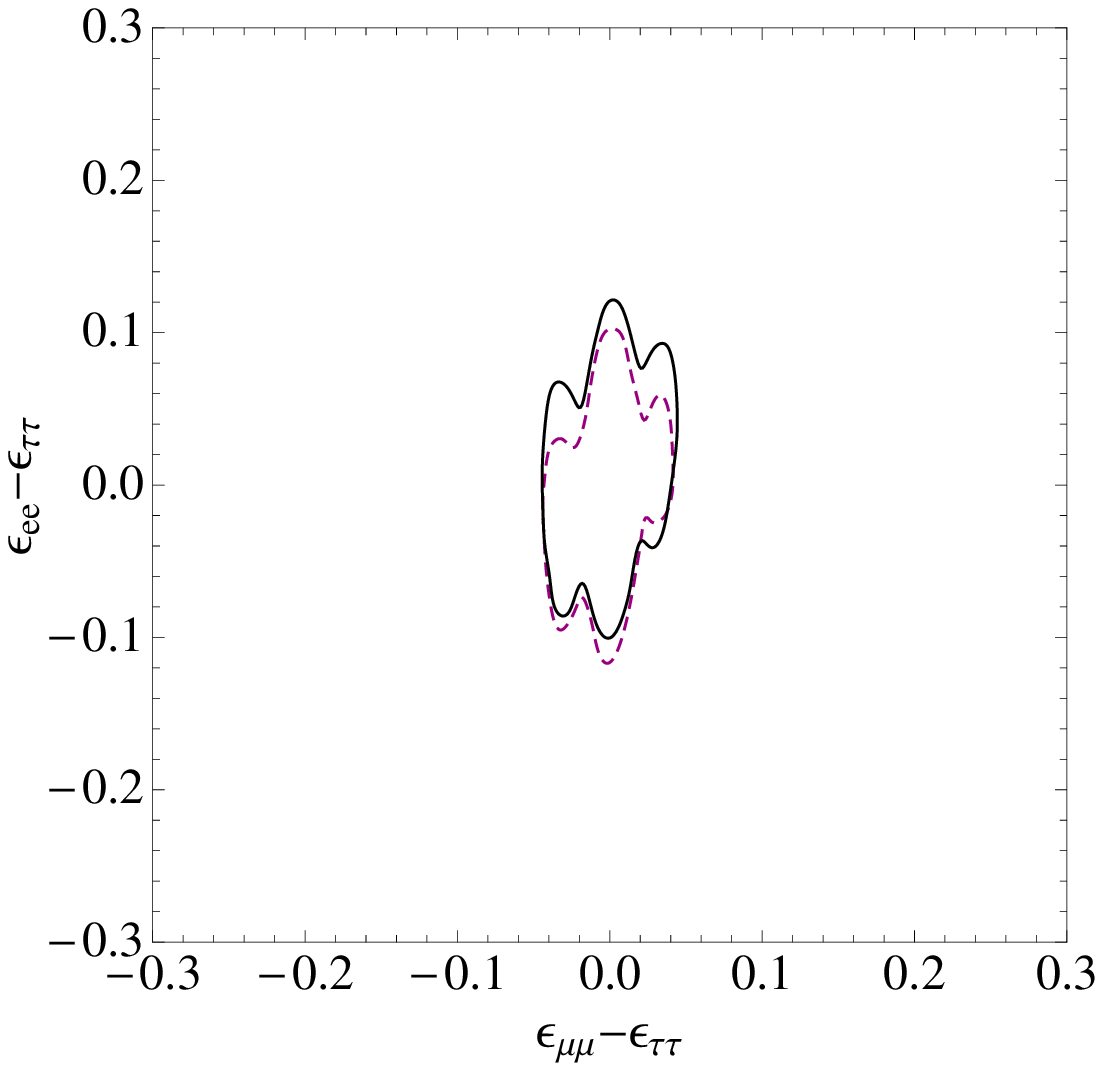}
\end{tabular}
\caption{\label{fig:diag} Sensitivity to (\epsee-\epstautau) and (\epsmumu-\epstautau) at the IDS50 setup.Top panels: $\bar \theta_{13} = 0$; bottom panels: $\bar \theta_{13} = 3^\circ$. 
Left panels: 68\%, 90\% and 95\% CL contours for $\bar\theta_{23} = 45^\circ$; 
right panels: 95\% CL contour for $\bar\theta_{23} = 43^\circ$ (purple dotted line) and $\bar\theta_{23} = 47^\circ$ (black solid line).
Marginalization has been performed over $\theta_{23}$, $\Delta m^2_{31}$, the matter density, $\theta_{13}$ and $\delta$. }
\end{figure}

In Fig.~\ref{fig:diag} we show the sensitivities to the NSI diagonal parameters obtained with the IDS50 setup for different input values of $\bar\theta_{13}$ and $\bar\theta_{23}$.
As for the \epsmutau case, the results are extremely similar for the other two setups under study, and hence their results are not shown. 
Top panels correspond to $\bar \theta_{13} = 0$, the bottom ones to $\bar \theta_{13} = 3^\circ$. Left panels are obtained for $\bar\theta_{23} = 45^\circ$, in which the red, green and the blue lines stand for the 68\%, 90\% and 95\% CL contours, respectively.
In the right panels only the 95\% CL contours are drawn, for $\bar\theta_{23} = 43^\circ$ by the purple dashed lines and $\bar\theta_{23} = 47^\circ$ by the black solid lines.
The standard marginalization is adopted here too.

First of all, we see that all panels show that the sensitivity to $(\epsilon_{ee} - \epsilon_{\tau\tau})$ is about an order of magnitude worse than the sensitivity to 
$(\epsilon_{\mu\mu}-\epsilon_{\tau\tau})$. This behaviour is in agreement with the fact that the leading dependence on the latter combination appears at $\mathcal{O}(\varepsilon^2)$ in the oscillation probabilities, while for the former one it appears at $\mathcal{O}(\varepsilon^3)$ (see \cite{Kikuchi:2008vq}).
Notice, nonetheless, that the sensitivity to (\epsee - \epstautau) (approximately $\simeq 10\%$ and $\simeq 20\%$ at 95\%CL, 2 d.o.f.'s,  for $\bar\theta_{13}=3^\circ$ and 0, respectively) is
better than  the sensitivity  achieved at any other facilities considered in the literature. 
It improves as $\bar \theta_{13}$ increases, probably due to the effect of the golden channel, 
as we can see by comparing top and bottom panels. On the other hand we have observed, in agreement 
with the discussion in Sec.~\ref{sec:NSI}, that the (\epsee - \epstautau) sensitivity is mainly 
limited by the matter uncertainty, which has been set to $5\%$ in our simulations.
In other words, unless the PREM error on the matter density is improved, a $\sim 10\%$ sensitivity to (\epsee - \epstautau) would be the limiting accuracy that could be reached\footnote{
An alternative method to constrain $\epsilon_{ee}$, which is free from this problem and is complementary to our method, is to use solar neutrinos, whose sensitivity to $\epsilon_{ee}$ appears to reach $\sim 20$\% at 1 $\sigma$CL (1 d.o.f.) \cite{Minakata:2010be}.}.

The impact of a non-maximal atmospheric mixing angle can be seen in the right panels in Fig.~\ref{fig:diag}: two narrow strips appear at both sides of the central region. 
By looking into the disappearance probability, $P_{\mu\mu}$ in Eq.~(\ref{Pmumu}), 
it is easy to realise that the sensitivity in the central region of the plots is driven through the term proportional to $(\epsilon_{\mu\mu} - \epsilon_{\tau\tau})^2$ in the disappearence channel, being the only one which does not vanish to order $\varepsilon^2$ for maximal mixing in the atmospheric sector. 
The narrow bands at both sides appear as a consequence of the non-maximal input for the atmospheric mixing angle, and they are driven by the terms proportional to $\delta \theta_{23} (\epsilon_{\mu\mu} - \epsilon_{\tau\tau})$ in $P_{\mu\mu}$. 
Indeed, the fact that these two strips appear both for $\bar\theta_{23}=43^\circ$ and $\bar\theta_{23}=47^\circ$ indicates the existence of an ``octant'' degeneracy between $\delta \theta_{23}$ and (\epsmumu - \epstautau). 

It is also remarkable that no significant correlations among $\epsilon_{\alpha\alpha}$, \epsemu and \epsetau have 
been found, as expected from the results presented in Sec.~\ref{subsec:emuetau}.

\section{Discovery potential for CP violation in the $(\phi_{e\mu}, \phi_{e\tau}, \delta)$ space}
\label{sec:CP}

One of the most interesting aspects of any system involving NSI is the possible existence of multiple sources of CP violation. 
It is important to understand characteristic features of CP violation such as correlations between the phases or possible degeneracies arising between them. 
In this paper we focus on the study of CP violation associated with the two NSI CP-violating phases \phiemu and \phietau together with the standard $\nu$SM phase $\delta$, 
which appear in the golden and silver channels. 
Depending on the values of $\theta_{13},|\epsilon_{e\mu}|$ and $|\epsilon_{e\tau}|$, strong and complicated correlations are expected to take place between these three phases. 
The third NSI phase, $\phi_{\mu\tau}$, only appears in $P_{\mu\mu}$ and $P_{\mu\tau}$ (see Appendix~\ref{sec:expandedP}) and is 
uncorrelated to the rest of the CP-phases. For this reason effects of CP violation in the $\mu-\tau$ sector will not be 
studied in this paper.

For any realistic NP model giving rise to NSI at low energies, the effects are generally not expected to be larger than 
$\mathcal{O}(10^{-2})$. For this reason, we will focus on ``reasonable'' values for $|\epsilon_{e\mu}|$ and $|\epsilon_{e\tau}|$, in the range $|\epsilon_{\alpha\beta}| \in [10^{-3},10^{-2}]$, for which the correlations with $\theta_{13}$ can still be large\footnote{
The range of values considered for the NSI moduli in this paper differ significantly from that in Ref.~\cite{Winter:2008eg}, where values for $|\epsilon_{e\tau}|$ as large as unity were considered. In addition, the number of NSI parameters in the analysis is different as well. As a consequence, the comparison of the results obtained is not straightforward. 
Some qualitative features of the results obtained in Ref.~\cite{Winter:2008eg} have been recovered, though.
}.
%
For NSI moduli smaller than $10^{-3}$, the effect of $\delta$ always dominates and it is very 
difficult to detect CP violation due to NSI. On the other hand, we will see that a very interesting structure arises in  the discovery potential in the three-dimensional parameter space 
for the input values we have considered for the NSI moduli. In particular, 
we have studied three cases: (a) both moduli are ``small'', 
$|\bar\epsilon_{e\mu}|=|\bar\epsilon_{e\tau}|=10^{-3}$,  
(b) both of them are  ``large'', $10^{-2}$, and finally (c) an ``asymmetric'' case where 
$|\bar\epsilon_{e\mu}|=10^{-3};|\bar\epsilon_{e\tau}|=10^{-2}$. Within these three cases, 
the last one is particularly interesting. Let us remind that, in the golden channel, \epsemu plays a leading role while \epsetau is subdominant (Sec.~\ref{sec:golden}). Therefore, the difference by a factor of ten in their order of magnitudes triggers interesting three-fold correlations between $\delta, \phi_{e\mu}$ and $\phi_{e\tau}$. 

\subsection{Non-standard CP violation in the absence of $\nu$SM CP violation}
\label{subsec:CPnostd}

The first question we address is  whether it is possible to detect a new CP-violating signal due to NSI in the absence of standard CP violation. 
We study, therefore, the CP discovery potential (defined in Sec.~\ref{subsec:CPfrequentist}) in the case where the input value for the standard CP-violating phase, $\bar\delta$, is set to zero or $\pi$.  
The two cases defined 
in Sec.~\ref{subsec:CPfrequentist} are considered. The first possibility stands for a relatively large value of $\theta_{13}$, $\bar \theta_{13} = 3^\circ$, which would be already measured by the time the HENF is built. In this case, we can safely use Eq.~(\ref{eq:chi2CPCmin}) to study the 
CP-discovery potential. The second possibility arises when no signal for a non-vanishing 
$\theta_{13}$ is found by the ongoing and soon-coming neutrino oscillation experiments. 
In this case, we can only conclude that $\theta_{13} \lsim 3^\circ$, and we have to use Eq.~(\ref{eq:chi2CPCth13th13barmin}) instead. 
These two cases yield the best and the worst results\footnote{
We have also checked that for $\bar \theta_{13} > 3^\circ$ our results do not change dramatically. 
} 
for the CP discovery potential in the ($\phi_{e\mu},\phi_{e\tau}$) space for $\delta=0$ and $180^\circ$.

 \begin{figure}[h!]
  \includegraphics[width=1\textwidth,angle=0]{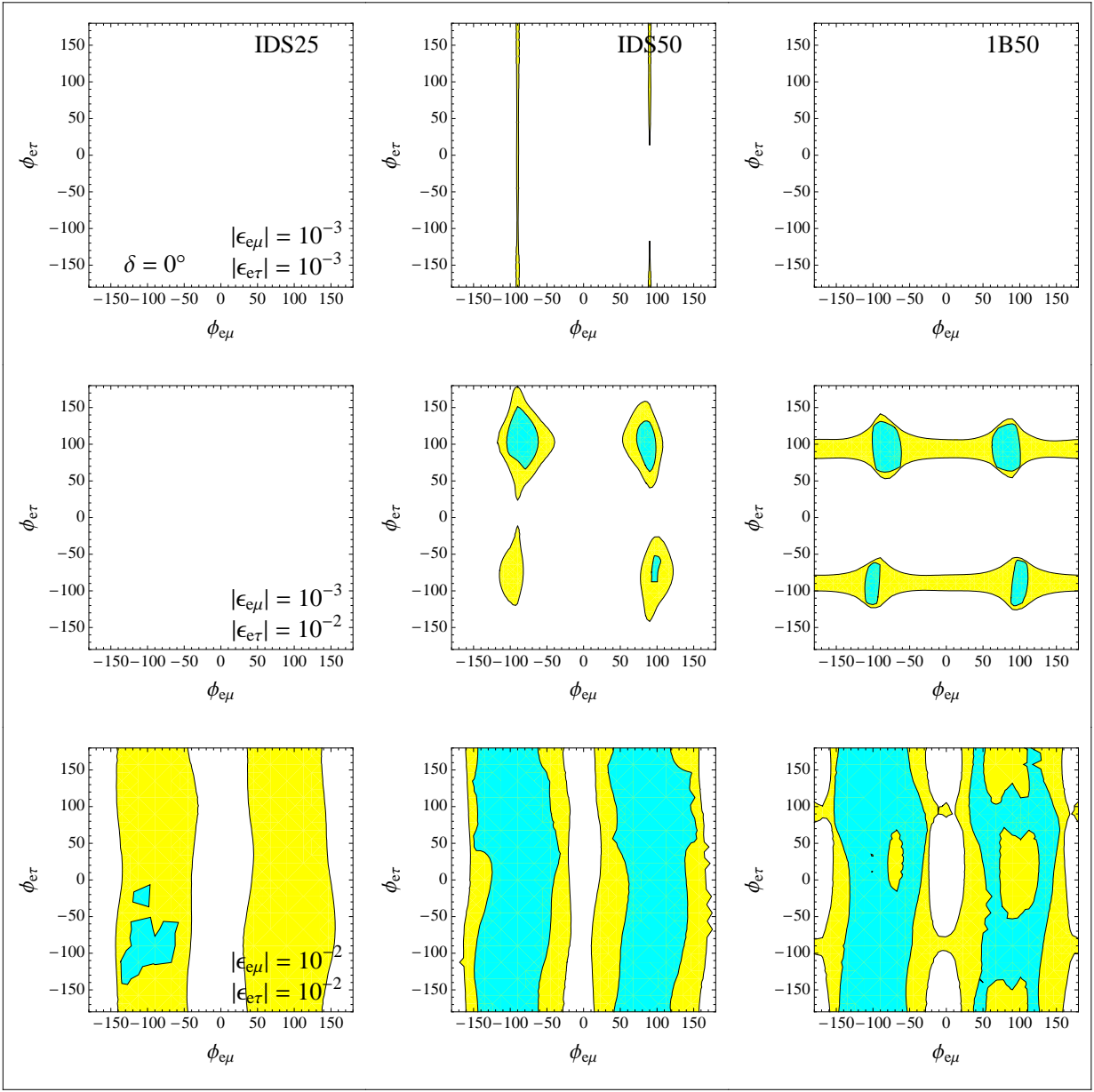}
\caption{\label{fig:NSICPviolcomp} The 99\% CL (3 d.o.f.'s) CP discovery potential in the $(\phi_{e\mu},\phi_{e\tau})$ plane for $\bar\delta = 0$. 
From top to bottom: $(|\bar\epsilon_{e\mu}|, |\bar\epsilon_{e\tau}|) = (10^{-3}, 10^{-3})$, $ (10^{-3}, 10^{-2})$ and $(10^{-2}, 10^{-2})$.
The yellow (light gray) regions  have been obtained for $\theta_{13} = \bar \theta_{13} = 3^\circ$, Eq.~(\ref{eq:chi2CPCmin}).
The cyan (dark gray) regions have been obtained after searching for intrinsic degeneracies in $\theta_{13}$ and then marginalizing over $\bar \theta_{13}$,
Eq. ~(\ref{eq:chi2CPCth13th13barmin}). 
}
\end{figure}
%
\begin{figure}[h!]
  \includegraphics[width=1\textwidth,angle=0]{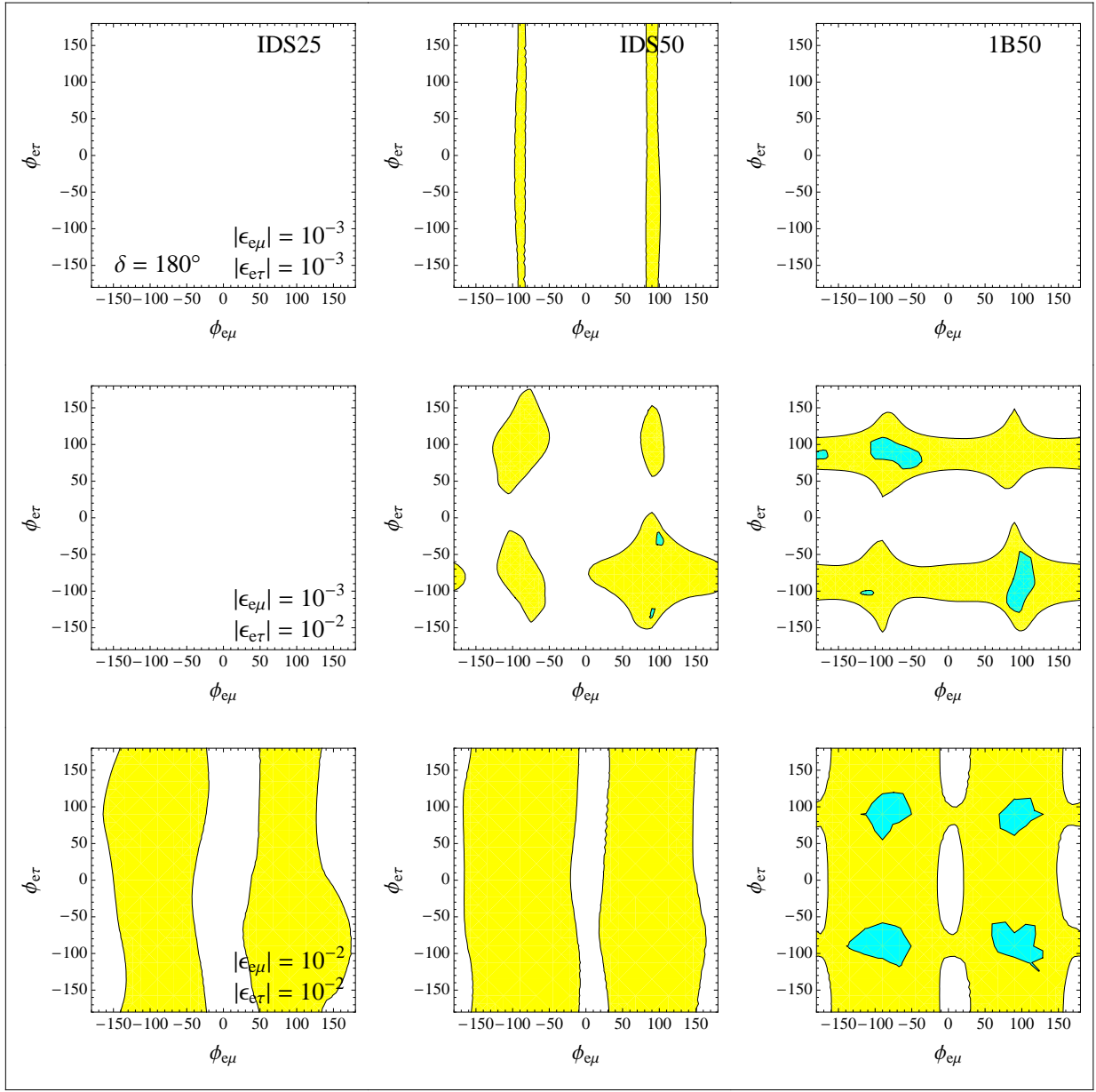}
\caption{\label{fig:NSICPviolPicomp} 
The same as Fig.~\ref{fig:NSICPviolcomp} but with $\bar\delta = 180^\circ$. 
}
\end{figure}

We show in Figs.~\ref{fig:NSICPviolcomp} and \ref{fig:NSICPviolPicomp} the CP discovery potential in the $(\phi_{e\mu},\phi_{e\tau})$ plane for $\bar\delta = 0$ and $180^\circ$, respectively. 
From left to right we show results for the IDS25, IDS50 and 1B50 setups, respectively. From top to bottom we present the CP discovery potential for the three choices of the two NSI moduli input values, $(|\bar\epsilon_{e\mu}|, |\bar\epsilon_{e\tau}|) = (10^{-3}, 10^{-3})$, $ (10^{-3}, 10^{-2})$, and $(10^{-2}, 10^{-2})$, respectively. 
 The shaded regions represent the area of the parameter space in which CP violation can be distinguished from CP conservation at the 99\% CL (3 d.o.f.). The yellow (light gray) regions have been obtained for $\bar \theta_{13} = 3^\circ$ using Eq.~(\ref{eq:chi2CPCmin}). The cyan (dark gray) regions, on the other hand, have been obtained after searching for intrinsic degeneracies in $\theta_{13}$ and then marginalizing over $\bar \theta_{13}$, using Eq. (\ref{eq:chi2CPCth13th13barmin}). 

We first discuss the case of $\bar \theta_{13} = 3^\circ$ (yellow regions). The results obtained are quite similar for $\bar\delta = 0$ and $180^\circ$. In both cases the CP discovery potential for small NSI parameters corresponding to $(|\bar\epsilon_{e\mu}|, |\bar\epsilon_{e\tau}|) = (10^{-3}, 10^{-3})$ (top row) vanishes for the IDS25 and the 1B50 setups, and it is non-vanishing only for two very small regions at $|\phi_{e\mu}| \simeq 90^\circ$, independently of the value of $\phi_{e\tau}$, for the IDS50. The IDS25 setup presents no CP discovery potential at all for the case where $(|\bar\epsilon_{e\mu}|, |\bar\epsilon_{e\tau}|) = (10^{-3}, 10^{-2})$ either. 
For this choice of NSI parameters, on the other hand, the IDS50 and the 1B50 setups show different performances: 
the former is able to discover CP violation due to NSI when both NSI phases are nearly maximal, $(|\phi_{e\mu}|, |\phi_{e\tau}|) \sim (90^\circ, 90^\circ)$; 
the latter can establish NSI-induced CP violation for $|\phi_{e\tau}| \sim 90^\circ$ (regardless of the value of $\phi_{e\mu})$. 
In the case of ``large'' NSI parameters, $(|\bar\epsilon_{e\mu}|, |\bar\epsilon_{e\tau}|) = (10^{-2}, 10^{-2})$, the three setups yield a very good CP discovery potential for $\bar \theta_{13} = 3^\circ$. In this case we observe a similar pattern of the sensitivity regions for all three setups:  most of the space is covered, apart from two strips around $\phi_{e\mu} = 0$ and $180^\circ$ whose widths mildly vary for differing setups.

As it was already mentioned in Sec.~\ref{subsec:emuetau}, the dependence on \epsemu and \epsetau in the $\nu_e \rightarrow \nu_\mu$ oscillation probability is quite different: 
the coefficient of the \epsemu term is roughly one order of magnitude larger than the corresponding coefficient of the \epsetau term in the considered range of energy and baselines.
Therefore, the behaviour of the CP discovery potential due to NSI at the IDS25 and IDS50 are primarily determined by \epsemu if $|\epsilon_{e\mu}|$ and $|\epsilon_{e\tau}|$ are comparable. This is clearly seen by the vertical bands shown in the top and bottom panels in Figs.~\ref{fig:NSICPviolcomp} and~\ref{fig:NSICPviolPicomp}: the CP discovery potential presents practically no dependence at all on \epsetau, its behaviour being dominated by \epsemu, since for these panels both parameters are of the same order of magnitude
(and, therefore, the latter is suppressed by its coefficient). In the asymmetric case, however, $(|\bar\epsilon_{e\mu}|,|\bar\epsilon_{e\tau}|)=(10^{-3},10^{-2})$ (middle row), these vertical bands disappear since now both parameters are competitive. 
A particularly interesting feature is seen in the right panels, which correspond to the performance of the 1B50 setup. 
In this case, due to the presence of the silver channel, the roles of \epsemu and \epsetau are interchanged:
two horizontal bands appear at $|\phi_{e\tau}|=90^\circ$, showing that the behaviour of the CP discovery potential is dominated by \epsetau in spite of the low statistics of the silver channel. 
This is a consequence of the enhanced role of the silver channel due to the asymmetric choice of \epsemu and \epsetau and by the size of their corresponding coefficients
(inverted with respect to the golden channel).
For $|\bar\epsilon_{e\mu}| = 10^{-3}$ we see no significant difference for $\bar \delta = 0$ or $180^\circ$ in the case $\bar \theta_{13} = 3^\circ$ (yellow regions).

Now we discuss the CP discovery potential for the case of $\theta_{13}\lsim 3^\circ$, depicted as the cyan regions in Figs.~\ref{fig:NSICPviolcomp} and~\ref{fig:NSICPviolPicomp}. The first thing we notice is that, when marginalization over $\theta_{13}$ and $\bar \theta_{13}$ is performed, the CP discovery potential is partially lost for all the 
setups under study. This effect is due to the presence of interference terms in the form $\theta_{13} |\epsilon_{e\alpha}| \times \exp{i(\delta + \phi_{e\alpha})}$ in the probabilities. 
Notice that this effect is much worse in the case of $\bar\delta=180^\circ$ than for $\bar\delta = 0$. In the former case, only small islands in the ($\phi_{e\mu},\phi_{e\tau}$) space survive 
at around the maximally CP-violating values of the NSI phases only for the 1B50 setup, while the IDS25 and the IDS50 setups show no discovery potential at all for any of the considered input values of the NSI moduli. The marked difference between the shaded regions 
 for $\bar\delta = 0$ and $\bar\delta =  180^\circ$ illuminates very well how complicated the interplay among the three CP-phases is; once we marginalize over $\theta_{13}$, flipping the sign of $e^{i \delta}$ leads to a cancellation between the standard and non-standard CP-violating contribution, which results in a heavy loss of the CP discovery potential of the facilities. This cancellation is less effective when the 1B50 setup is considered, as the silver channel is enhanced when the golden channel gets depleted and viceversa.

It is remarkable to see that high enough neutrino energies turn out to be of key importance in order to observe NSI-induced CP violation; the IDS50 setup always performs better than the IDS25, for all the input values we have considered for the NSI moduli.

To conclude this section, we point out the following two features:
firstly, as this analysis has been performed with fixed mass hierarchy, it is expected that the CP discovery potential could become worse when the sign-$\Delta m^2$ degeneracies are taken into account\footnote{
Notice, however, that in the two-baselines setups for measuring the $\nu$SM parameters the magic baseline detector is able to solve the sign degeneracy in most of the parameter space \cite{Bandyopadhyay:2007kx}. It is an intriguing question to examine to what extent it continues to hold with NSI. This question has been investigated in \cite{Gago:2009ij} but only partially
and needs to be examined further. }.
%
Secondly, a more refined analysis of the correlations between the three phases with a proper treatment of backgrounds, systematic errors and marginalization over atmospheric 
parameters should be performed to confirm robustness of the observed features.

\subsection{$\delta_{\rm CP}$ fraction:  Non-standard CP violation in presence of $\nu$SM CP violation}
\label{sec:dCPfraction}

In the previous section, we have presented the two dimensional slice of the three-dimensional ``CP sensitivity volume'' at the very particular points $\bar\delta=0$ or $180^\circ$. 
For different values of $\bar\delta$, the CP discovery potential changes dramatically due to the correlations between the three phases. We have indeed found very different features depending on the considered setup, the choice of the input values for the NSI moduli $(|\bar\epsilon_{e\mu}|,|\bar\epsilon_{e\tau}|)$, and the mixing angle $\bar \theta_{13}$; in particular, many ``holes'' appear indicating regions inside the three-dimensional parameter space for which we are not able to distinguish a CP-violating input from the CP-conserving points.  
The existence and position of these holes change for the three setups and for the considered choices of $(|\bar\epsilon_{e\mu}|,|\bar\epsilon_{e\tau}|)$ and $\bar \theta_{13}$. 
Therefore, to repeat the procedure adopted in Figs.~\ref{fig:NSICPviolcomp} and \ref{fig:NSICPviolPicomp} and draw infinitely many slices of the CP discovery potential
for different values of $\bar \delta$ would neither be practical nor shed any useful light over the intimate structure of the correlations.

\begin{figure}[h!]
  \includegraphics[width=1\textwidth,angle=0]{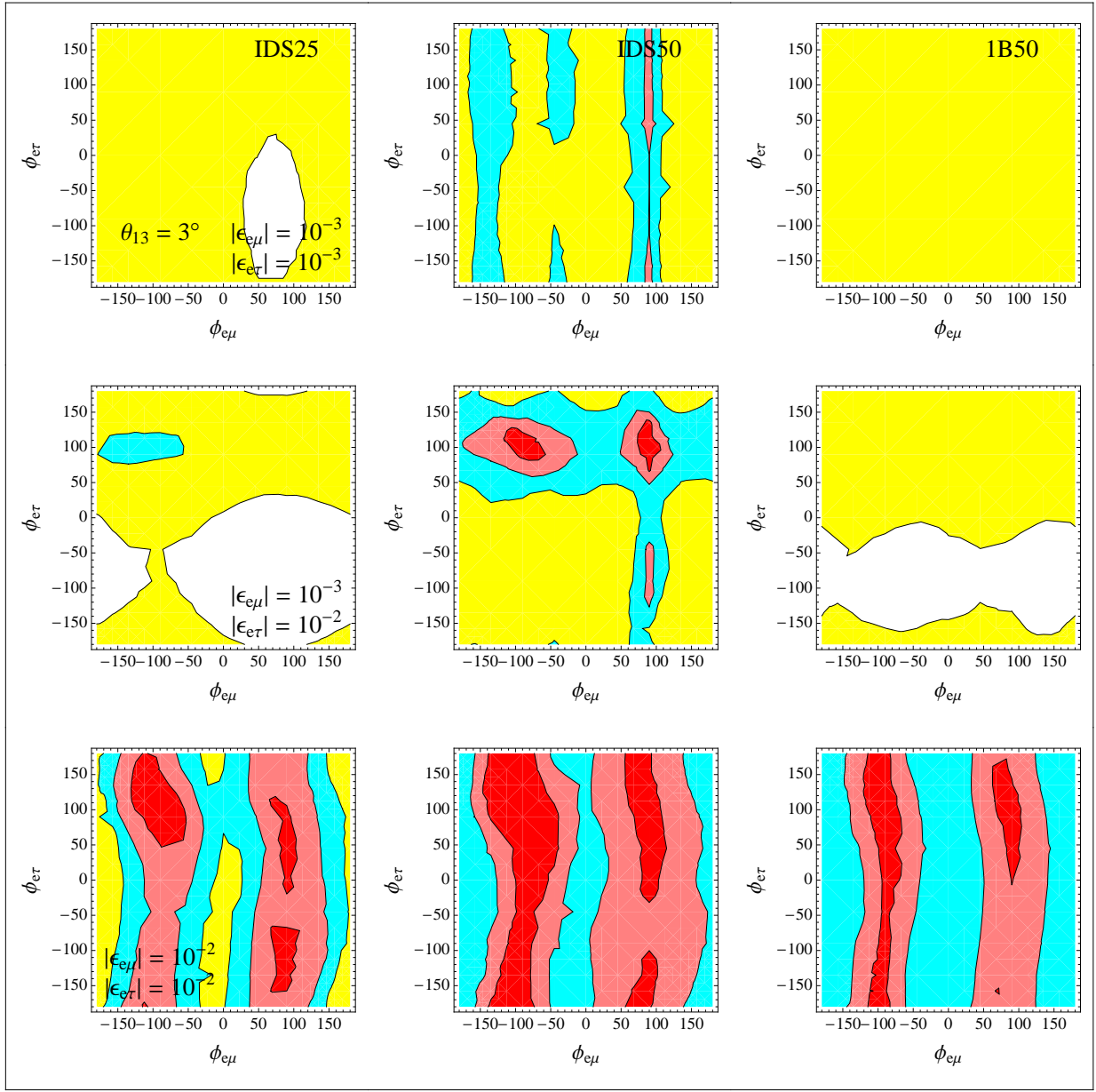}
\caption{\label{fig:NSIdCPfractionth133} Contours for the $\delta_{\rm CP}$-fraction $F_\delta$ in the $(\phi_{e\mu},\phi_{e\tau})$ plane (as defined in the text)
for $\bar \theta_{13} = 3^\circ$, obtained using Eq.~(\ref{eq:chi2CPCmin}). 
From left to right: results for the IDS25, IDS50 and 1B50 setups. From top to bottom: 
$(|\bar\epsilon_{e\mu}|, |\bar\epsilon_{e\tau}|) = (10^{-3}, 10^{-3})$, $ (10^{-3}, 10^{-2})$ and $(10^{-2}, 10^{-2})$.
The white, yellow, cyan, pink and red regions correspond to $F_\delta (\phi_{e\mu},\phi_{e\tau}) \leq 40\%,\,60\%,\,80\%,\,90\%$ and $95\%$, respectively.
}
\end{figure}

\begin{figure}[h!]
  \includegraphics[width=0.7\textwidth,angle=0]{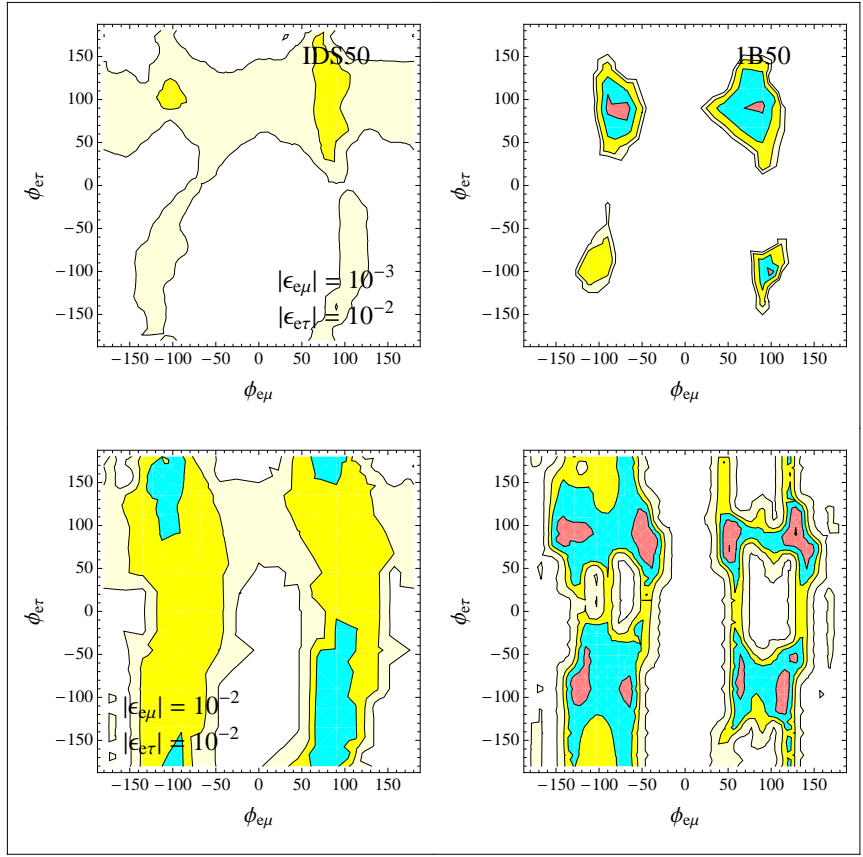}
\caption{\label{fig:NSIdCPfractionth13marg} 
Contours for the $\delta_{\rm CP}$-fraction $F_\delta$ in the $(\phi_{e\mu},\phi_{e\tau})$ plane (as defined in the text)
after searching for intrinsic degeneracies in $\theta_{13}$ and marginalizing over $\bar \theta_{13}$, using Eq.~(\ref{eq:chi2CPCth13th13barmin}). 
Left (right) panels: results for the IDS50 (1B50) setup. Top panels: $(|\bar\epsilon_{e\mu}|, |\bar\epsilon_{e\tau}|) =  (10^{-3}, 10^{-2})$; 
bottom panels: $(|\bar\epsilon_{e\mu}|, |\bar\epsilon_{e\tau}|) =  (10^{-2}, 10^{-2})$. 
The white, light yellow, yellow, cyan, pink and red regions correspond to 
$F_\delta (\phi_{e\mu},\phi_{e\tau}) \leq 5\%,\,40\%,\,60\%,\,80\%,\,90\%$ and $95\%$, respectively.
}
\end{figure}

We therefore introduce a new quantity, the ``{\it $\delta_{\rm CP}$-fraction} contour in the $(\phi_{e\mu},\phi_{e\tau})$ space'', to condense the information. 
This quantity,  denoted as $F_\delta (\phi_{e\mu},\phi_{e\tau})$, is defined as the fraction of possible values of $\bar\delta$ which fall into the region where CP violation can be established at the 99\% CL (3 d.o.f.) for a certain point 
in the $(\phi_{e\mu}, \phi_{e\tau})$ space. Notice that this is nothing but the usual CP-fraction redefined on the 
two-dimensional plane\footnote{
The concept of CP-fraction was introduced in Refs.~\cite{Huber:2004gg,Huber:2005jk} to compare in a condensed form the performances of different proposals regarding the measurement of a given observable. It is defined as the fraction of the $\delta$-parameter space (\emph{i.e.}, the fraction of $2\pi$) for which a given setup is able to perform a given task. 
} 
%
(\phiemu,\phietau). Such contours are shown for $\bar\theta_{13}=3^\circ$ in Fig.~\ref{fig:NSIdCPfractionth133}, using Eq.~(\ref{eq:chi2CPCmin}), and in Fig.~\ref{fig:NSIdCPfractionth13marg}, using Eq.~(\ref{eq:chi2CPCth13th13barmin}). 

In Fig.~\ref{fig:NSIdCPfractionth133}, the white, yellow, cyan, pink and red regions correspond to $F_\delta (\phi_{e\mu},\phi_{e\tau}) \leq 40\%,\, 60\%,\,80\%,\,90\%$ and $95\%$, respectively, for fixed $\theta_{13}=\bar\theta_{13}=3^\circ$. 
Globally, the IDS50 setup has the best CP discovery potential, while the IDS25 and 1B50 setups yield comparable results (the 1B50 performance being slightly better). The top and bottom panels, which correspond to $(|\bar\epsilon_{e\mu}|, |\bar\epsilon_{e\tau}|) = (10^{-3}, 10^{-3})$ and $(10^{-2}, 10^{-2})$, respectively, display similar features as those we have seen for 
 $\bar\delta = 0$ or $180^\circ$. $F_\delta (\phi_{e\mu},\phi_{e\tau})$ larger than 40\% (60\%) is achieved in almost the whole plane for all the setups in the case of ``small'' (``large'') NSI parameters. A $F_\delta (\phi_{e\mu},\phi_{e\tau})$ greater than 80\% is achieved for all the setups under study around $|\phi_{e\mu}| \sim 90^\circ$ for ``large'' NSI input values.

On the other hand, the result for asymmetric NSI moduli, $(|\bar\epsilon_{e\mu}|, |\bar\epsilon_{e\tau}|) = (10^{-3}, 10^{-2})$, is peculiar.
We can see that the CP discovery potential for the IDS25 and the 1B50 setups is worse than for ``small'' NSI moduli, since a 
larger white region with $F_\delta (\phi_{e\mu},\phi_{e\tau}) \leq 40\% $ arises for $\phi_{e\tau} \sim -90^\circ$. At the same time, the discovery potential for the IDS50 setup is maximal at $\phi_{e\tau} \sim 90^\circ$. The fact that the main features of $F_\delta (\phi_{e\mu}, \phi_{e\tau})$ are determined by $\phi_{e\tau}$ rather than by $\phi_{e\mu}$ appears to be again a consequence of the asymmetric choice of NSI parameters. 
Having $|\bar\epsilon_{e\tau}|$ an order of magnitude larger than $|\bar\epsilon_{e\mu}|$, the two parameters play an equally important role and huge correlations arise between them, which could explain the absence of a peak in sensitivity at $\phi_{e\mu} \simeq \phi_{e\tau}  \simeq -90^\circ$ for the IDS50 setup. When the two parameters are of the same order, $\epsilon_{e\mu}$ always dominates over $\epsilon_{e\tau}$, in agreement with our previous results. A comparison between left panels in Fig.~\ref{fig:NSIdCPfractionth133} and, for instance, Fig.12 of Ref.~\cite{Bernabeu:2010rz} indicates that marginalization over NSI hurts the CP sensitivity in a significant way.

In Fig.~\ref{fig:NSIdCPfractionth13marg} we show contours corresponding to $F_\delta (\phi_{e\mu},\phi_{e\tau}) \leq  5\%, 40\%, 60\%, 80\%$ and $95\%$ in white, light yellow, yellow, cyan, pink and red, respectively, after marginalization over $\theta_{13}$ and $\bar\theta_{13}$ below $3^\circ$, using Eq.~(\ref{eq:chi2CPCth13th13barmin}). 
Results for the IDS25 setup are not presented because we have found vanishing $F_\delta$ for any choice of $|\bar\epsilon_{e\mu}|$ and $|\bar\epsilon_{e\tau}|$ under consideration. 
Similarly, we do not show any results corresponding to $(|\bar\epsilon_{e\mu}|, |\bar\epsilon_{e\tau}|) = (10^{-3}, 10^{-3})$ either, since we have found vanishing $F_\delta$ for all the three setups.  
Notice that we have included in this figure an additional contour for $F_\delta < 5\% $ in order to achieve a better resolution in small $F_\delta$ regions. 
When we compare Fig.~\ref{fig:NSIdCPfractionth13marg} to Fig.~\ref{fig:NSIdCPfractionth133} we observe some contrived features. 
That is, the CP-fraction obtained for the ISD50 is not always better than that for 1B50.  In certain limited regions the CP-fraction is actually larger for the 1B50 setup, 
but the area covered by the regions where some sensitivity to CP violation is achieved (light yellow) is larger for the IDS50 setup. 
It appears that this feature is an outcome of the complicated correlations between these phases. 

\subsection{CP volume fraction}
\label{sec:cpcoverage}

If we focus on the most general possible case in which CP violation comes both from the $\nu$SM as well as from NSI, it would be interesting to understand its global features. 
For this purpose we define a new quantity which we call the ``{\it CP volume fraction}''. This is defined as the fraction of volume in which CP-violating signal can be distinguished, at a given CL, from a CP-conserving one in the three-dimensional parameter space spanned by $\delta, \phi_{e\mu}$ and $\phi_{e\tau}$. We use 99\% CL (3 d.o.f.) to define the CP volume fraction presented in the figures in this subsection. 

\begin{figure}[h!]
  \includegraphics[width=1\textwidth,angle=0]{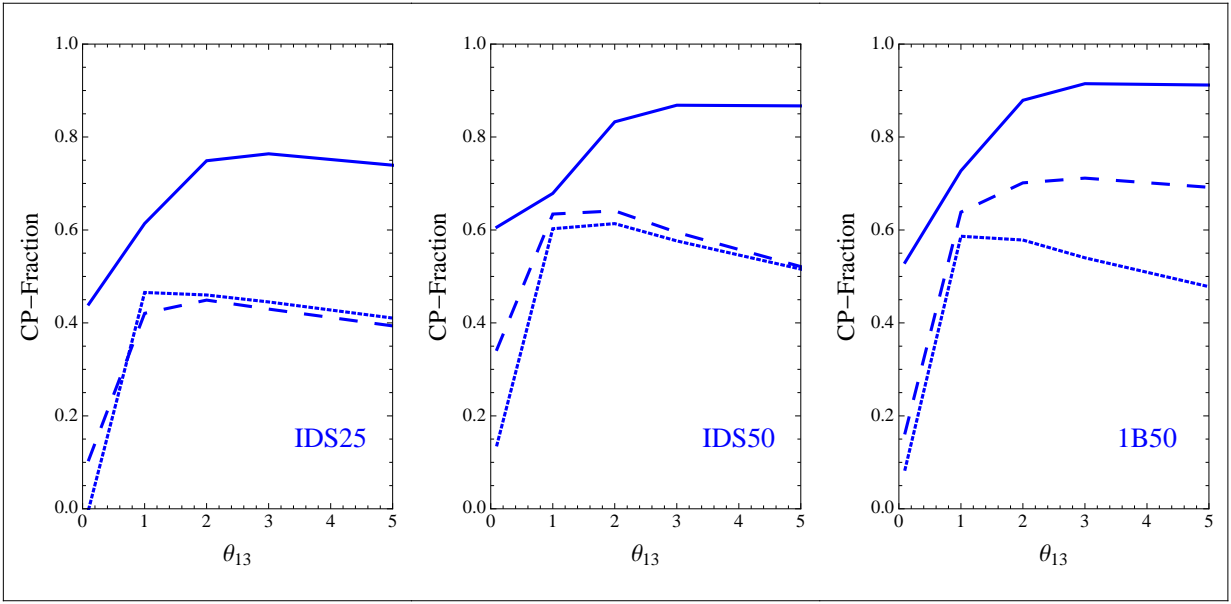}
\caption{\label{fig:comparisoncpvolume} Comparison of the CP volume fractions (as defined in the text)  as a function of $\bar\theta_{13}$, using Eq.~(\ref{eq:chi2CPCmin}).
From left to right: IDS25, IDS50 and 1B50 setups.
Results for  $(|\bar\epsilon_{e\mu}|, |\bar\epsilon_{e\tau}|) = (10^{-3},10^{-3}), (10^{-3},10^{-2})$ and $(10^{-2},10^{-2})$ are represented by dotted, dashed and solid lines, respectively.
 }
\end{figure}

In Fig.~\ref{fig:comparisoncpvolume}, we present the results of the CP volume fraction computed using Eq.~(\ref{eq:chi2CPCmin}) for the three setups and the three choices (``small'', ``asymmetric'' and ``large'', as usual) as a function of $\bar\theta_{13}$.
Consider first the IDS25 and IDS50 setups: we can see that, at both setups, the CP volume fraction achievable for $(|\bar\epsilon_{e\mu}|, |\bar\epsilon_{e\tau}|) = (10^{-3},10^{-3})$ and $ (10^{-3},10^{-2})$ are extremely similar. This means that the CP discovery potential for both setups is dominated by $|\bar\epsilon_{e\mu}|$
and the effect of $|\bar\epsilon_{e\tau}|$ is marginal, in agreement with the results obtained in previous sections. On the other hand, we see that when $|\bar\epsilon_{e\mu}| = 10^{-2}$ (solid line), 
the CP volume fraction rises abruptly to approximately 80\% or above for $\bar \theta_{13} \geq 2^\circ$.

This is not the case for the 1B50 setup: we can see that in the case of ``small'' NSI parameters (dotted line) it shows a similar (but slightly worse) sensitivity  than that of the IDS50. 
The CP volume fraction is approximately 50\% for $\bar \theta_{13} \geq 2^\circ$. However, as soon as $|\bar\epsilon_{e\tau}| = 10^{-2}$, it rises to 70\%, and reaches 90\% for larger $|\bar\epsilon_{e\mu}| = 10^{-2}$ at larger $\theta_{13}$.
The behaviour clearly indicates that the 1B50 setup is taking advantage of the combination of both channels in an efficient way.

Another interesting observation that we can draw from Fig.~\ref{fig:comparisoncpvolume} concerns the $\theta_{13}$-dependence of the CP volume fraction. Firstly, the CP volume for $\theta_{13}=0$ reflects the ability of each setup of observing CP violation exclusively due to NSI. While for the IDS25 setup the CP volume is nonzero only when $|\bar\epsilon_{e\mu}|$ is ``large'', the higher energy setups would be able to measure a CP violating signal even in the case of ``small'' NSI parameters. This confirms the results presented in Sec.~\ref{subsec:CPnostd} (Figs.~\ref{fig:NSICPviolcomp} and \ref{fig:NSICPviolPicomp}).
Secondly, we can see in the ``small'' NSI parameters case, and also in the ``asymmetric'' case for the IDS25 and IDS50 setups, that the CP volume fraction decreases at ``large'' $\theta_{13}$, reaching a maximum for 
$\bar \theta_{13} \in [1^\circ,2^\circ]$. It means that, when the NSI parameters are rather small, correlations with a ``large'' $\theta_{13}$ can actually reduce the CP-discovery potential of those facilities. This is no longer true when the NSI parameters are set to be ``large": in this case, the CP volume fraction increases for increasing $\theta_{13}$ value but remains practically unchanged for $\theta_{13} \geq 2^\circ$. No destructive correlations arise when the NSI parameters are sufficiently ``large". 

\begin{figure}[h!]
  \includegraphics[width=1\textwidth,angle=0]{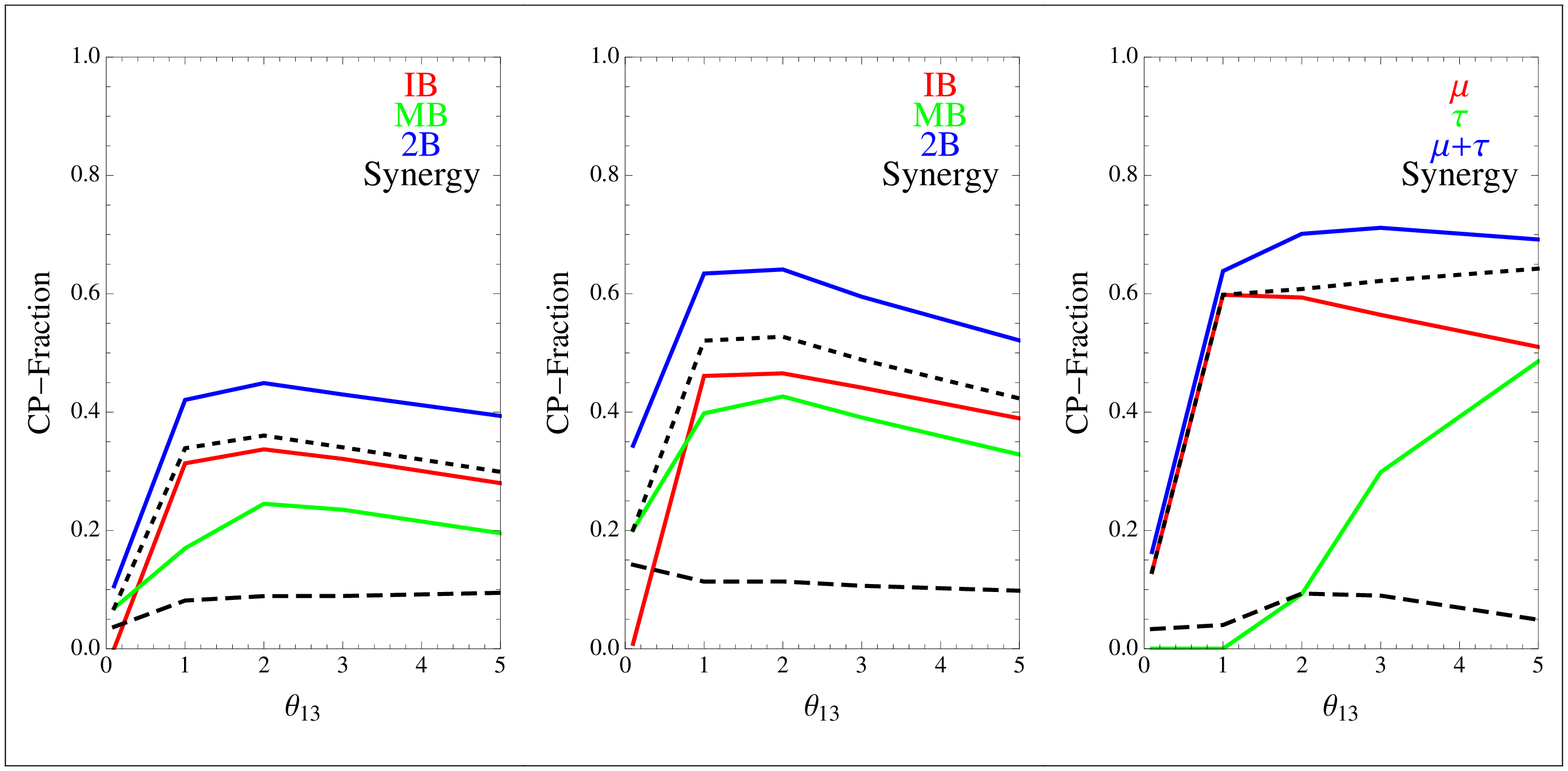}
\caption{\label{fig:synergies} Comparison of the CP volume fractions (as defined in the text) as a function of $\bar \theta_{13}$ 
for $(|\bar\epsilon_{e\mu}|, |\bar\epsilon_{e\tau}|) =  (10^{-3},10^{-2})$. 
From left to right: IDS25, IDS50 and 1B50 setups.
Results for the first detector (MIND at the Intermediate Baseline) and for the second detector (either MIND at the Magic Baseline or MECC at the Intermediate Baseline)
are shown in red and green, respectively. Combination of the two detectors ($V_{1+2}$, see text) is shown in blue. 
The dotted black line corresponds to the simple sum of the CP volume fractions of the separate baselines or detectors, $V_{1,2}$ (see text). 
The dashed black lines represent the {\it synergy}, \emph{i.e.} $V_{1+2} - V_{1,2}$ (see text).
 }
\end{figure}

To conclude this section, we address here the issue of the relative importance of the different detectors and different baselines, and of the possible synergies between them. 
We present in Fig.~\ref{fig:synergies} the CP volume fractions computed for each detector and their combinations for the three setups. 
Only the case of ``asymmetric" NSI parameters, $(|\bar\epsilon_{e\mu}|,|\bar\epsilon_{e\tau}|) = (10^{-3},10^{-2})$, is shown. 
Notice that it is precisely the case where the synergy between channels/detectors is most important because the two NSI parameters are equally relevant in the golden channel, 
which presents the largest statistics.

Let $V_i$ be the CP volume fraction obtained by analyzing data given by a detector $i$, where $i = 1$ refers to the MIND detector at $L = 4000$ km and $i = 2$
to the MIND detector at $L  = 7500$ km (left and middle panels) or to the MECC detector at $L = 4000$ km (right panel), respectively. The volume $V_1$ is represented by
red lines, $V_2$ by green lines in all panels. The CP volume fraction obtained combining two detectors,  $V_{1+2}$, is represented by the blue lines.
Notice that $V_{1+2}$ is not the simple sum of the CP volume fractions of the two detectors, $V_1 + V_2$. 
When we combine the CP volume fractions of two detectors, we must take into account that some part of the parameter space can be covered by both detectors at the same time.
We call the latter volume fraction as $V_{\rm overlap}$, the fraction of the three-dimensional parameter space that is covered simultaneously by both detectors. 
Then the correct definition of the simple sum of the CP volume fractions of two detectors is $V_{1,2} = V_1 + V_2 - V_{\rm overlap}$. 
The volume $V_{1,2}$ is shown by the black dotted lines in Fig.~\ref{fig:synergies}. 
We can see that, in general, $V_{1,2}$ does not coincide with $V_{1+2}$, because the combination of the data from two detectors could in principle cover regions of the 
parameter space that are not accessible to each detector separately. 
This is nothing but the effect of the {\it synergy} between two detectors, an increase in the CP volume fraction due to simultaneous analysis of the data sets of two detectors, 
a procedure different from the simple sum of the sensitivities achieved by each one separately. 
The {\it synergy} $V_{1+2} - V_{1,2} $ is represented by  the black dashed lines in Fig.~\ref{fig:synergies}.

The first thing to be noticed is that the {\it synergy} is never much larger than approximately 10\% for any of the considered setups.
In the IDS25 and IDS50 setups, $V_{1,2}$ (black solid lines) almost coincides with $V_{1}$ (red lines for the intermediate detector). This means that most of the CP volume fraction due to the magic baseline detector is already covered by the intermediate detector.  
In the case of 1B50 setup the situation is different. 
For $\bar \theta_{13} \leq 1^\circ$ the MECC detector does not contribute at all, as the red, black and blue lines coincide. For $\bar \theta_{13} \geq 1^\circ$, the MECC contribution grows linearly and it starts to cooperate with the MIND detector, as expected. However, we see that $V_{1,2}$ (black) and $V_{1+2}$ (blue) are closer compared to the cases of IDS25/50. This means that the overlap region $V_{\rm overlap}$ is relatively small in this case. The two detectors are complementary, as they test different regions of the parameter space.

\section{Summary and Conclusions}
\label{sec:conclusions}

In this paper, we have performed a complete study of the effects of NSI in neutrino propagation at the High Energy Neutrino Factory (HENF). 
We paid particular attention to the correlations among the whole set of oscillation parameters, the $\nu$SM and the NSI ones. 
Our analysis is the first one which takes into account all the NSI parameters at the same time in the simulations. 
Among all the new facilities proposed and discussed in the literature, we have chosen HENF due to its high energies and very long baselines, 
as it  is likely to be best suited to study effects of NSI in propagation. 
We have examined the three setups defined in Sec.~\ref{subsec:NF-setups}:
(a) IDS25, the standard IDS setup;  (b) IDS50, a 50 GeV upgraded version of IDS25; 
and (c) 1B50, a one baseline 50 GeV setup with a composite detector capable of detecting $\nu_\mu$ as well as $\nu_{\tau}$. 
The comparison among the three different setups has been performed keeping in mind that a possible optimization of HENF 
could be performed in order to search for New Physics (NP) in case  that $\theta_{13}$ turns out to be at reach by the ongoing (or forthcoming) neutrino experiments.

Our analyses have been divided into two parts: in the first part (Secs.~\ref{sec:golden} and \ref{sec:disappearance}) the 
sensitivities to NSI parameters have been studied after a brief discussion of effects of NSI on the measurement of $\nu$SM parameters; the second part (Sec.~\ref{sec:CP}) is devoted to the analysis of the CP violation discovery potential of the setups under study associated to the three phases, $\delta$, \phiemu, and \phietau. 

Results for the first part can be summarized as follows:
\begin{itemize}
\item 
Significant correlations between $\theta_{13}$, \epsemu and \epsetau have been found at HENF setups. 
Such effects can be reduced placing a detector at the magic baseline, but they cannot be eliminated. 
The ultimate sensitivity to $\theta_{13}$ is worsened by a factor 3, 5 and 10 for the IDS50, IDS25 and 1B50 setups, respectively, 
when NSI are included in the marginalization procedure compared to the case without NSI (see Fig.~\ref{fig:sensdelta_0}). 
On the other hand, no correlations with $\epsilon_{\mu\tau}$ and only a marginal effect of $\epsilon_{\alpha\alpha}$ have been observed.
\item
The sensitivities to $\vert \epsilon_{e \mu} \vert$ and $\vert \epsilon_{e \tau} \vert$ essentially 
come from $\nu_e \rightarrow \nu_\mu$ (and, to a lesser extent, $\nu_e \rightarrow \nu_\tau$)
oscillations channels,
while the sensitivities to $\vert \epsilon_{\mu \tau} \vert$ and the diagonal NSI parameters are achieved mostly through the $\nu_\mu \rightarrow \nu_\mu$ disappearance channel. 
We have numerically checked that these two sectors are practically decoupled, since no significant correlations on their sensitivities have been found. 
The marginalization over $\epsilon_{e\tau}$ does not affect significantly the sensitivity to $\epsilon_{e\mu}$, however the converse is not true. 

\item
A higher sensitivity to $\vert \epsilon_{e \mu} \vert$ is achieved by the higher energy setups, IDS50 and 1B50.
For these setups, the sensitivity to $\vert \epsilon_{e \mu} \vert$  at 95\% CL (2 d.o.f.'s) for $\bar \theta_{13} = 0 \, ( 3^\circ)$ is in the range 
$[0.5, 1.7] \times 10^{-3}\,  ([0.7,2.5]\times 10^{-3})$, depending upon $\phi_{e\mu}$. 
The sensitivity obtained by the IDS25 setting is worse by a factor of 3 (see Fig.~\ref{fig:epsemu}).

\item 
The sensitivity to $\vert \epsilon_{e \tau} \vert$, which is worse than the sensitivity to $\vert \epsilon_{e \mu} \vert$, varies significantly between the different setups, although generally better 
for the two baseline setups. 
The smallest value of $\vert \epsilon_{e \tau} \vert$ that can be excluded at 95\% CL (2 d.o.f.'s) for $\bar \theta_{13} = 0 \, ( 3^\circ)$, depending on $\phi_{e\tau}$, 
is in the range 
$[4, 5.5] \times 10^{-3}\,  ([5.5,12]\times 10^{-3})$ for the IDS25, 
$[2, 3.3] \times 10^{-3}\,  ([4,6]\times 10^{-3})$ for the IDS50 and
$[5, 10] \times 10^{-3}\,  ([6,17]\times 10^{-3})$ for the 1B50 (see Fig.~\ref{fig:epsetau}).

\item 
The sensitivities to the diagonal parameters and to $\epsilon_{\mu\tau}$, on the other hand, are quite independent from the setup under consideration. 
The sensitivity to $(\epsilon_{ee}-\epsilon_{\tau\tau})$ is quite limited, $\sim 0.2 \, (\sim 0.1)$ for $\bar\theta_{13}=0 \, (3^\circ)$, since this parameter
shows up only at $\mathcal{O}(\varepsilon^3)$ in all the oscillation channels. The sensitivity to this combination of parameters is strongly limited by the uncertainty on the matter density in the earth. 

The sensitivity to $(\epsilon_{\mu\mu}-\epsilon_{\tau\tau})$ is approximately one order of magnitude better and reaches
$\sim 0.03 \, (\sim 0.05)$  for $\theta_{23} = 45^\circ \, (\theta_{23} = 43^\circ$ or $47^\circ$), see Fig.~\ref{fig:diag}.
The sensitivity to $|\epsilon_{\mu\tau}|$ shows a remarkable dependence on the corresponding phase \phimutau (much larger than in the case of $|\epsilon_{e\mu}|$ and $|\epsilon_{e\tau}|$). The values that can be tested range from $\mathcal{O}(10^{-2})$ to $\mathcal{O}(10^{-3})$ for $|\phi_{\mu\tau}|=90^\circ $ and $|\phi_{\mu\tau}|=0,\, 180^\circ $, respectively 
(see Fig.~\ref{fig:epsmutau}).
\end{itemize}

In the second part of this work, we have studied the potential of the HENF to discover CP violation due to the $\nu$SM phase and/or 
the NSI phases $\phi_{e\mu},\phi_{e\tau}$. This has been done by computing the discovery potential in the three-dimensional space, \emph{i.e.} the region of sensitivity to 
CP violation at the $99\% $ CL (3 d.o.f.'s) in the $(\delta,\phi_{e\mu},\phi_{e\tau})$ space.
Since \phimutau is not expected to be correlated to the above mentioned CP-phases we have not studied the 
correlations between \phimutau and them in this work.\footnote{
Notice that our results for the CP discovery potential constitutes only a 
first step toward more careful studies of the complicated correlations arising between $\delta$, \phiemu and $\phi_{e\tau}$. 
Analyses which include a proper treatment of backgrounds, systematic errors and marginalization over other parameters (with particular emphasis, in this case, 
on the sign of the mass hierarchy) should be performed to confirm the results presented here.
}

We have distinguished two different cases depending on the value of $\theta_{13}$:  either $\theta_{13}$ is at reach in the currently running (or in the next generation) neutrino experiments, or it is out of reach. In the former case $\theta_{13}$ has been fixed to be $3^\circ$ as a typical value, and in the latter, we have marginalized over $\theta_{13}$ in the allowed range, $\theta_{13}\lsim 3^\circ$.  We have chosen ``reasonable'' values for the NSI moduli, $(|\bar\epsilon_{e\mu}|,|\bar\epsilon_{e\tau}|)\in \left[ 10^{-3},\, 10^{-2}\right] $, 
paying special attention to the particular case where $|\bar\epsilon_{e \tau}|$ is an 
order of magnitude larger than $|\bar\epsilon_{e \mu}|$. In this case effects of both parameters are competitive in the golden channel 
and interesting correlations arise between them. Results obtained in this section can be summarized as follows:
\begin{itemize}
\item 
If $\delta$ turns out to be at a CP-conserving value (\textit{i.e.}, 
when CP violation is exclusively due to NSI) a strong dependence on the true values of $\theta_{13}$ and $\delta$ (either $0$ or $180^\circ$) is observed. 
For $\bar\theta_{13}=3^\circ$, CP violation can be established in most of the parameter space only for ``large'' NSI parameters, 
$(|\bar\epsilon_{e\mu}|, |\bar\epsilon_{e\tau}|) =  (10^{-2},10^{-2})$ for both $\delta = 0$ and $180^\circ$ in all the three setups. 
When marginalizing over $\bar \theta_{13}$, however, CP violation can be established in a 
significant amount of the parameter space only for $\bar\delta=0$ and only for the IDS50 and 1B50 setups. For $\bar\delta= 180^\circ$ destructive interference occurs 
and CP violation can be distinguished from CP conservation only in the points where $(|\phi_{e\mu}|,\, |\phi_{e\tau}|)=(90^\circ,\, 90^\circ)$.
\item 
For generic values of the $\nu$SM CP phase, the $\delta_{CP}$ {\it fraction} $F_{\delta}$ (defined in Sec.~\ref{sec:dCPfraction}) is larger than $60\%$ for $\bar\theta_{13}=3^\circ$ and ``large'' NSI parameters in all (most) of the phase space for the IDS50 and 1B50 (IDS25) setups.  Yet, the regions in which  $F_{\delta} > 40\%$ extend to most of the phase space even for ``small'' NSI parameters, $(|\bar\epsilon_{e\mu}|, |\bar\epsilon_{e\tau}|) =  (10^{-3},10^{-3})$, for all the setups. When marginalizing over $\bar\theta_{13}$ the CP sensitivity coverage 
becomes much poorer, though. Vanishing CP sensitivity ($F_\delta=0$) 
is found for the IDS25 setup for all the input values of the moduli. The same result is obtained  for the IDS50 and the 1B50 setups for ``small'' NSI parameters, 
and non-vanishing $F_\delta$ is found only in some regions for the ``asymmetric'' and ``large'' choices of the NSI moduli.
\item
We have also analyzed the  {\it CP volume fraction}, defined in Sec.~\ref{sec:cpcoverage}.
The relevant features are as follows: 
(1) The high energy setups have higher CP volume fraction than the IDS25 for all cases considered; 
(2) The CP volume fraction at the IDS25 and the IDS50 setups is determined by the value of $|\bar\epsilon_{e\mu}|$, whereas $|\bar\epsilon_{e\tau}|$ has no impact;
(3) For ``small'' values of the NSI parameters, the CP volume fraction shows a 
maximum at  $\theta_{13} \sim 1^\circ$ at all setups (the same is true also for ``asymmetric'' NSI in the IDS25 and IDS50 setups). For $|\epsilon_{e\tau}| = 10^{-2}$ the 1B50 setup 
has the largest CP volume fraction among the three setups, $\sim 70 \% \, (90 \%)$  for $\theta_{13} \geq 2^\circ$ for $|\bar\epsilon_{e\mu}| = 10^{-3} (10^{-2})$.
\item
Finally, we have also studied the synergy between the two detectors located at different baselines (IDS25 and IDS50) and between the golden and silver channels (1B50) for ``asymmetric'' NSI parameters, since the synergy may be most prominent in this case. We found the synergy to be most significant for the IDS50 setup, though it is still small (roughly speaking, 10\%).
\end{itemize}

To summarize the outcome of our comparison between the three setups we make the following remarks: 
(1) generally speaking, the high energy setups, IDS50 and 1B50, are better than IDS25, a naturally expected result since NSI behave as a generalized matter effect; 
(2) the former two settings have their own merit and demerit: 
their sensitivities to $|\epsilon_{e\mu}|$ are comparable, while the one to $|\epsilon_{e\tau}|$ is higher for the IDS50. An accurate measurement of $\theta_{13}$ is more robust 
to a possible obstruction by NSI for the IDS50. 
The CP violation discovery potential is higher for the 1B50 setup, 
in particular for large $\theta_{13} \sim 5^\circ$. However, the IDS50 setup may yield a better global performance than the 1B50 because of its robustness thanks to the detector located at the magic baseline.

As a final remark, it should be noted that the sensitivities achieved at the high energy setups studied in this paper are remarkable, and close to the edge
of the effects produced by some neutrino models of New Physics. If New Physics is at the TeV scale, it is quite likely that new sources of CP violation beyond that of $\nu$SM exist. 
If one interprets our exercise as an example for such generic cases, apparently, the lessons we have learned are that interplay between these phases are highly nontrivial, and any regularities between them or knowledge of the right model would be of crucial importance. 
What we have definitely learnt in this study is that higher neutrino energies (such as 50 GeV) have proven to be crucial in order to pursue these elusive NSI effects in neutrino oscillation experiments.

\vspace{1cm}
{\bf NOTE ADDED}

After submission of this paper, an excess of $\nu_{e}$ appearance events far above background was reported by the T2K \cite{T2K-new} and the MINOS \cite{MINOS-new} experiments, indicating a large value of $\theta_{13}$. A global fit which includes these new data gives $\theta_{13} \simeq 8.3^\circ$ \cite{bari-new}.
Therefore, we have performed a first scrutiny of our results under the premises of such a large value of $\theta_{13}$, which is larger than what we examined during our analysis, i.e., $\theta_{13}$ up to $5^\circ$.
The main outcomes of this preliminary analysis are the following:
\begin{enumerate}
\item
 the maximal sensitivity to $\epsilon_{e\mu}$ and $\epsilon_{e\tau}$ is not affected significantly at any of the considered setups although the results are, in general, slightly worsened; in addition, the sensitivity minima are slightly displaced in the \phietau axis with respect to the $\theta_{13} = 3^\circ$ case;
\item
 the sensitivity to $\epsilon_{\mu\tau}$ and to the diagonal parameters $\epsilon_{\alpha\alpha}$ is not affected significantly, either; the only exception being the sensitivity to 
$(\epsilon_{\mu\mu} - \epsilon_{\tau\tau})$ at the 1B50 setup, which gets worse by a factor $\sim 3$;
\item 
 from a very preliminary scan of the CP discovery potential, we get slightly worse results since a large $\theta_{13}$ gives stronger
correlations between the NSI phases and $\delta$, (as mentioned above). These preliminary results are in reasonable agreement with the extrapolation of Fig.~\ref{fig:synergies}.
\end{enumerate}

Since full understanding of effects of NSI in propagation in the case of large $\theta_{13}$ requires new theoretical machinery and the analysis can go beyond what we have presented in this paper, we leave the detailed discussion of this case to a future communication. 

\section*{Acknowledgements}

We are extremely grateful to Enrique Fern\'andez Mart\'inez and Mattias
Blennow for their help and advice related to the use of the MonteCUBES
software. We would also like to thank Osamu Yasuda and Walter Winter for
illuminating discussions.
This work was initiated in the mutual visits between us while A.D., H.M.
and J.L.P. were supported by the JSPS-CSIC Bilateral Joint Projects
(Japan-Spain) 2007-2008. P.C. and A.D. would like to acknowledge support
from Comunidad Aut\'onoma de Madrid under project
HEPHACOS-S2009/ESP-1473. P.C., A.D. and J.L.P. would like to acknowledge
support from project FPA2009-09017 (DGI del MCyT, Spain), from the
Spanish Government under the Consolider-Ingenio 2010 programme CUP,
``Canfranc Underground Physics'', project number CSD00C-08-44022, and
from the European Community under the European Comission Framework
Programme 7, Design Study: EUROnu, Project Number 212372 (the EU is not
liable for any use that may be made of the information contained
herein). J.L.P. would also like to acknowledge partial support from the European Community  
under the European Commission Framework Programme 7 Design Studies: EuCARD (European Coordination for  
Accelerator Research and Development, Grant Agreement number 227579), the UK India
Education and Research Initiative under project SA (06-07)-68 and LAGUNA (Project Number 212343). H.M.
thanks IFIC, University of Valencia, for support and hospitality
extended to him during visits in 2009 and 2010. P.C. would also like to
acknowledge finantial support from Comunidad Aut\'onoma de Madrid, and
thanks the Max Planck Institut in Munich and the Institute for Particle
Physics Phenomenology in Durham where part of this work was completed.
A.D. would also like to thank IFIC in Valencia where part of this work
was completed.

This document is an output from the UKIERI (UK India Education and
Research Initiative) project funded by the British Council, the UK
Department for Education and Skills (DfES), Office of Science and
Innovation, the FCO , Scotland, Northern Ireland, Wales, GSK, BP, Shell
and BAE for the benefit of the India Higher Education Sector and the UK
Higher Education Sector. The views expressed are not necessarily those
of the funding bodies.

\appendix
\label{appendix}

\section{Expressions of Expanded oscillation probabilities}
\label{sec:expandedP}

In this Appendix, we show the oscillation probabilities $P_{\alpha \beta}$ in matter with constant density, in presence of NSI affecting only to propagation in matter. We start from the oscillation probability expansions derived in~\cite{Kikuchi:2008vq} where $\epsilon_{\alpha\beta}$, $\theta_{13}$, and $\Delta m^2_{31}/\Delta m^2_{21}$ are considered the expansion parameters\footnote{Remember that, in the set-up considered in this paper, $A\sim\Delta_{31}$.}. Here we expand also on $\delta\theta_{23}\equiv\theta_{23}-\pi/4$, considering therefore:

\[
\varepsilon: \left\lbrace \epsilon_{\alpha\beta},\, \theta_{13},\, \Delta m^2_{31}/\Delta m^2_{21},\, \delta\theta_{23}\right\rbrace 
\]
as the order $\varepsilon$ expansion parameters. 

The oscillation probabilities for golden and silver channels at $O(\varepsilon^2)$ are given by the following formulae:
\begin{eqnarray}
\hspace{-10mm}
P_{e\mu} &=& 
\Biggl | 
A^{SM}_{e\mu}
+ \epsilon_{e\mu } 
\left[ \sin \left( \frac{AL}{2} \right) e^{-i \frac{\Delta_{31}L}{2}} 
+ \biggl (\frac{A}{\Delta_{31} -A}\biggr )\sin \left( \frac{\Delta_{31} -A}{2}L \right) \right] 
\nonumber \\
&& \hspace{20mm} 
- \epsilon_{e\tau } 
\left[ \sin \left( \frac{AL}{2} \right) e^{-i \frac{\Delta_{31}L}{2}} - 
\biggl (\frac{A}{\Delta_{31} -A}\biggr )\sin \left( \frac{\Delta_{31} -A}{2}L \right) \right]
\Biggr |^2 + O(\varepsilon^3)
\nonumber \\
&=&
\Biggl | \sqrt{2}\ci \si \frac{\Delta_{21}}{A}+ \epsilon_{e\mu }- \epsilon_{e\tau }\biggr |^2\sin ^2\frac{AL}{2} 
\nonumber \\
&+&\biggl |\sqrt{2}\st e^{-i\delta }\frac{\Delta_{31}}{A}+\epsilon_{e\mu }+\epsilon_{e\tau } \biggr |^2\biggl (\frac{A}{\Delta_{31} -A}\biggr )^2\sin ^2\frac{\Delta_{31} -A}{2}L 
\nonumber \\
&+&4\text{ Re } \biggl[ \left(\ci \si \frac{\Delta_{21}}{A}+ \frac{1}{\sqrt{2}}\left( \epsilon_{e\mu} - \epsilon_{e\tau }\right) \right)\left(\st e^{i\delta }\frac{\Delta_{31}}{A}+\frac{1}{\sqrt{2}}\left( \epsilon_{e\mu }^*+ \epsilon_{e\tau }^*\right)\right)\biggr ] 
\nonumber \\
&&\hspace{56mm} \times \frac{A}{\Delta_{31} -A}\sin \frac{AL}{2}\cos \frac{\Delta_{31} L}{2}\sin \frac{\Delta_{31} -A}{2}L 
\nonumber \\
&+&4 \text{ Im } \biggl[ \left(\ci \si \frac{\Delta_{21}}{A}+ \frac{1}{\sqrt{2}}\left( \epsilon_{e\mu }- \epsilon_{e\tau }\right) \right)\left(\st e^{i\delta }\frac{\Delta_{31}}{A}+\frac{1}{\sqrt{2}}\left(\epsilon_{e\mu }^*+ \epsilon_{e\tau }^*\right)\right)\biggr ]  
\nonumber \\
&&\hspace{56mm}\times \frac{A}{\Delta_{31} -A}\sin \frac{AL}{2}\sin \frac{\Delta_{31} L}{2}\sin \frac{\Delta_{31} -A}{2}L + O(\varepsilon^3) \, , \nonumber \\
\label{Pemu}
\end{eqnarray}
%
\begin{eqnarray}
\hspace{-10mm}
P_{e\tau}
&=&
\Biggl | 
A^{SM}_{e\tau}
+ \epsilon_{e\tau } 
\left[ \sin \left( \frac{AL}{2} \right) e^{-i \frac{\Delta_{31}L}{2}} 
+ \biggl (\frac{A}{\Delta_{31}-A}\biggr )\sin \left( \frac{\Delta_{31}-A}{2}L \right) \right] 
\nonumber \\
&& \hspace{20mm} 
- \epsilon_{e\mu } 
\left[ \sin \left( \frac{AL}{2} \right) e^{-i\frac{\Delta_{31}L}{2}} - 
\biggl (\frac{A}{\Delta_{31}-A}\biggr )\sin \left( \frac{\Delta_{31}-A}{2}L \right) \right]
\Biggr |^2 + O(\varepsilon^3)
\nonumber\\
&=&
|\sqrt{2}\ci \si \frac{\Delta_{21}}{A}+ \epsilon_{e\mu }- \epsilon_{e\tau }\biggr |^2\sin ^2\frac{AL}{2} 
\nonumber \\
&+&\biggl |\sqrt{2}\st e^{-i\delta }\frac{\Delta_{31}}{A}+\epsilon_{e\mu }+\epsilon_{e\tau } \biggr |^2\biggl (\frac{A}{\Delta_{31} -A}\biggr )^2\sin ^2\frac{\Delta_{31} -A}{2}L 
\nonumber \\
&-&4\text{ Re } \biggl[ \left(\ci \si \frac{\Delta_{21}}{A}+ \frac{1}{\sqrt{2}}\left( \epsilon_{e\mu} - \epsilon_{e\tau }\right) \right)\left(\st e^{i\delta }\frac{\Delta_{31}}{A}+\frac{1}{\sqrt{2}}\left(\epsilon_{e\mu }^*+ \epsilon_{e\tau }^*\right)\right)\biggr ] 
\nonumber \\
&&\hspace{56mm} \times \frac{A}{\Delta_{31}-A}\sin \frac{AL}{2}\cos \frac{\Delta_{31}L}{2}\sin \frac{\Delta_{31}-A}{2}L 
\nonumber \\
&-&4 \text{ Im } \biggl[ \left(\ci \si \frac{\Delta_{21}}{A}+ \frac{1}{\sqrt{2}}\left( \epsilon_{e\mu }- \epsilon_{e\tau }\right)\right) \left(\st e^{i\delta }\frac{\Delta_{31}}{A}+\frac{1}{\sqrt{2}}\left(\epsilon_{e\mu }^*+ \epsilon_{e\tau }^*\right)\right)\biggr ] 
\nonumber \\
&&\hspace{56mm}\times \frac{A}{\Delta_{31}-A}\sin \frac{AL}{2}\sin \frac{\Delta_{31}L}{2}\sin \frac{\Delta_{31}-A}{2}L + O(\varepsilon^3)  \,,
\nonumber \\
\label{Petau}
\end{eqnarray}
where $A^{SM}_{\alpha\beta}$ stands for the standard oscillation amplitude, $A$ for the matter density, $\Delta_{ij}=(\Delta m^2_{ij}/2E)$, and $s_{ij}$ and $c_{ij}$ stand for $\sin \theta_{ij} $ and $\cos\theta_{ij} $, respectively.

On the other hand, since the sensitivities to $\epsilon_{\e\mu}$ and $\epsilon_{e\tau}$ are mainly achieved through the golden and silver channels, for the $\nu_\mu-\nu_\tau$ sector we show here only the dependence on $\epsilon_{\alpha\alpha}$ and $\epsilon_{\mu\tau}$ of the relevant oscillation probabilities, which will be called $P^{NSI}_{\alpha\beta}$:  

\begin{eqnarray}
P^{NSI}_{\mu\mu} &=& - P^{NSI}_{\mu\tau} 
\nonumber\\
&=&
-\left\lbrace \delta \theta_{23}\left(\epsilon_{\mu\mu}-\epsilon_{\tau\tau}\right)+\text{Re}\left(\epsilon_{\mu\tau} \right)   \right\rbrace \left( AL\right)\sin\left(\Delta_{31}L\right)
\nonumber\\
&+&
\left\lbrace 4\delta \theta_{23}\left(\epsilon_{\mu\mu}-\epsilon_{\tau\tau}\right)\frac{A}{\Delta_{31}}+\left( \epsilon_{\mu\mu}-\epsilon_{\tau\tau}\right)^2\left( \frac{A}{\Delta_{31}}\right)^2\right\rbrace \sin^2\dfrac{\Delta_{31}L}{2}
\nonumber\\
&-&\frac{1}{2}\left(\text{Re}(\epsilon_{\mu\tau})\right)^2\left(AL \right)^2\cos\left( \Delta_{31}L\right)-\left( \text{Im}(\epsilon_{\mu\tau})\right)^2\frac{A}{\Delta_{31}}(AL)\sin\left(\Delta_{31}L\right)\,. + O(\varepsilon^3) \nonumber\\ 
\label{Pmumu}
\end{eqnarray}

The complete oscillation probabilities at quadratic order in $\varepsilon$ can be found in~\cite{Kikuchi:2008vq}.

\end{document}